\documentclass[12pt,a4paper]{article}
\usepackage{amsmath,amsfonts,amsthm}
\usepackage{authblk}

\usepackage[normalem]{ulem}

\allowdisplaybreaks[3]
\numberwithin{equation}{section}

\newtheorem{notation}{Notation}

\usepackage[body={16cm,23cm}]{geometry}
\usepackage[usenames,dvipsnames]{color}

\newcommand{\beqa}{\begin{eqnarray}}
\newcommand{\eeqa}{\end{eqnarray}}
\newcommand{\as}{{\mathbf a}}
\newcommand{\ads}{{\mathbf a}^{\dagger}}
\newcommand{\ns}{{\mathbf n}}
\newcommand{\fs}{{\mathbf f}}
\newcommand{\es}{{\mathbf e}}
\newcommand{\hs}{{\mathbf h}}
\newcommand{\Rf}{{\mathbf R}}
\newcommand{\Lf}{{\mathbf L}}
\newcommand{\Lfc}{\Check{\mathbf L}}
\newcommand{\Lfb}{\overline{\mathbf L}}
\newcommand{\Lfcb}{\Check{\overline{\mathbf L}}}
\newcommand{\Kf}{{\mathbf K}}
\newcommand{\Kfb}{\overline{\mathbf K}}
\newcommand{\Kfc}{\Check{\mathbf K}}
\newcommand{\Kfcb}{\Check{\overline{\mathbf K}}}
\newcommand{\co}{{\mathsf \Delta}}
\newcommand{\cob}{\overline{\mathsf \Delta}}

\newcommand{\g}{{\mathbf g}}
\begin{document}
\title{
Asymptotic representations of 
augmented $q$-Onsager algebra and boundary $K$-operators related to 
 Baxter Q-operators
}
\author[1]{Pascal Baseilhac}
\affil[1]{
Laboratoire de Math\'ematiques et Physique Th\'eorique CNRS/UMR 7350,
 F\'ed\'eration Denis Poisson FR2964,
Universit\'e de Tours,
Parc de Grammont, 37200 Tours, 
France
}

\author[2]{Zengo Tsuboi\footnote{additional post member at 
Osaka City University Advanced Mathematical Institute  
(since 1 February 2018)}}
\affil[2]{
Laboratoire de physique th\'eorique, D\'epartement de physique 
de l'ENS, \'Ecole normale sup\'erieure, PSL Research University, 
Sorbonne Universit\'es, UPMC Univ. Paris 06, CNRS, 75005 Paris, France
}
\date{}
\maketitle
\begin{abstract}
We consider intertwining relations of the augmented $q$-Onsager algebra introduced by Ito and Terwilliger, 
and obtain  generic (diagonal) boundary $K$-operators in terms of the Cartan element of 
$U_{q}(sl_2)$. These $K$-operators solve reflection equations. 
Taking appropriate limits of these $K$-operators in Verma modules, we derive 
$K$-operators for Baxter Q-operators and corresponding reflection equations.
\end{abstract}
\\[6pt]
Nuclear Physics B 929 (2018) 397-437
\\
https://doi.org/10.1016/j.nuclphysb.2018.02.017
\section{Introduction}
In the context of quantum integrable systems with periodic boundary conditions,
 the Baxter Q-operator \cite{Bax72} 
 is an important object. 
 It contains the information about the eigenfunctions and Bethe roots
 of the Hamiltonian and transfer matrix that are studied within a Bethe ansatz approach. 
 Importantly, a key ingredient in the construction of Baxter Q-operators are L-operators. In the context of the representation theory, the L-operators that are suitable for Baxter Q-operators are obtained as certain homomorphic images of the universal  R-matrix for a given quantum affine algebra. The auxiliary spaces of these L-operators are q-oscillator representations 
 of one of the Borel subalgebras of the quantum affine algebra. 
 This `q-oscillator construction' of the Q-operators was proposed by Bazhanov, Lukyanov and 
 Zamolodchikov \cite{BLZ97}, and developed by a number of authors 
 (for example, \cite{BHK02,Kulish:2005qc,BT08,T12,BGKNR10,KT14} and references therein
 \footnote{See also \cite{BLMS10,RT15} for the rational ($q=1$) case, 
 and \cite{Derkachov,DKM03} for a different approach.}). 
 It is known
\footnote{This might be a sort of common knowledge or folklore among a part of the experts. 
  Thus, there were cases where detailed explanations on limits were omitted and only 
  the final expressions of q-oscillator representations were explicitly written in papers. 
  Set aside Q-operators, a relation between a q-oscillator algebra and a limit of a highest
   weight representation of $U_{q}(sl_{2})$ was discussed in \cite{Chaichian:1989rq}.}
  (cf.\ \cite{BLZ97,BHK02}) that these q-oscillator representations of one of the Borel subalgebras 
 are given as limits of representations of them 
 (they are sometimes called `asymptotic representations'). 
 A systematic study of this from the point of view of the representation theory 
 was done in \cite{HJ11}. 
 Thus, taking limits 
 of representations of one of the Borel algebras
  is basically enough to derive the L-operators for Q-operators associated 
  with integrable systems with periodic boundary condition. 
 However, it is important to stress that these representations cannot be straightforwardly
 \footnote{ As for elliptic 
 quantum groups, where the notion of the Borel subalgebras is obscure, 
 an extension to the whole algebra might  be necessary 
  (cf.\ \cite{Zhang17}).}
  extended to those of the whole quantum affine algebra. Instead, the extended representations could be  interpreted \cite{T12} as representations of contracted algebras of the original quantum affine algebra. 

By analogy, for models with non-periodic integrable boundary conditions \cite{Skly,MN} the explicit construction of Baxter Q-operators is an interesting problem that deserves to be further studied
 \cite{DKM03,YNZ05,FS15}
 \footnote{Among these, \cite{FS15} is the only paper relevant to us in the sense
 that it deals with an 
 `oscillator construction' ($q=1$ case) of Q-operators.}. 
From the algebraic point of view, for these models the relevant algebras are the reflection equation algebras and related coideal subalgebras of quantum affine algebras \cite{MRS}. Given a certain representation, integrable boundary conditions are classified according 
to solutions - the so-called K-matrices - to the reflection  and dual reflection equations \cite{Skly,MN}. In order to
construct Q-operators for those integrable models with boundaries, a crucial step is the construction of K-operators associated with q-oscillator representations.

In the present paper, we focus on a class of K-operators associated with integrable models with diagonal boundary conditions and $U_q(\widehat{sl_2})$ R-matrix. In this case, the coideal subalgebras of $U_q(\widehat{sl_2})$ that are relevant in the analysis are related with the so-called augmented q-Onsager algebra \cite{IT} (see also \cite{BB1}), and its contracted versions introduced in this paper. For certain homomorphic images of two different coideal subalgebras
 of $U_q(\widehat{sl_2})$, we determine the K-operators. Certain limits are then considered, from which K-operators for Q-operators of models with diagonal boundary conditions can be 
derived.   
In contrast to the periodic boundary conditions case, we have to deal with 
limits of representations of the whole $U_q(\widehat{sl_2})$ in the spirit of \cite{T12} 
since the augmented q-Onsager algebra 
is realized by the generators of the whole $U_q(\widehat{sl_2})$ rather than one of the 
Borel subalgebras. 
 The intertwining relations of our K-operators for Q-operators are no longer 
 the ones for the augmented q-Onsager algebra but the ones for a contracted version of it. 

The paper is organized as follows. In Section 2, we recall the 
definitions of the quantum algebra $U_q(\widehat{sl_2})$, $U_q(sl_2)$ 
and two q-oscillator algebras that will be needed
for our purpose. L-operators and their limits are recalled in Section 3. In Section 4, we introduce the augmented q-Onsager algebra though
generators and relations. Two different types of realizations of the augmented q-Onsager algebra are considered, namely as a right or left coideal subalgebra $U_q(\widehat{sl_2})$. The two corresponding
 intertwining relations are considered and solved, giving an explicit expression for K-operators. The reflection and dual reflection equations they satisfy are displayed. Certain limits of those K-operators are considered, that are required for the construction of Q-operators. 
 The rational limit ($q \to 1$) of 
 these K-operators for Q-operators correspond to the K-operators found in \cite{FS15}. 
 In Appendices, we give some material that is needed for the main discussion. In Appendix A, to make the text self-contained we   give a brief review on the universal R-matrix. In Appendix B, the contracted algebras associated with $U_q(\widehat{sl_2})$ and corresponding L-operators are reviewed. 
 A universal form of the intertwining relations among L-operators for Q-operators is presented. 
 In Appendix C, contractions of the augmented q-Onsager algebra are introduced and corresponding K-operators are described. In Appendices D,E,F, miscellaneous results are collected.
 In Appendix \ref{ApG}, definitions of universal T-and Q-operators 
 in terms K-operators text are explained. 
Throughout this paper, we will work on the general gradation of $U_q(\widehat{sl_2})$. 
This does not produce particularly new results since the L-operators in the general 
gradation can easily be obtained from the ones in a particular gradation by similarity 
transformation and rescaling of the spectral parameter. 
However, we expect that this may clarify some relations rather ambiguously treated in literatures.
\begin{notation} In the text, $q\in {\mathbb C}$ is assumed not to be a root of unity. We introduce the $q-$commutator $[X,Y]_q=XY-qYX$. In particular, we denote $[X,Y]=[X,Y]_1$. 
Also, we use the notation:
\beqa
 [n]_q=(q^n-q^{-n})/(q-q^{-1}), 
 \qquad 
 (a;q)_{\infty}=\prod_{j=0}^{\infty}(1-aq^{j}). \nonumber
\eeqa 
\end{notation}

\section{Quantum algebras}
In this section, basic definitions that will be used in the next sections are introduced. Successively, 
we recall the definitions of the quantum affine algebra $U_{q}(\widehat{sl_2})$, the quantum algebra $U_{q}(sl_2)$ and two q-oscillator algebras through generators and relations. Coproduct, automorphisms and certain finite dimensional representations are also displayed. 
We will follow the style of presentation  in \cite{KT14}. 
 \vspace{1mm}

\subsection{The quantum affine algebra $U_{q}(\widehat{sl_2})$}
The quantum affine algebra $U_{q}(\widehat{sl_2})$ is a Hopf algebra
generated by the generators
 $e_{i},f_{i},h_{i},d$, where
$i \in \{0,1 \}$.
For $i,j \in \{0,1\}$, the
 defining relations of the algebra $U_{q}(\widehat{sl_2})$ are
given by
\begin{align}
& [h_{i},h_{j}]=0, \quad [h_{i}, e_{j} ] =a_{ij} e_{j}, \quad
[h_{i}, f_{j} ] =-a_{ij} f_{j}, 
\label{sl2h-def1}
\\[6pt]
&[e_{i},f_{j}]=\delta_{ij} \frac{q^{h_{i}} -q^{-h_{i}} }{q-q^{-1}},
\label{sl2h-def2}
\\[6pt]
&[e_{i},[e_{i},[e_{i},e_{j}]_{q^{2}}]]_{q^{-2}}=
[f_{i},[f_{i},[f_{i},f_{j}]_{q^{-2}}]]_{q^{2}}=0
\qquad i \ne j ,
\label{sl2h-def3}
\end{align}
where $(a_{ij})_{0 \le i,j\le 1}$ is the
Cartan matrix
\begin{align}\nonumber
(a_{ij})_{0 \le i,j\le 1}=
\begin{pmatrix}
2& -2 \\
-2 & 2
\end{pmatrix}
.
\end{align}
The algebra has automorphisms $\sigma$ and $\tau $  defined by 
\begin{align}
\begin{split}
& \sigma (e_{0})=e_{1}, \qquad \sigma (f_{0})=f_{1}, \qquad \sigma (h_{0})=h_{1},  
\\[6pt]
& \sigma (e_{1})=e_{0}, \qquad \sigma (f_{1})=f_{0}, \qquad \sigma (h_{1})=h_{0}. 
\end{split}
\label{auto1}
\end{align}
and 
\begin{align}
& \tau (e_{i})=f_{i}, \qquad \tau (f_{i})=e_{i}, \qquad \tau (h_{i})=-h_{i},  
\qquad i=0,1. 
\label{auto2}
\end{align}
We use the following co-product
  $ \Delta : U_{q}(\widehat{sl_2}) \to U_{q}(\widehat{sl_2}) \otimes U_{q}(\widehat{sl_2})$:
\begin{align}
\Delta (e_{i})&=e_{i} \otimes 1 + q^{-h_{i}} \otimes e_{i}, \nonumber\\
\Delta (f_{i})&=f_{i} \otimes q^{h_{i}} + 1 \otimes f_{i},\label{copro-h} \\
\Delta (h_{i})&=h_{i} \otimes 1 + 1 \otimes h_{i}. \nonumber
\end{align}
We will also utilize an opposite co-product defined by
\begin{align}
\Delta'=\mathfrak{p} \circ \Delta,\qquad \mathfrak{p}\circ
(X\otimes Y)=
Y\otimes X,\qquad X,Y\in U_{q}(\widehat{sl_2}).
\end{align}
The automorphisms \eqref{auto1} and \eqref{auto2} are related to 
the co-product as
\begin{align}
(\sigma \otimes \sigma)\circ \Delta =\Delta \circ \sigma ,
\qquad 
(\tau \otimes \tau)\circ \Delta =\Delta^{\prime} \circ \tau . 
 \label{copro-auto}
\end{align}
We always assume 
that the central element $h_{0}+h_{1}$ is zero.
Anti-pode, co-unit and grading element $d$ are not explicitly used  in this
paper.

The Borel subalgebra ${\mathcal B}_{+}$
(resp. ${\mathcal B}_{-}$) is generated by
$e_{i}, h_{i} $ (resp. $f_{i},h_{i}$), where
$i \in \{0,1 \}$.
For complex numbers $c_{i} \in {\mathbb C}$ which obey the relation
$\sum_{i=0}^{1} c_{i} =0$, the  transformation
\begin{align}
\tau_{c_{1}}(h_{i})=  h_{i} +c_{i}, \qquad  i = 0,1,
 \label{shiftauto}
\end{align}
gives the shift automorphism  of $ {\mathcal
B}_{+} $ (or of ${\mathcal B}_{-} $). Here we omit the unit element
multiplied by the above complex numbers.

There exists a unique element \cite{Dr85,KT92-1}
${\mathcal R}$ in a completion of $ {\mathcal B}_{+} \otimes {\mathcal B}_{-} $
called the universal R-matrix which satisfies the following
relations
\begin{align}
\Delta'(a)\ {\mathcal R}&={\mathcal R}\ \Delta(a)
\qquad \text{for} \quad \forall\ a\in U_{q}(\widehat{sl_2})\,   ,\nonumber\\
(\Delta\otimes 1)\,
{\mathcal R}&={\mathcal R}_{13}\, {\mathcal R}_{23}\, ,\label{R-def}\\
(1\otimes \Delta)\, {\mathcal R}&={\mathcal R}_{13}\,
{\mathcal R}_{12}\,\nonumber
\end{align}
where
${\cal R}_{12}={\cal R}\otimes 1$, ${\cal R}_{23}=1\otimes {\cal R}$,
${\cal R}_{13}=(\mathfrak{p} \otimes 1)\, {\cal R}_{23}$. 
We will use the relation 
\begin{align}
(\xi^{ h_{1}} \otimes \xi^{ h_{1}}){\cal R}=
{\cal R}(\xi^{ h_{1}} \otimes \xi^{ h_{1}})
\quad \text{for} \quad \xi \in \mathbb{C}\setminus \{0 \}, 
\label{com-Cartan}
\end{align}
which follows from the first relation in \eqref{R-def}. 
The Yang-Baxter equation
\begin{align}
{\mathcal R}_{12}{\mathcal R}_{13}{\cal R}_{23}=
{\mathcal R}_{23}{\mathcal R}_{13}{\mathcal R}_{12}\ ,\label{YBE}
\end{align}
is a corollary of these relations \eqref{R-def}. 
For $\overline{\mathcal R}={\mathcal R}_{21}
=(\mathfrak{p} \otimes 1) {\mathcal R}_{12}$, the  relations \eqref{R-def} 
become
\begin{align}
\Delta(a)\ \overline{\mathcal R}&=\overline{\mathcal R}\ \Delta^{\prime}(a)
\qquad \text{for} \quad \forall\ a\in U_{q}(\widehat{sl_2})\,   ,\nonumber\\
(\Delta\otimes 1)\,
\overline{\mathcal R}&=\overline{\mathcal R}_{23}\, \overline{\mathcal R}_{13}\, ,\label{Rb-def}\\
(1\otimes \Delta)\, \overline{\mathcal R}&=\overline{\mathcal R}_{12}\,
\overline{\mathcal R}_{13}\,\nonumber
\end{align}
One can also check that ${\mathcal R}^{-1}_{21}$ (resp.\ ${\mathcal R}^{-1}$ ) 
satisfies \eqref{R-def} (resp.\ \eqref{Rb-def}).
The universal R-matrix can be written in the form
\begin{align}
{\mathcal R}=\tilde{{\mathcal R}}\ q^{\mathcal K},
\qquad {\mathcal
  K}=\frac{1}{2}h_{1}\otimes h_{1}. \label{R-red}
\end{align}
Here $\tilde{{\mathcal R}}$  is the reduced universal $R$-matrix, which
is a series in $e_j\otimes 1$ and $1 \otimes f_j$
and does not contain Cartan elements.
Thus the reduced universal R-matrix is unchanged under the shift automorphism  $\tau_{c_{1}}$ of
 $ {\mathcal B}_{+}$, see (\ref{shiftauto}), while
the prefactor $\mathcal{K}$ is shifted as
\begin{align}
{\mathcal K}  \mapsto
{\mathcal K}   +
\frac{c_{1}}{2} (1 \otimes h_{1}).
\label{unRshift}
\end{align}
The universal R-matrix is invariant under $ \sigma \otimes \sigma $: 
\begin{align}
(\sigma \otimes \sigma ) \mathcal{R} = \mathcal{R}
\end{align}
Then we will use the following relation, which follows from this: 
\begin{align}
(\sigma \otimes 1 ) \mathcal{R} = (1 \otimes \sigma^{-1} ) \mathcal{R}.
\end{align}
\subsection{The quantum algebra $U_{q}(sl_2)$}
The algebra $U_{q}(sl_2)$ is  generated by the elements $E, F, H$.
The defining relations  are
\begin{align}
&
[H,E]=2E, \qquad [H,F]=-2F,
\nonumber
\\ &
[E, F] = \frac{q^{H} - q^{-H} }{q-q^{-1}} .
\label{HEF-sl2}
\end{align}
The following elements are central in $U_{q}(sl_2)$:
\begin{align}
C=FE+\frac{q^{H+1} + q^{-H-1}}{(q-q^{-1})^{2}} 
=EF+\frac{q^{H-1} + q^{-H+1}}{(q-q^{-1})^{2}}.
\end{align}
Note that 
the following map gives an automorphism of the algebra.
\begin{align}
\nu: \quad 
E \mapsto F,
\qquad 
F \mapsto E,
\qquad 
H \mapsto -H.
\end{align}
We will also use an anti-automorphism defined by
\begin{align}
& 
^{t}: \quad 
E \mapsto q^{-H-1}F,
\qquad 
F \mapsto Eq^{H+1},
\qquad 
H \mapsto H, 
 \label{t-sl2}
\end{align}
where $(ab)^{t}=b^{t}a^{t}$ holds for any $a,b \in U_{q}(sl_2)$.
There is an evaluation map $\mathsf{ev}_{x}$:
$U_{q}(\widehat{sl_2}) \mapsto U_{q}(sl_2)$:
\begin{align}
\begin{split}
& e_{0} \mapsto x^{s_{0}}F,  \qquad
 f_{0} \mapsto x^{-s_{0}} E, \qquad
 h_{0} \mapsto -H,
\\
& e_{1} \mapsto x^{s_{1}} E,  \qquad
f_{1} \mapsto x^{-s_{1}} F, \qquad
h_{1} \mapsto H,
\end{split}
\label{eva}
\end{align}
where $x  \in {\mathbb C}$ is a spectral parameter and $s_{0},s_{1} \in {\mathbb C}$. 
We set $s=s_{0}+s_{1}$. 
If we apply the similarity transformation 
$\mathsf{ev}_{x}(a) \mapsto  
x^{\frac{s_{0}-s_{1}}{4}H}\mathsf{ev}_{x}(a) x^{-\frac{s_{0}-s_{1}}{4}H} $ 
(resp.\ $\mathsf{ev}_{x}(a) \mapsto  
x^{-\frac{s_{1}}{2}H}\mathsf{ev}_{x}(a) x^{\frac{s_{1}}{2}H} $ )
for $a \in U_{q}(\widehat{sl_2})$ and 
the rescaling of the spectral parameter 
$x \mapsto x^{\frac{2}{s}}$ 
(resp.\ $x \mapsto x^{\frac{1}{s}}$), 
we will obtain the principal gradation $s_{0}=s_{1}=1$ 
(resp.\ the homogeneous gradation $s_{0}=1, s_{1}=0$). 
Let $\pi_{\mu}^{+}$  be the Verma module over $U_{q}(sl_2)$ with the highest weight $\mu$. In a basis $\{v_{n} | n \in {\mathbb Z}_{\ge 0} \}$, we have
\begin{align}
H v_{n} =(\mu -2n)v_{n}, \quad
E v_{n}=[n]_{q}[\mu -n+1]_{q} v_{n-1},
\quad
 F v_{n}=v_{n+1} .
 \label{hwvglmn}
\end{align}
For $ \mu \in {\mathbb Z}_{\ge 0}$, 
 the finite dimensional irreducible module $\pi_{\mu} $
with the highest weight $\mu$ is given as quotient of Verma modules: 
\begin{align}
\pi_{\mu}^{+} /\pi_{-\mu-2}^{+} \simeq \pi_{\mu}. 
 \label{quotient}
\end{align}
 In particular,
$\pi_{1}(E)=E_{12}$, $\pi_{1}(F)=E_{21}$ and $\pi_{1}(H)=E_{11}-E_{22}$ gives the fundamental representation of $U_{q}(sl_2)$, where
$E_{ij}$ is a $2 \times 2$ matrix unit whose
$(k,l)$-element is $\delta_{i,k}\delta_{j,l}$. 
In this case, \eqref{t-sl2} coincides with transposition of matrices.

Then the compositions
 $\pi_{\mu }^{+}(x)=\pi_{\mu }^{+} \circ \mathsf{ev}_{x}$  and
$\pi_{\mu }(x)=\pi_{\mu } \circ \mathsf{ev}_{x}$
give evaluation representations of $U_{q}(\widehat{sl_2})$.
%
%
\subsection{The q-oscillator algebras}
We introduce two kinds of oscillator algebras $\mathrm{Osc}_{i}$ ($i=1,2$).
 They are generated by  the elements $\hs_{i},\es_{i}, \fs_{i}$ which 
 obey the following relations:
\begin{align}
\begin{split}
& [\hs_{1}, \hs_{1}]=0,
\qquad [\hs_{1},\es_{1}]=2\es_{1},
\qquad [\hs_{1},\fs_{1}]=-2\fs_{1},
\\
& \fs_{1}\es_{1}=q\frac{1-q^{\hs_{1}}}{( q-q^{-1})^{2} } ,
\qquad
\es_{1}\fs_{1}=q\frac{1-q^{\hs_{1}-2 }}{( q-q^{-1})^{2} },
\end{split}
\label{osc1}
\end{align}
\begin{align}
\begin{split}
& [\hs_{2}, \hs_{2}]=0,
\qquad [\hs_{2},\es_{2}]=2\es_{2},
\qquad [\hs_{2},\fs_{2}]=-2\fs_{2},
\\
& \fs_{2}\es_{2}=q^{-1}\frac{1-q^{-\hs_{2}}}{( q-q^{-1})^{2} } ,
\qquad
\es_{2}\fs_{2}=q^{-1}\frac{1-q^{-\hs_{2}+2 }}{( q-q^{-1})^{2} },
\end{split}
\label{osc2}
\end{align}
Note that $\mathrm{Osc}_{1}$ and  $\mathrm{Osc}_{2}$ can be swapped  by the transformation
$q \mapsto q^{-1} $.
One can prove the following corollaries of  \eqref{osc1} and \eqref{osc2}:
\begin{align}
& [\es_{1},\fs_{1}]=\frac{q^{\hs_{1} }}{q-q^{-1}},
\qquad
[\es_{2},\fs_{2}]=-\frac{q^{-\hs_{2} }}{q-q^{-1}},
\label{comosc1}
\\
& [\es_{1},\fs_{1}]_{q^{-2}}= \frac{1}{q-q^{-1}},
\qquad
[\es_{2},\fs_{2}]_{q^{2}} = -\frac{1}{q-q^{-1}}.
\label{comosc2}
\end{align}
We will use anti-automorphisms of $\mathrm{Osc}_{i}$, 
which  are analogues of \eqref{t-sl2}, 
defined by
\begin{align}
& 
^{t}: \quad 
\es_{i} \mapsto q^{-\hs_{i}-1}\fs_{i},
\qquad 
\fs_{i} \mapsto \es_{i}q^{\hs_{i}+1},
\qquad 
\hs_{i} \mapsto \hs_{i}, 
 \label{t-osc}
\end{align}
where $(ab)^{t}=b^{t}a^{t}$ holds for any $a,b \in \mathrm{Osc}_{i}$, $i=1,2$.
Relations \eqref{comosc1} are nothing but contractions
\footnote{Contractions of a quantum algebra and its relation to
a q-oscillator algebra was discussed in \cite{Chaichian:1989rq}.}
 of the commutation relation
 \eqref{HEF-sl2}.
On the other hand,
 the relations \eqref{comosc2}
are conditions that central elements take constant values.
 The following  limits of the generators of $U_{q}(sl_2)$ for the Verma module
 \footnote{for \eqref{limit1}: on the renormalized basis $v_{n}^{\prime}=q^{-\frac{\mu s_{0}n}{s}}v_{n}$; 
 for \eqref{limit2}: on the renormalized basis $v_{n}^{\prime}=q^{\frac{\mu s_{0}n}{s}}v_{n}$}
  $\pi^{+}_{\mu}$
\begin{align}
\begin{split}
& \lim_{q^{-\mu} \to 0}\pi^{+}_{\mu} (H-\mu)=\hs_{1},
\qquad
\lim_{q^{-\mu} \to 0} \pi^{+}_{\mu}(q^{-H-\mu})=0,  \\
& \lim_{q^{-\mu} \to 0}\pi^{+}_{\mu}( Fq^{-\frac{s_{0}\mu }{s}})=\fs_{1},
\qquad
 \lim_{q^{-\mu} \to 0} \pi^{+}_{\mu}(Eq^{-\frac{s_{1}\mu }{s}})=\es_{1},
\end{split}
\label{limit1}
\end{align}
\begin{align}
\begin{split}
& \lim_{q^{\mu} \to 0} \pi^{+}_{\mu}(H-\mu)= \hs_{2},
\qquad
\lim_{q^{\mu} \to 0} \pi^{+}_{\mu}(q^{H+\mu})=0,  \\
& \lim_{q^{\mu} \to 0} \pi^{+}_{\mu}(Fq^{\frac{s_{0}\mu }{s}})=\fs_{2},
\qquad
 \lim_{q^{\mu} \to 0} \pi^{+}_{\mu}(Eq^{\frac{s_{1}\mu }{s}})=\es_{2},
\end{split}
  \label{limit2}
\end{align}
realize the q-oscillator algebras $\mathrm{Osc}_{1}$ and
$\mathrm{Osc}_{2}$, respectively. 
Once these limits are taken in formulas, one may forget about the representations,  
and consider only algebraic relations defined in 
\eqref{osc1} or \eqref{osc2}.

\section{L-operators and limits}
In this section, we first recall the definition of the Lax operators which follows from the universal $R-$matrix. 
In the context of quantum integrable systems, 
it is known that certain limits (cf.\ \cite{BLZ97,BHK02}) of $L-$operators 
provide the basic ingredient for the construction of Baxter Q-operators associated 
with the Yang-Baxter algebra 
(see \cite{Boos07,BT08,BGKNR10,T12} for examples of L-operators for Q-operators, 
and \cite{BLMS10} for examples of the rational case). 
By analogy, here we consider different limits 
of $L-$operators that will be useful in the construction of $Q-$operators 
associated with the reflection equation algebra.  
Technically, we follow the presentation of \cite{KT14}. 
From now on we denote $\lambda=q-q^{-1}$. We set
\begin{align}
\Lf (x)& =
\begin{pmatrix}
q^{\frac{H}{2}} -q^{-1} x^{s} q^{-\frac{H}{2}} & \lambda x^{s_{0}}F q^{-\frac{H}{2} } \\
\lambda x^{s_{1}} E q^{\frac{H}{2} } & q^{- \frac{H}{2}} -q^{-1} x^{s} q^{\frac{H}{2}}
 \end{pmatrix} ,
 \label{Lgen1}
 \\[6pt]
\overline{\Lf} (x)& =
\begin{pmatrix}
q^{\frac{H}{2}} -q^{-1} x^{-s} q^{-\frac{H}{2}} & \lambda x^{-s_{1}}F q^{-\frac{H}{2} } \\
\lambda x^{-s_{0}} E q^{\frac{H}{2} } & q^{- \frac{H}{2}} -q^{-1} x^{-s} q^{\frac{H}{2}}
 \end{pmatrix} .
 \label{Lgen2}
\end{align}
These are images of the universal R-matrix (see Appendix \ref{ApA}) 
\begin{align}
\Lf (x/y)&= \Lf (x,y)=
\phi(x/y) (\mathsf{ev}_{x} \otimes \pi_{1}(y) ) \mathcal{R},
\\[6pt]
\overline{\Lf} (x/y)&= \overline{\Lf} (x,y)=
\phi(y/x)  (\mathsf{ev}_{x} \otimes \pi_{1}(y) ) \mathcal{R}_{21},
\quad 
x,y \in {\mathbb C}, 
\end{align}
where the overall factor is defined by 
 $\phi(x)=e^{-\Lambda(x^{s}q^{-1})} $, 
$\Lambda (x)=\sum_{k=1}^{\infty}\frac{C_{k}}{k(q^{k}+q^{-k})}x^{k}$ 
and \eqref{highercas}. 
One can check
\begin{align}
& \Lf (x)\overline{\Lf} (x)=\overline{\Lf} (x)\Lf (x)=q^{-1}(\lambda^2C-x^{s}-x^{-s}) , 
\label{LL=C}
\\[6pt]
& g_{2}\Lf (xq^{\frac{4}{s}})^{t_{2}} g_{2}^{-1}\overline{\Lf} (x)^{t_{2}}=
\g_{1}^{-1}\Lf (xq^{\frac{4}{s}})^{t_{2}} \g_{1}\overline{\Lf} (x)^{t_{2}}=
\nonumber 
\\[6pt]
& \qquad =\overline{\Lf} (x)^{t_{2}} g_{2}\Lf (xq^{\frac{4}{s}})^{t_{2}} g_{2}^{-1}=
\overline{\Lf} (x)^{t_{2}} \g_{1}^{-1}\Lf (xq^{\frac{4}{s}})^{t_{2}} \g_{1}=
q (\lambda^2C-q^{2}x^{s}-q^{-2}x^{-s}) ,
\label{LtLt=C}
\end{align}
where $\g_{1}=\g \otimes 1$,  $\g=q^{\frac{(s_{0}-s_{1})H}{s}}$, 
$g_{2}=1 \otimes g$, $g=\pi_{1}(\g)=\mathrm{diag}(q^{\frac{s_{0}-s_{1}}{s}},q^{-\frac{s_{0}-s_{1}}{s}})$, and $^{t_{2}}$ is the transposition in the second component of the tensor product.
Evaluating the first space of these L-operators for the fundamental representation, 
we obtain R-matrices of the 6-vertex model.
\begin{align}
R(x)=
q^{\frac{1}{2}}(\pi_{1} \otimes 1)\Lf (x)& =
\begin{pmatrix}
q-q^{-1}x^{s} & 0 & 0 & 0 \\
0 & 1-x^{s} &  \lambda x^{s_{1}} & 0 \\
0 & \lambda x^{s_{0}} & 1-x^{s}  & 0 \\
 0 & 0 & 0 & q-q^{-1}x^{s}
 \end{pmatrix} ,
 \label{Rmat1}
 \\[6pt]
\overline{R}(x)=
q^{\frac{1}{2}}(\pi_{1} \otimes 1)\overline{\Lf} (x)& =
\begin{pmatrix}
q-q^{-1}x^{-s} & 0 & 0 & 0 \\
0 & 1-x^{-s} &  \lambda x^{-s_{0}} & 0 \\
0 & \lambda x^{-s_{1}} & 1-x^{-s}  & 0 \\
 0 & 0 & 0 & q-q^{-1}x^{-s}
 \end{pmatrix} ,
 \label{Rmat2}
\end{align}
\subsection{L-operators for Q-operators}
Applying the limits \eqref{limit1} and \eqref{limit2} 
to  \eqref{Lgen1} and  \eqref{Lgen2}, we define 
four type of L-operators as 
\footnote{The factor $q^{-\frac{\mu \otimes
\pi_{1}(1)(h_{1})}{2}}$ came from \eqref{unRshift} for $c_{1}=-\mu$.} 
\footnote{
We could also use automorphisms of $U_{q}(\widehat{sl_2})$ or $U_{q}(sl_2)$ 
to derive 
various L-operators: for example, 
\begin{align}
\Lf ^{(2)\prime }(x)& = \lim_{q^{-\mu } \to 0} (\pi^{+}_{\mu} \circ \sigma \otimes 1)
\Lf (xq^{-\frac{\mu}{s}})q^{-\frac{\mu \otimes \pi_{1}(1)(h_{0})}{2}}
=
\begin{pmatrix}
q^{- \frac{\hs_{1}}{2}} -q^{-1} x^{s} q^{\frac{\hs_{1}}{2}}&\lambda x^{s_{0}} \es_{1} q^{\frac{\hs_{1}}{2} } \\
 \lambda \fs_{1} x^{s_{1}}q^{-\frac{\hs_{1}}{2} } & q^{\frac{\hs_{1}}{2}}  
 \end{pmatrix}. 
 \label{LQ2p}
\end{align}
This could be a substitute of \eqref{LQ2} 
(cf. \cite{BGKNR10}). 
Instead of using automorphisms, we will use Chevalley like generators of 
the q-oscillator algebras and  take various different limits of the L-operators (as already demonstrated in \cite{KT14}). 
A merit for this is that the resultant L-operators become just 
reductions of the original L-operators  
(for example, compare \eqref{Lgen1} with \eqref{LQ1}, \eqref{LQ1-an} and 
\eqref{LQ2-an}). 
This is also the case with the intertwining relations and the K-operators. 
}
\begin{align}
\Lf^{(1)}(x)& = \lim_{q^{-\mu } \to 0} (\pi^{+}_{\mu} \otimes 1)
\Lf (xq^{-\frac{\mu}{s}})q^{-\frac{\mu \otimes \pi_{1}(1)(h_{1})}{2}}
=
\begin{pmatrix}
q^{\frac{\hs_{1}}{2}}  & \lambda x^{s_{0}} \fs_{1} q^{-\frac{\hs_{1}}{2} } \\
\lambda x^{s_{1}} \es_{1} q^{\frac{\hs_{1}}{2} } & q^{- \frac{\hs_{1}}{2}} -q^{-1} x^{s} q^{\frac{\hs_{1}}{2}}
 \end{pmatrix} ,
 \label{LQ1}
\\[6pt]
\Lf^{(2)}(x)& = \lim_{q^{\mu } \to 0} (\pi^{+}_{\mu} \otimes 1)
\Lf (xq^{\frac{\mu}{s}})q^{-\frac{\mu \otimes \pi_{1}(1)(h_{1})}{2}}
=
\begin{pmatrix}
q^{\frac{\hs_{2}}{2}} -q^{-1} x^{s} q^{-\frac{\hs_{2}}{2}}
 & \lambda x^{s_{0}} \fs_{2} q^{-\frac{\hs_{2}}{2} } \\
\lambda x^{s_{1}} \es_{2} q^{\frac{\hs_{2}}{2} } & q^{- \frac{\hs_{2}}{2}} 
 \end{pmatrix} ,
 \label{LQ2}
\\[6pt]
\overline{\Lf}^{(1)}(x)
& = \lim_{q^{-\mu } \to 0} (\pi^{+}_{\mu} \otimes 1)
\left(
\left(q^{\frac{(s_{0}-s_{1})\mu H}{2s}}\otimes 1 \right)
\overline{\Lf} (xq^{\frac{\mu}{s}})
\left(q^{-\frac{(s_{0}-s_{1})\mu H}{2s}}\otimes 1 \right)
q^{-\frac{\mu \otimes \pi_{1}(1)(h_{1})}{2}}
\right)
\nonumber
\\[6pt]
&=
\begin{pmatrix}
q^{\frac{\hs_{1}}{2}}  & \lambda x^{-s_{1}} \fs_{1} q^{-\frac{\hs_{1}}{2} } \\
\lambda x^{-s_{0}} \es_{1} q^{\frac{\hs_{1}}{2} } & q^{- \frac{\hs_{1}}{2}} -q^{-1} x^{-s} q^{\frac{\hs_{1}}{2}}
 \end{pmatrix} ,
 \label{LhQ1}
\\[6pt]
\overline{\Lf}^{(2)}(x)
& = \lim_{q^{\mu } \to 0} (\pi^{+}_{\mu} \otimes 1)
\left(
\left(q^{\frac{(s_{1}-s_{0})\mu H}{2s}}\otimes 1 \right)
\overline{\Lf} (xq^{-\frac{\mu}{s}})
\left(q^{-\frac{(s_{1}-s_{0})\mu H}{2s}}\otimes 1 \right)
q^{-\frac{\mu \otimes \pi_{1}(1)(h_{1})}{2}}
\right)
\nonumber 
\\[6pt]
&=
\begin{pmatrix}
q^{\frac{\hs_{2}}{2}} -q^{-1} x^{-s} q^{-\frac{\hs_{2}}{2}}
 & \lambda x^{-s_{1}} \fs_{2} q^{-\frac{\hs_{2}}{2} } \\
\lambda x^{-s_{0}} \es_{2} q^{\frac{\hs_{2}}{2} } & q^{- \frac{\hs_{2}}{2}} 
 \end{pmatrix} .
 \label{LhQ2}
\end{align}
As mentioned above, the L-operators \eqref{LQ1} and \eqref{LQ2} are essential ingredients in the construction of Baxter Q-operators associated with the Yang-Baxter algebra. For instance, for a spin chain with periodic boundary conditions, 
the corresponding Q-operators are defined as trace over product of these L-operators: 
\begin{multline}
{\mathbb Q}^{(a)}(x) 
=
({\mathbb Z}^{(a)})^{-1}
(\mathrm{tr}_{W_{a}} \otimes 1^{\otimes L})
\left(
q^{\alpha \hs_{a}}
{\mathbf L}^{(a)}_{0L}\left(x \xi_{L}^{-1} \right)
\cdots 
{\mathbf L}^{(a)}_{02}\left(x \xi_{1}^{-1} \right)
{\mathbf L}^{(a)}_{01}\left(x \xi_{1}^{-1} \right)
\right)
, 
\quad a=1,2, 
 \label{Q-op-period}
\end{multline}
where $\xi_{1},\dots ,\xi_{L} \in {\mathbb C} \setminus \{0 \}$ are inhomogeneities on the spectral parameter in the quantum space; $\alpha \in {\mathbb C} $; 
the trace is taken over the auxiliary space (a Fock space $W_{a}$ for $\mathrm{Osc}_{a}$ denoted as $0$). Here the normalization 
operator ${\mathbb Z}^{(a)}$ is defined by 
\begin{align}
{\mathbb Z}^{(a)}=
(\pi_{1}(\xi_{1}) \otimes \cdots \otimes \pi_{1}(\xi_{L}))
\Delta^{\otimes (L-1)}
(\mathrm{tr}_{W_{a}} \otimes 1)
(1 \otimes z)^{\hs_{a} \otimes 1}, 
\quad 
z=
q^{\alpha + \frac{1}{2} h_{1}}. 
 \label{normQ}
\end{align}
In Appendix \ref{ApG}, we will propose Q-operators Q-operators associated 
with  different types of reflection  equation algebra (cf. eqs. \eqref{refeqlim1st}, \eqref{refeqlim2nd}). 
Such operators are useful in the analysis of spin chains with open diagonal boundary conditions.
 For this purpose, we need additional L-operators 
\eqref{LhQ1} and \eqref{LhQ2}  that are introduced as follows.
 Observe that the pair of L-operators \eqref{LQ1} and  \eqref{LhQ1} 
(or \eqref{LQ2} and  \eqref{LhQ2}) no longer satisfies  
relations corresponding to 
\eqref{LL=C} and \eqref{LtLt=C}. For this reason, consider the following L-operators:
\begin{align}
\Lfc^{(1)}(x)
%
 &= \lim_{q^{-\mu } \to 0} (\pi^{+}_{\mu} \otimes 1)
 q^{\frac{\mu \otimes \pi_{1}(1)(h_{1})}{2}} 
\left(q^{\frac{(s_{0}-s_{1})\mu H}{2s}}\otimes 1 \right)
\Lf (xq^{\frac{\mu}{s}})
\left(q^{-\frac{(s_{0}-s_{1})\mu H}{2s}}\otimes 1 \right)
q^{-\mu} 
\nonumber 
\\[6pt]
&=
\begin{pmatrix}
q^{\frac{\hs_{1}}{2}}   -q^{-1} x^{s} q^{-\frac{\hs_{1}}{2}}& \lambda x^{s_{0}} \fs_{1} q^{-\frac{\hs_{1}}{2} } \\
\lambda x^{s_{1}} \es_{1} q^{\frac{\hs_{1}}{2} } &  -q^{-1} x^{s} q^{\frac{\hs_{1}}{2}}
 \end{pmatrix} ,
 \label{LQ1-an}
\\[6pt]
\Lfcb^{(1)}(x)& = \lim_{q^{-\mu } \to 0} (\pi^{+}_{\mu} \otimes 1)
\left(
q^{\frac{\mu \otimes \pi_{1}(1)(h_{1})}{2}}
\overline{\Lf} (xq^{-\frac{\mu}{s}}) q^{-\mu}
\right)
\nonumber
\\[6pt]
&=
\begin{pmatrix}
q^{\frac{\hs_{1}}{2}}   -q^{-1} x^{-s} q^{-\frac{\hs_{1}}{2}} 
& \lambda x^{-s_{1}} \fs_{1} q^{-\frac{\hs_{1}}{2} } \\
\lambda x^{-s_{0}} \es_{1} q^{\frac{\hs_{1}}{2} } & 
 -q^{-1} x^{-s} q^{\frac{\hs_{1}}{2}}
 \end{pmatrix} ,
 \label{LhQ1-an}
 \\[6pt]
\Lfc^{(2)}(x)
%
& = \lim_{q^{\mu } \to 0} (\pi^{+}_{\mu} \otimes 1)
q^{\frac{\mu \otimes \pi_{1}(1)(h_{1})}{2}} 
\left(q^{\frac{(s_{1}-s_{0})\mu H}{2s}}\otimes 1 \right)
\Lf (xq^{-\frac{\mu}{s}})
\left(q^{-\frac{(s_{1}-s_{0})\mu H}{2s}}\otimes 1 \right)
q^{\mu} 
\nonumber 
\\[6pt]
&=
\begin{pmatrix}
   -q^{-1} x^{s} q^{-\frac{\hs_{2}}{2}}& \lambda x^{s_{0}} \fs_{2} q^{-\frac{\hs_{2}}{2} } \\
\lambda x^{s_{1}} \es_{2} q^{\frac{\hs_{2}}{2} } & q^{-\frac{\hs_{2}}{2}} -q^{-1} x^{s} q^{\frac{\hs_{2}}{2}}
 \end{pmatrix} ,
 \label{LQ2-an}
\\[6pt] 
\Lfcb^{(2)}(x)& = \lim_{q^{\mu } \to 0} (\pi^{+}_{\mu} \otimes 1)
\left(
q^{\frac{\mu \otimes \pi_{1}(1)(h_{1})}{2}}
\overline{\Lf} (xq^{\frac{\mu}{s}}) q^{\mu}
\right)
\nonumber
\\[6pt]
&=
\begin{pmatrix}
  -q^{-1} x^{-s} q^{-\frac{\hs_{2}}{2}} 
& \lambda x^{-s_{1}} \fs_{2} q^{-\frac{\hs_{2}}{2} } \\
\lambda x^{-s_{0}} \es_{2} q^{\frac{\hs_{2}}{2} } & 
q^{-\frac{\hs_{2}}{2}}  -q^{-1} x^{-s} q^{\frac{\hs_{2}}{2}}
 \end{pmatrix} .
 \label{LhQ2-an}
 \end{align}
Then the limits of \eqref{LL=C}-\eqref{LtLt=C} are given by 
\begin{align}
\Lf^{(1)}(x)\Lfcb^{(1)}(x)&=\Lfcb^{(1)}(x)\Lf^{(1)}(x)=q^{-1}(q-x^{-s}),
\label{LLcb1=c}
 \\[6pt]
 \Lfc^{(1)}(x)\Lfb^{(1)}(x)&=\Lfb^{(1)}(x)\Lfc^{(1)}(x)=q^{-1}(q-x^{s}),
 \label{LcLb1=c}
 \\[6pt]
\Lf^{(2)}(x)\Lfcb^{(2)}(x)&=\Lfcb^{(2)}(x)\Lf^{(2)}(x)=q^{-1}(q^{-1}-x^{-s}),
\label{LLcb2=c}
 \\[6pt]
\Lfc^{(2)}(x)\Lfb^{(2)}(x)&=\Lfb^{(2)}(x)\Lfc^{(2)}(x)=q^{-1}(q^{-1}-x^{s}), 
\label{LcLb2=c}
 \\[6pt]
g_{2}\Lf^{(1)} (xq^{\frac{4}{s}})^{t_{2}} g_{2}^{-1}\Lfcb^{(1)} (x)^{t_{2}}&=
\Lfcb^{(1)} (x)^{t_{2}} g_{2}\Lf^{(1)} (xq^{\frac{4}{s}})^{t_{2}} g_{2}^{-1}=
q (q-q^{-2}x^{-s}) ,
 \label{LcbL1s=c}
 \\[6pt]
g_{2}\Lfc^{(1)} (xq^{\frac{4}{s}})^{t_{2}} g_{2}^{-1}\Lfb^{(1)} (x)^{t_{2}}&=
\Lfb^{(1)} (x)^{t_{2}} g_{2}\Lfc^{(1)} (xq^{\frac{4}{s}})^{t_{2}} g_{2}^{-1}=
q (q-q^{2}x^{s}) ,
\label{LbLc1s=c}
 \\[6pt]
g_{2}\Lf^{(2)} (xq^{\frac{4}{s}})^{t_{2}} g_{2}^{-1}\Lfcb^{(2)} (x)^{t_{2}}&=
\Lfcb^{(2)} (x)^{t_{2}} g_{2}\Lf^{(2)} (xq^{\frac{4}{s}})^{t_{2}} g_{2}^{-1}=
q (q^{-1}-q^{-2}x^{-s}) , 
\label{LcbL2s=c}
 \\[6pt]
g_{2}\Lfc^{(2)} (xq^{\frac{4}{s}})^{t_{2}} g_{2}^{-1}\Lfb^{(2)} (x)^{t_{2}}&=
\Lfb^{(2)} (x)^{t_{2}} g_{2}\Lfc^{(2)} (xq^{\frac{4}{s}})^{t_{2}} g_{2}^{-1}=
q (q^{-1}-q^{2}x^{s}) ,
\label{LbLc2s=c}
\end{align}
where the following relations are used:
\begin{align}
\lim_{q^{-\mu} \to 0}\pi^{+}_{\mu}( Cq^{-\mu})
 &= \es_{1}\fs_{1} +\frac{q^{\hs_{1}-1}}{\lambda^2} =\frac{q}{\lambda^{2}},
 \label{caslimi1}
 \\[6pt]
 \lim_{q^{\mu} \to 0}\pi^{+}_{\mu}(  Cq^{\mu})
 &= \es_{2}\fs_{2} +\frac{q^{-\hs_{2}+1}}{\lambda^2} =\frac{q^{-1}}{\lambda^{2}}.
 \label{caslimi2}
\end{align}
 These relations are among the conditions that are necessary to establish the commutativity of T- and Q-operators.

The intertwining relations for these L-operators have unusual form (cf.\ \cite{T12}). 
For example, \eqref{LQ1} satisfies
\begin{align}
\begin{split}
& \left(\hs_{1} \otimes 1+1 \otimes (E_{11}-E_{22}) \right) \Lf^{(1)}(x/y)
=\Lf^{(1)}(x/y)
\left(\hs_{1} \otimes 1+1 \otimes (E_{11}-E_{22}) \right),
\\[6pt]
& \left(1 \otimes y^{s_{0}}E_{21}+x^{s_{0}}\fs_{1} \otimes q^{E_{11}-E_{22}}\right)\Lf^{(1)}(x/y)
=\Lf^{(1)}(x/y)
\left( x^{s_{0}}\fs_{1} \otimes 1  + q^{\hs_{1}}\otimes y^{s_{0}} E_{21} \right),
\\[6pt]
& \left(1 \otimes y^{s_{1}} E_{12}+x^{s_{1}}\es_{1} \otimes q^{-E_{11}+E_{22}}\right)\Lf^{(1)}(x/y)
=\Lf^{(1)}(x/y)
\left( x^{s_{1}}\es_{1} \otimes 1  + q^{-\hs_{1}}\otimes y^{s_{1}} E_{12} \right),
\\[6pt]
& \left(x^{-s_{0}}\es_{1} \otimes 1\right)\Lf^{(1)}(x/y)
=\Lf^{(1)}(x/y)
\left( x^{-s_{0}}\es_{1} \otimes q^{-E_{11}+E_{22}} +1 \otimes y^{-s_{0}} E_{12}\right),
\\[6pt]
& \left(q^{\hs_{1}} \otimes y^{-s_{1}}E_{21}+x^{-s_{1}}\fs_{1} \otimes 1\right)\Lf^{(1)}(x/y)
=\Lf^{(1)}(x/y)
\left( x^{-s_{1}}\fs_{1} \otimes q^{E_{11}-E_{22}} \right).
\end{split}
\end{align}
They are derived from the first relation in \eqref{R-def} or \eqref{Rb-def} 
by taking the limits involved in the definitions \eqref{LQ1}-\eqref{LhQ2-an} and 
\eqref{LQ1-an}-\eqref{LhQ2}.  
For more details, see Appendix \ref{ApB}. 
\section{The augmented $q-$Onsager algebra}
In this section, we first recall the definition of the augmented $q-$Onsager algebra  \cite{IT,BB1} through generators and relations. Realizations of the augmented $q-$Onsager algebra as either right or left coideal subalgebras of $U_q(\widehat{sl_2})$ are then introduced, and co-actions maps are given. Correspondingly,  two different intertwiners of the augmented $q-$Onsager algebra are constructed explicitly. They solve a reflection equation and dual reflection in $U_q(sl_2)\otimes U_q(sl_2)$.  Under the specialization $\pi_1$, known results are recovered.\vspace{1mm}

The augmented $q-$Onsager algebra - denoted below ${\cal O}_q^{aug}$ -
 is generated by four generators ${\textsf K}_0, {\textsf K}_1,{\textsf Z}_1,\tilde{\textsf Z}_1$ subject to the defining relations \cite{BB1}:
\begin{align}
\begin{split}
[ {\textsf K}_0, {\textsf K}_1]&=0\ ,\\
 {\textsf K}_0{\textsf Z}_1&=
  q^{-2} {\textsf Z}_1{\textsf K}_0\ ,\qquad {\textsf K}_0\tilde{\textsf Z}_1=q^{2}\tilde{\textsf Z}_1{\textsf K}_0\ ,
\\
{\textsf K}_1{\textsf Z}_1&=
 q^{2}{\textsf Z}_1{\textsf K}_1\ ,\ \ 
 \qquad{\textsf K}_1\tilde{\textsf Z}_1=q^{-2}\tilde{\textsf Z}_1{\textsf K}_1\ ,
\\
\big[{\textsf Z}_1,\big[{\textsf Z}_1,\big[{\textsf Z}_1,\tilde{\textsf
Z}_1\big]_{q^{2}}\big]_{q^{-2}}\big]&=
\rho_{\mathrm{diag}}{\textsf Z}_1(\,{\textsf K}_1{\textsf K}_1-\,{\textsf K}_0{\textsf K}_0){\textsf Z}_1,
\\
\big[\tilde{\textsf Z}_1,\big[\tilde{\textsf Z}_1,\big[\tilde{\textsf Z}_1,{\textsf
Z}_1\big]_{q^{2}} \big]_{q^{-2}}\big]&=
  \rho_{\mathrm{diag}}\tilde{\textsf Z}_1({\textsf K}_0{\textsf K}_0-{\textsf K}_1{\textsf K}_1)\tilde{\textsf Z}_1\ 
 \end{split}
 \label{Tauggen}
\end{align}
with
\beqa
\rho_{\mathrm{diag}}=\frac{(q^3-q^{-3})(q^2-q^{-2})^3}{q-q^{-1}}\  .\label{rhodiag}
\eeqa
%
%
%
%
 This algebra can be embedded into  $U_{q}(\widehat{sl_2})$. Below, we will introduce two different realizations  of the algebra ${\cal O}_q^{aug}$.  They are related 
each other via the automorphism \eqref{auto2} of $U_{q}(\widehat{sl_2})$.
\subsection{The first realization}
In this subsection,  the augmented $q$-Onsager algebra is realized as a right coidal subalgebra of $U_q(\widehat{sl_2})$. According to the coaction map, an intertwiner $\Kf(x)$ is explicitly constructed.

\subsubsection{Right coideal subalgebra of $U_{q}(\widehat{sl_2})$ and the intertwiner ${\Kf}(x)$}
A realization of the augmented $q-$Onsager algebra ${\cal O}_q^{aug}$, 
as a right coideal subalgebra of $U_q(\widehat{sl_2})$ is known \cite{BB1}. Let $\epsilon_\pm$ be non-zero scalars. It is given by
\footnote{$({\textsf K}_0, {\textsf K}_1,  {\textsf Z}_1 , \tilde{\textsf Z}_1 )$ 
in \eqref{realopaug} corresponds to 
$(\overline{\textsf K}_0, \overline{\textsf K}_1, 
-\overline{\tilde{\textsf Z}}_1, -\overline{\textsf Z}_1  ) $ 
 in eq.\ (3.24) in \cite{BB1} under the transformations
  $q \mapsto q^{-1}$ and $\epsilon_{\pm} \mapsto \bar{\epsilon}_{\mp}$.}:
\begin{align}
\begin{split}
{\textsf K}_0&= \epsilon_+q^{-h_0}  \ ,\qquad {\textsf K}_1= \epsilon_- q^{-h_1} \ , \\
{\textsf Z}_1&=      (q^2-q^{-2})\big( \epsilon_-q f_1q^{-h_1}  + \epsilon_+ e_0\big) \ ,
\\
\tilde{\textsf Z}_1&= (q^2-q^{-2})\big( \epsilon_- e_1  + \epsilon_+ qf_0 q^{-h_0}\big) \ . 
 \end{split}
 \label{realopaug}
\end{align}
Note that 
the automorphism \eqref{auto1} of $U_{q}(\widehat{sl_2})$ also gives the automorphism of ${\cal O}_q^{aug}$,
\begin{align}
\sigma : \quad 
{\textsf K}_{0} \mapsto {\textsf K}_{1} ,
\quad 
{\textsf K}_{1} \mapsto {\textsf K}_{0} ,
\quad 
{\textsf Z}_1 \mapsto \tilde{\textsf Z}_1 ,
\quad 
\tilde{\textsf Z}_1 \mapsto {\textsf Z}_1 ,
\end{align}
 under the condition  $\sigma(\epsilon_{\pm})=\epsilon_{\mp}$. The co-action map $\Delta : {\cal O}_q^{aug} \mapsto {\cal O}_q^{aug} \otimes U_q(\widehat{sl_2})$ that is compatible with the relations (\ref{Tauggen}) corresponds to the restriction of the co-product \eqref{copro-h} of  $U_q(\widehat{sl_2})$  to ${\cal O}_q^{aug}$ under the realization \eqref{realopaug}.It is such that:
\begin{align}
\begin{split}
\Delta({\textsf K}_0)&=  {\textsf K}_0 \otimes q^{-h_0}\ ,
\qquad \Delta({\textsf K}_1)= {\textsf K}_1 \otimes q^{-h_1} \ ,
 \\
\Delta({\textsf Z}_1)&= 
{\textsf Z}_1  \otimes 1 + (q^2-q^{-2})\left({\textsf K}_1 \otimes  qf_1q^{-h_1} + {\textsf K}_0 \otimes  e_0 \right), \\ 
\Delta(\tilde{\textsf Z}_1)&=   
 \tilde{\textsf Z}_1 \otimes 1 + (q^2-q^{-2})\left({\textsf K}_1 \otimes  e_1 + {\textsf K}_0 \otimes  qf_0q^{-h_0}\right).
 \end{split}
\label{deltadefaug}
\end{align}
On the other hand, the restriction of the opposite co-product 
$\Delta^{\prime}$ of  $U_q(\widehat{sl_2})$ 
 to ${\cal O}_q^{aug}$ gives 
the co-action map 
$\Delta^{\prime}: {\cal O}_q^{aug} \mapsto U_q(\widehat{sl_2})   \otimes {\cal O}_q^{aug} $. 
\vspace{2mm}

The restriction of the evaluation map \eqref{eva} to ${\cal O}_q^{aug}$ 
under \eqref{realopaug}  produces 
the evaluation map  ${\cal O}_q^{aug} \mapsto U_q({sl_2}) $: 
\begin{align}
\mathsf{ev}_{x}(\textsf{K}_{0})& =\epsilon_{+}q^{H},
\qquad 
\mathsf{ev}_{x}(\textsf{K}_{1}) =\epsilon_{-}q^{-H},
\nonumber 
\\
\mathsf{ev}_{x}(\textsf{Z}_{1})&= (q^2-q^{-2})F
 \big( \epsilon_-x^{-s_1}q^{1-H} + \epsilon_+x^{s_0}\big),\nonumber\\
\mathsf{ev}_{x}(\tilde{\textsf{Z}}_{1})&=
 (q^2-q^{-2}) E \big( \epsilon_-x^{s_1} + \epsilon_+x^{-s_0}q^{H+1} \big) .\nonumber
\end{align}

Let us now consider the following  intertwining relations associated with the first realization of the augmented $q-$Onsager algebra. They read:
\begin{align}
\mathsf{ev}_{x^{-1}}(a) \Kf(x) &= 
 \Kf(x) \mathsf{ev}_{x}(a) 
 \qquad 
 \text{for any } 
 a \in 
 \{ \textsf{K}_{0},\textsf{K}_{1} ,\textsf{Z}_{1},\tilde{\textsf{Z}}_{1} \}. 
 \label{intertao0-1}
\end{align}
The  equations for $ a \in 
 \{ \textsf{K}_{0},\textsf{K}_{1}\}$ 
 imply 
  that $[\Kf(x),q^H]=0$. The  equations for $ a \in 
 \{\textsf{Z}_{1},\tilde{\textsf{Z}}_{1} \}$ give:
\begin{align}
F ( \epsilon_{+} x^{-s_{0}} + \epsilon_{-} x^{s_{1}} q^{-H+1} ) 
 \Kf(x)
&=
\Kf(x)
F ( \epsilon_{+} x^{s_{0}} + \epsilon_{-} x^{-s_{1}}q^{-H+1}  ) , 
 \label{intertao1-1}
 \\[6pt]
E ( \epsilon_{-} x^{-s_{1}} + \epsilon_{+} x^{s_{0}}q^{H+1}  ) 
 \Kf(x)
&=
\Kf(x)
E ( \epsilon_{-} x^{s_{1}} + \epsilon_{+} x^{-s_{0}} q^{H+1} ) . 
 \label{intertao1-2}
\end{align}
\subsubsection{Solutions of the intertwining relations}
According to the intertwining relations \eqref{intertao1-1} and  \eqref{intertao1-2}, solutions are defined up to a function $f(H)$ of the Cartan element ($f(x)$ is a function 
of $x\in \mathbb{C} $ with $f(x+2)=f(x)$).
 We find various different solutions  with 
different non-trivial prefactors. Here we present two typical examples of them: 
\footnote{After we obtained these solutions, we were informed by 
S.\ Belliard that he obtained a solution 
for the rational case $Y(sl_2)$. 
} 
\begin{align}
 \Kf(x)&=  
 x^{s_{0}H}
 \frac{
      \left(-\frac{\epsilon_{-}}{\epsilon_{+}}x^{s}q^{-H-1};q^{-2}
   \right)_{\infty}
   }{
       \left(-\frac{\epsilon_{-}}{\epsilon_{+}}x^{-s}q^{-H-1};q^{-2}
   \right)_{\infty}
   }
   \qquad \text{for} \quad |q|>1, 
 \label{sol6}
   \\[6pt]
&= x^{s_{0}H}
 \frac{
      \left(-\frac{\epsilon_{-}}{\epsilon_{+}}x^{-s}q^{-H+1};q^{2}
   \right)_{\infty}
   }{
       \left(-\frac{\epsilon_{-}}{\epsilon_{+}}x^{s}q^{-H+1};q^{2}
   \right)_{\infty}
   }
   \qquad \text{for} \quad |q|<1, 
  \label{sol6-2}
\end{align} 
and
\begin{align}
 \Kf(x)&=  x^{-s_{1}H}
 \frac{
      \left(-\frac{\epsilon_{+}}{\epsilon_{-}}x^{s}q^{H-1};q^{-2}
   \right)_{\infty}
   }{
       \left(-\frac{\epsilon_{+}}{\epsilon_{-}}x^{-s}q^{H-1};q^{-2}
   \right)_{\infty}
   }
   \qquad \text{for} \quad |q|>1, 
   \label{sol7}
   \\[6pt]
   &=  x^{-s_{1}H}
 \frac{
      \left(-\frac{\epsilon_{+}}{\epsilon_{-}}x^{-s}q^{H+1};q^{2}
   \right)_{\infty}
   }{
       \left(-\frac{\epsilon_{+}}{\epsilon_{-}}x^{s}q^{H+1};q^{2}
   \right)_{\infty}
   }
   \qquad \text{for} \quad |q|<1.
   \label{sol7-2}
\end{align} 
These solutions \eqref{sol6}-\eqref{sol7-2} satisfy
\begin{align}
 \Kf(x) \Kf(x^{-1})= \Kf(x^{-1}) \Kf(x)=1, \qquad  \Kf(1)=1.
\end{align}
Note that other expressions are given in Appendix \ref{ApC}. 
\subsubsection{Reflection equations}
Let us define the R-operators in $U_{q}(sl_2) \otimes U_{q}(sl_2)$ by 
$ \Rf_{12}(x,y)=(\mathsf{ev}_{x} \otimes \mathsf{ev}_{y}) \mathcal{R}$ 
and $ \Rf_{21}(x,y)=(\mathsf{ev}_{x} \otimes \mathsf{ev}_{y}) \mathcal{R}_{21}$. 
Then the first relations in 
\eqref{R-def} and \eqref{Rb-def} produce the following intertwining relations
\begin{align}
\begin{split}
((\mathsf{ev}_{x} \otimes \mathsf{ev}_{y}) \Delta^{\prime}(a) ) \Rf_{12}(x,y)
&=
\Rf_{12}(x,y) ((\mathsf{ev}_{x} \otimes \mathsf{ev}_{y}) \Delta(a) ),
\\
((\mathsf{ev}_{x} \otimes \mathsf{ev}_{y}) \Delta(a) ) \Rf_{21}(x,y)
&=
\Rf_{21}(x,y) ((\mathsf{ev}_{x} \otimes \mathsf{ev}_{y}) \Delta^{\prime}(a) )
\quad 
\text{for any }
a \in U_{q}(\widehat{sl_2}). 
 \label{R-intert}
\end{split}
\end{align}
The intertwining relations \eqref{intertao0-1} and \eqref{R-intert}
imply the following 
reflection equation in $U_{q}(sl_2) \otimes U_{q}(sl_2)$:
%
%
\begin{align}
\Rf_{12}(x^{-1},y^{-1}) \Kf_{1}(x) \Rf_{21}(x,y^{-1}) \Kf_{2}(y) 
=\Kf_{2}(y) \Rf_{12}(x^{-1},y)  \Kf_{1}(x) \Rf_{21}(x,y) ,
\label{un-refeq0}
\end{align}
where we set $ \Kf_{1}(x) =\Kf(x) \otimes 1$, $ \Kf_{2}(y) =1\otimes \Kf(y) $.
In fact, the intertwining relations 
$\mathbf{r}_{i}((\mathsf{ev}_{x} \otimes \mathsf{ev}_{y})\Delta^{\prime}(a))=
((\mathsf{ev}_{x^{-1}} \otimes \mathsf{ev}_{y^{-1}})\Delta^{\prime}(a)) \mathbf{r}_{i}$ 
for any $a \in {\cal O}_q^{aug} $ 
follow
\footnote{This is not a substitute of a proof of \eqref{un-refeq0}. 
One will be able to prove this on the level of irreducible representations of 
${\cal O}_q^{aug}$ by using the Schur's lemma 
(which fixes $\mathbf{r}_{1}=\mathrm{scalar} \times \, \mathbf{r}_{2}$) and 
an assumption on the behavior of $\mathbf{r}_{i}$ with respect to 
the spectral parameters $x,y$ (which determines $\mathrm{scalar}=1$). 
We do not have a universal K-matrix relevant to our discussion. Thus 
we do not have a proof of \eqref{un-refeq0} on the level of the algebra. 
On the other hand, we have a proof of the reflection equation for the L-operators \eqref{refeq2}, 
which follows from this generic reflection equation \eqref{un-refeq0}.} 
from \eqref{intertao0-1} and \eqref{R-intert}, 
where the right hand side and the left hand side of 
\eqref{un-refeq0} are denoted as 
$\mathbf{r}_{1} $ and $\mathbf{r}_{2} $, respectively. 
Evaluating \eqref{un-refeq0} for $1 \otimes \pi_{1}$, we obtain the following 
reflection equation for the L-operators
\begin{align}
\Lf \left(\frac{y}{x}\right) \Kf_{1}(x) \overline{\Lf} \left( xy \right) 
 K_{2}(y) 
=K_{2}(y) 
 \Lf \left(\frac{1}{xy} \right)  \Kf_{1}(x) \overline{\Lf} \left(\frac{x}{y}\right),
\label{refeq2}
\end{align}
where we set $K (x) =\pi_{1}(\Kf(x))$ 
and used the difference property with respect to the spectral parameters. 
Expanding \eqref{refeq2} with respect to the spectral parameter $y$,  one recognizes  the intertwining relations \eqref{intertao0-1}. 

\paragraph{  Specialization to $\pi_{1}$:}
Evaluating \eqref{refeq2} further  for $\pi_{1}\otimes 1$, we obtain the following 
reflection equation for the R-matrices.
\begin{align}
R \left(\frac{y}{x}\right) K_{1}(x) \overline{R} \left( xy \right) 
 K_{2}(y) 
=K_{2}(y) 
 R \left(\frac{1}{xy} \right)  K_{1}(x) \overline{R} \left(\frac{x}{y}\right). 
\label{refeq0}
\end{align}
The solution of \eqref{refeq0} is given by 
\begin{align}
K (x)& =\pi_{1}(\Kf(x))=
\kappa(x)
\begin{pmatrix}
x^{s_{0}}\epsilon_{+} + x^{-s_{1}} \epsilon_{-} & 
0 \\
0  & 
x^{-s_{0}}\epsilon_{+} + x^{s_{1}} \epsilon_{-}
 \end{pmatrix} .
 \label{K-mat1}
 \end{align}
 Here $ \kappa(x)$ is an overall factor.  In case one uses  \eqref{sol6} for $|q|>1$, it reads 
 \begin{align}
  \kappa(x)=
 \frac{
      \left(-\frac{\epsilon_{-}}{\epsilon_{+}}x^{s}q^{-2};q^{-2}
   \right)_{\infty}
   }{
       \epsilon_{+} \left(-\frac{\epsilon_{-}}{\epsilon_{+}}x^{-s};q^{-2}
   \right)_{\infty}
   }.  
   \label{overallK}
 \end{align}
Note that the solution (\ref{K-mat1}) is a special case of the most general scalar solution
\footnote{Here the word `scalar' means the matrix elements of the solution are not 
operators but scalar quantities.}
 \eqref{K-mat1-genfun} of the reflection equation (\ref{refeq0}) \cite{DeV,GZ}. In the context of quantum integrable systems,  it characterizes systems with arbitrary diagonal boundary conditions.
\subsection{The second realization}
Next, the augmented $q$-Onsager algebra is realized as a left coidal subalgebra of $U_q(\widehat{sl_2})$. An intertwiner $\overline{\Kf}(x)$ is explicitly constructed.

\subsubsection{Left coideal subalgebra of $U_{q}(\widehat{sl_2})$ and the intertwiner $ \overline{\Kf}(x)$}
Using the automorphism \eqref{auto2} of $U_q(\widehat{sl_2})$, 
%
%
%
a second realization of the augmented $q-$Onsager algebra ${\cal O}_q^{aug}$, now as a left-coideal subalgebra of $U_q(\widehat{sl_2})$. Let $\overline{\epsilon}_\pm$ be non-zero scalars. 
It is given by
\footnote{$(\overline{\textsf K}_0, \overline{\textsf K}_1, 
 \overline{\textsf Z}_1 ,\tilde{\overline{\textsf Z}}_1 ) $ 
in \eqref{realopaugbar} corresponds to 
$({\textsf K}_0, {\textsf K}_1,  -\tilde{\textsf Z}_1 , -{\textsf Z}_1)$ 
 in eq.\ (3.21) in \cite{BB1} under the transformations $q \mapsto q^{-1}$ 
 and $ \overline{\epsilon}_{\pm} \mapsto \epsilon_{\mp} $.
In addition, instead of using the automorphism \eqref{auto2}, one can also use an anti-automorphism 
$\bar{\tau}$ defined by 
\begin{align}
& \bar{\tau} (e_{i})=-f_{i}, \qquad \bar{\tau} (f_{i})=-e_{i}, \qquad \bar{\tau} (h_{i})=h_{i} 
, \qquad \bar{\tau} (q)=q^{-1}, 
\qquad i=0,1. 
\label{anti-auto}
\end{align}
}:
\begin{align}
\begin{split}
\overline{\textsf K}_0&= \tau({\textsf K}_0)= \overline{\epsilon}_-q^{h_0}  \ ,
\qquad \overline{\textsf K}_1= \tau({\textsf K}_1)  =\overline{\epsilon}_+ q^{h_1} ,
\\
\overline{\textsf Z}_1&= \tau({\textsf Z}_1) = (q^2-q^{-2})\big( \overline{\epsilon}_+ q e_1q^{h_1} + \overline{\epsilon}_- f_0 \big)  ,
\\
\tilde{\overline{\textsf Z}}_1&=  \tau(\tilde{\textsf Z}_1)  =  (q^2-q^{-2})\big( \overline{\epsilon}_+ f_1 + \overline{\epsilon}_-q e_0q^{h_0} \big)
 \ . 
 \end{split}
 \label{realopaugbar}
\end{align}
where $\tau(\epsilon_{\pm})=\overline{\epsilon}_{\mp}$ is assumed. 
%
%
%
%
Note that 
the automorphism \eqref{auto1} of $U_{q}(\widehat{sl_2})$ also gives the automorphism of ${\cal O}_q^{aug}$,
\begin{align}
\sigma : \quad 
\overline{\textsf K}_{0} \mapsto \overline{\textsf K}_{1} ,
\quad 
\overline{\textsf K}_{1} \mapsto \overline{\textsf K}_{0} ,
\quad 
\overline{\textsf Z}_1 \mapsto \tilde{\overline{\textsf Z}}_1 ,
\quad 
\tilde{\overline{\textsf Z}}_1 \mapsto \overline{\textsf Z}_1 ,
\end{align}
 under the condition  $\sigma(\overline{\epsilon}_{\pm})=\overline{\epsilon}_{\mp}$. 

The co-action map $\Delta : {\cal O}_q^{aug} \mapsto U_q(\widehat{sl_2}) \otimes {\cal O}_q^{aug} $, that is compatible with the relations (\ref{Tauggen}) corresponds to the restriction of the co-product \eqref{copro-h} of  $U_q(\widehat{sl_2})$ 
 to ${\cal O}_q^{aug}$ under the realization \eqref{realopaugbar}. 
It is such that:
\begin{align}
\begin{split}
\Delta(\overline{\textsf K}_0)&=  q^{h_0} \otimes \overline{\textsf K}_0 \ ,\qquad \Delta(\overline{\textsf K}_1)= q^{h_1}\otimes \overline{\textsf K}_1  \ ,
\\
\Delta(\overline{\textsf Z}_1)&= 1 \otimes \overline{\textsf Z}_1   + (q^2-q^{-2})\left( qe_1q^{h_1} \otimes \overline{\textsf K}_1 + f_0\otimes \overline{\textsf K}_0 \right),\\
\Delta(\tilde{\overline{\textsf Z}}_1)&=   
  1 \otimes \tilde{\overline{\textsf Z}}_1 + (q^2-q^{-2})\left( f_1 \otimes \overline{\textsf K}_1  + qe_0q^{h_0} \otimes \overline{\textsf K}_0   \right).
\end{split}
\label{deltadefaugbar}
\end{align}

Recall the evaluation map (\ref{eva}). 
Define
\footnote{One may also interpret this as 
$\overline{\mathsf{ev}}_{x} ={\mathsf{ev}}_{x} 
\circ \sigma \circ \tau |_{\epsilon_{\pm}=\overline{\epsilon}_{\pm}} $, 
$\tau(\overline{\epsilon}_{\pm} )= \epsilon_{\mp}$.
} 
$\overline{\mathsf{ev}}_{x}=\mathsf{ev}_{x}|_{(s_0,s_1)\rightarrow (-s_1,-s_0)}$
. It follows:
\beqa
\overline{\mathsf{ev}}_{x}(\overline{\textsf{K}}_{0})&=& \overline{\epsilon}_-q^{-H}, \quad \overline{\mathsf{ev}}_{x}(\overline{\textsf{K}}_{1})= \overline{\epsilon}_+q^{H} ,\nonumber\\
\overline{\mathsf{ev}}_{x}(\overline{\textsf{Z}}_{1})&=& (q^2-q^{-2})E\big( \overline{\epsilon}_+x^{-s_0}q^{H+1} + \overline{\epsilon}_-x^{s_1} \big),\nonumber\\
\overline{\mathsf{ev}}_{x}(\tilde{\overline{\textsf{Z}}}_{1})&=& (q^2-q^{-2})F\big( \overline{\epsilon}_+x^{s_0} + \overline{\epsilon}_- x^{-s_1}q^{-H+1} \big) .\nonumber
\eeqa

We now consider the following 
 intertwining relations associated with the second realization of the augmented $q-$Onsager algebra: 
\begin{align}
\overline{\mathsf{ev}}_{x^{-1}q^{-\frac{2}{s}}}(a) \g \overline{\Kf}(x)^{t} &= 
\g \overline{\Kf}(x)^{t} \overline{\mathsf{ev}}_{xq^{\frac{2}{s}}}(a) 
 \qquad 
 \text{for any } 
 a \in 
 \{ \overline{\textsf{K}}_{0},\overline{\textsf{K}}_{1} ,
 \overline{\textsf{Z}}_{1},\tilde{\overline{\textsf{Z}}}_{1} \},
 \label{intertao0-1du}
\end{align}
where $\g=q^{H(s_{0}-s_{1})/s}$. 
Here $^{t}$ is the transposition. 
One may drop it from \eqref{intertao0-1du} 
since the K-operator here is a diagonal operator. 
The  equations for $ a \in 
 \{ \overline{\textsf{K}}_{0},\overline{\textsf{K}}_{1}\}$ 
 imply 
  that $[ \overline{\Kf}(x)^{t},q^H]=0$. The  equations for $ a \in 
 \{ \overline{\textsf{Z}}_{1},\tilde{ \overline{\textsf{Z}}}_{1} \}$ give:
\begin{align}
E (\bar{\epsilon}_{+}q x^{s_{0}} q^{H+1}+ \bar{\epsilon}_{-} q^{-1} x^{-s_{1}}  ) 
 \overline{\Kf}(x)^{t}
&=
\overline{\Kf}(x)^{t}
E(\bar{\epsilon}_{+}q^{-1} x^{-s_{0}} q^{H+1}+ \bar{\epsilon}_{-} q x^{s_{1}}  ) , 
 \label{intertao1-1du}
 \\[6pt]
F ( \bar{\epsilon}_{+}q^{-1} x^{-s_{0}} +\bar{\epsilon}_{-}q x^{s_{1}}q^{-H+1}  ) 
 \overline{\Kf}(x)^{t}
&=
\overline{\Kf}(x)^{t}
F ( \bar{\epsilon}_{+}q x^{s_{0}} +\bar{\epsilon}_{-}q^{-1} x^{-s_{1}}q^{-H+1}  ). 
 \label{intertao1-2du}
\end{align}
\subsubsection{Solutions of the intertwining relations}
The solutions of the intertwining relations \eqref{intertao0-1du} 
follow from the ones for the first realization \eqref{intertao0-1} under the identification
\begin{align}
\Kf(x) =\g \overline{\Kf}(xq^{-\frac{2}{s}})^{t}|_{\overline{\epsilon}_{\pm}=\epsilon_{\pm}} . 
 \label{id-du}
\end{align}
\subsubsection{Reflection equations}
The intertwining relations 
\eqref{intertao0-1du} and \eqref{R-intert}
imply the following dual 
reflection equation in $U_{q}(sl_2) \otimes U_{q}(sl_2)$:
\begin{multline}
\Rf_{12}(x^{-1}q^{-\frac{2}{s}},y^{-1}q^{-\frac{2}{s}})
\g_{1} \overline{\Kf}_{1}(x)^{t_{1}}
 \Rf_{21}(xq^{\frac{2}{s}},y^{-1}q^{-\frac{2}{s}})
 \g_{2}  \overline{\Kf}_{2}(y)^{t_{2}}  
 =
 \\
=\g_{2} \overline{\Kf}_{2}(y)^{t_{2}}  \Rf_{12}(x^{-1}q^{-\frac{2}{s}},yq^{\frac{2}{s}})  
\g_{1} \overline{\Kf}_{1}(x)^{t_{1}} 
 \Rf_{21}(xq^{\frac{2}{s}},yq^{\frac{2}{s}}) .
\label{un-refeq0du}
\end{multline}
This also follows from \eqref{un-refeq0} under the identification \eqref{id-du}. 
Evaluating \eqref{un-refeq0du} for $1 \otimes \pi_{1}$, we obtain 
the following dual reflection equation for the L-operators \cite{MN}:
\begin{align}
\Lf \left(\frac{y}{x}\right) \overline{\Kf}_{1}(x)^{t_{1}} 
g_{2}^{-1} \overline{\Lf}\left( xy q^{\frac{4}{s}}\right) g_{2}
\overline{K}_{2}(y)^{t_{2}} 
=\overline{K}_{2}(y)^{t_{2}}
g_{2} \Lf \left(\frac{q^{-\frac{4}{s}}}{xy}\right)  g_{2}^{-1}
\overline{\Kf}_{1}(x)^{t_{1}} \overline{\Lf}\left(\frac{x}{y}\right),
\label{refeq2dual}
\end{align}
where $g=\pi_1(\g^{-1})$ and $\overline{K}(x) =\pi_{1}(\overline{\Kf}(x))$.
Taking appropriate limits in the variable $y$ of the reflection equation (\ref{refeq2}), one recovers the intertwining relations \eqref{intertao1-1du}-\eqref{intertao1-2du}.\vspace{1mm}

\paragraph{  Specialization to $\pi_1$:} Specializing to the two-dimensional representation of $U_q(sl_2)$, the solution of the intertwining relations is unique 
(up to an overall factor). It reads:
\begin{align}
\overline{K}(x)& =\pi_{1}(\overline{\Kf}(x))=
\overline{\kappa}(x)
\begin{pmatrix}
qx^{s_{0}}\overline{\epsilon}_{+} + q^{-1}x^{-s_{1}}\overline{\epsilon}_{-} & 
0 \\
0 & 
q^{-1}x^{-s_{0}}\overline{\epsilon}_{+} + qx^{s_{1}}\overline{\epsilon}_{-}
 \end{pmatrix} .
 \label{Ksoldual}
 \end{align}
 Here $ \overline{\kappa}(x)$ is an overall factor.  In case one uses  \eqref{sol6} for $|q|>1$ 
 and \eqref{id-du}, it reads 
 \begin{align}
 \overline{\kappa}(x)= 
 \kappa(xq^{\frac{2}{s}})|_{\epsilon_{\pm}=\overline{\epsilon}_{\pm}}
 =
 \frac{
      \left(-\frac{\overline{\epsilon}_{-}}{\overline{\epsilon}_{+}}x^{s};q^{-2}
   \right)_{\infty}
   }{
      \overline{\epsilon}_{+} \left(-\frac{\overline{\epsilon}_{-}}{\overline{\epsilon}_{+}}x^{-s}q^{-2};q^{-2}
   \right)_{\infty}
   }.  
   \label{overallKbar}
 \end{align}
By construction, it solves the specialization of the dual reflection equation\footnote{One can modify this by 
the relations $ g_{1} \overline{R}\left( xy q^{-\frac{4}{s}}\right) g_{1}^{-1}=
g_{2}^{-1} \overline{R}\left( xy q^{-\frac{4}{s}}\right) g_{2}$,  
$g_{1}^{-1} R\left(\frac{q^{\frac{4}{s}}}{xy}\right)  g_{1}=
g_{2} R\left(\frac{q^{\frac{4}{s}}}{xy}\right)  g_{2}^{-1}$.} (\ref{refeq2dual}):
\begin{align}
R\left(\frac{y}{x}\right) \overline{K}_{1}(x)^{t_{1}} 
g_{1} \overline{R}\left( xyq^{\frac{4}{s}} \right) g_{1}^{-1}
\overline{K}_{2}(y)^{t_{2}} 
=\overline{K}_{2}(y)^{t_{2}}
g_{1}^{-1} R\left(\frac{q^{-\frac{4}{s}}}{xy}\right)  g_{1}
\overline{K}_{1}(x)^{t_{1}} \overline{R}\left(\frac{x}{y}\right).
\label{refeqdual0-2}
\end{align}
The solution (\ref{Ksoldual}) is a special case of the general scalar solution \eqref{K-matdual1-genfun} of the dual reflection equation. 
Note that the solutions of the reflection and dual reflection equations are related by the following transformation (see \cite{MN}):
\beqa
\overline{K}(x) = K^t\left(xq^{2/s}\right)g^{-1}|_{\overline{\epsilon}_{\pm}=\epsilon_{\pm}}. 
\nonumber
\eeqa

\subsection{Limit of intertwining relations and their solutions: K-operators for Q-operators}
In this subsection, we consider the limit  $q^{\mp \mu} \to 0$ 
of the intertwining relations and their solutions in the Verma module $\pi^{+}_{\mu}$. 
In order to avoid divergences, 
we have to renormalize the generators of the augmented q-Onsager algebra
 and the K-operators. The resulting K-operators 
serve as building blocks of Q-operators in that they solve 
reflection equations for the $L$-operators for Q-operators. 
Similar K-operators for the rational case can be found in \cite{FS15}.

\subsubsection{The first realization}

\paragraph{  The limit $q^{-\mu} \to 0$ under the shift $x \to xq^{\frac{\mu}{s}}$:}
Let us make a shift $x \to xq^{\frac{\mu}{s}}$ on the spectral parameter in 
\eqref{intertao1-1}-\eqref{intertao1-2} and multiply 
the factors $x^{-s_{0}\mu}q^{-\frac{\mu s_{0}H}{s}  -\frac{2s_{0}\mu}{s}} $ 
(for \eqref{intertao1-1}) and 
$x^{-s_{0}\mu}q^{-\frac{\mu s_{0}H}{s}  - \frac{2s_{1}\mu}{s}} $ 
(for \eqref{intertao1-2}) from the left. 
We find that 
the limit $q^{-\mu} \to 0$ for this in  $\pi^{+}_{\mu} $  produces 
\begin{align}
&\fs_{1}  ( \epsilon_{+} x^{-s_{0}} +\epsilon_{-} x^{s_{1}}  q^{-\hs_{1}+1} ) \Kf^{(1)}(x)
=
 \Kf^{(1)}(x) \fs_{1} (
 \epsilon_{+} x^{s_{0}} ), 
 \label{intertaolim1-1}
 \\[6pt]
& \es_{1}  ( \epsilon_{+} x^{s_{0}}q^{\hs_{1}+1}  ) \Kf^{(1)}(x)
=
 \Kf^{(1)}(x)\es_{1} (\epsilon_{+} x^{-s_{0}}q^{\hs_{1}+1}  +\epsilon_{-} x^{s_{1}}  ), 
 \label{intertaolim1-2}
\end{align}
where we set 
\begin{align}
\Kf^{(1)}(x)=
 \lim_{q^{-\mu} \to 0} x^{-s_{0}\mu}
  \pi^{+}_{\mu} ( q^{-\frac{s_{0}\mu H}{s}} 
\Kf(q^{\frac{\mu}{s}}x)).
  \label{Kmat-lim1}
\end{align}
Similarly, we derive the limit of the equations \eqref{intertao0-1} for the renormalized 
generators ${\textsf K}_{0}q^{-\mu}$ and ${\textsf K}_{1}q^{\mu}$:  
\begin{align}
q^{\pm \hs_{1}}\Kf^{(1)}(x)=\Kf^{(1)}(x) q^{\pm \hs_{1}}.
 \label{Cartlim1}
\end{align}
Then a solution
of \eqref{intertaolim1-1}, \eqref{intertaolim1-2} and \eqref{Cartlim1} is given
 by the limit of \eqref{sol6} or \eqref{sol6-2}:
\begin{align}
\begin{split}
 \Kf^{(1)}(x) 
 &=
  x^{s_{0}\hs_{1}}    \left(-\frac{\epsilon_{-}}{\epsilon_{+}}x^{s}q^{-\hs_{1}-1};q^{-2}
   \right)_{\infty}
   \quad \text{for} \quad |q|>1,
   \\[6pt]
   &=
  x^{s_{0}\hs_{1}}    \left(-\frac{\epsilon_{-}}{\epsilon_{+}}x^{s}q^{-\hs_{1}+1};q^{2}
   \right)_{\infty}^{-1}
   \quad \text{for} \quad |q|<1 . 
\end{split}   
   \label{sol6lim}
\end{align} 

\paragraph{  The limit $q^{-\mu} \to 0$ under the shift $x \to xq^{-\frac{\mu}{s}}$:}
Let us make a shift $x \to xq^{-\frac{\mu}{s}}$ on the spectral parameter in 
\eqref{intertao1-1}-\eqref{intertao1-2} and multiply 
the factor $x^{-s_{0}\mu}q^{\frac{\mu s_{0}H}{s}} $ (for \eqref{intertao1-1})
and $x^{-s_{0}\mu}q^{\frac{\mu s_{0}H}{s} -2\mu } $  (for \eqref{intertao1-2}) from the left. 
We find that 
the limit $q^{-\mu} \to 0$ for this in  $\pi^{+}_{\mu} $  produces 
\begin{align}
&\fs_{1}  ( \epsilon_{+} x^{-s_{0}}   ) \Kfc^{(1)}(x)
=
 \Kfc^{(1)}(x) \fs_{1} (
 \epsilon_{+} x^{s_{0}} +\epsilon_{-} x^{-s_{1}} q^{-\hs_{1}+1}), 
 \label{intertaolim1-1ch}
 \\[6pt]
& \es_{1}  (\epsilon_{-} x^{-s_{1}}  + \epsilon_{+} x^{s_{0}} q^{\hs_{1}+1}) \Kfc^{(1)}(x)
=
 \Kfc^{(1)}(x)\es_{1} (\epsilon_{+} x^{-s_{0}} q^{\hs_{1}+1} ), 
 \label{intertaolim1-2ch}
\end{align}
where we set 
\begin{align}
\Kfc^{(1)}(x)=
 \lim_{q^{-\mu} \to 0} x^{-s_{0}\mu}
  \pi^{+}_{\mu} ( q^{\frac{s_{0}\mu H}{s}} 
\Kf(q^{-\frac{\mu}{s}}x)).
  \label{Kmat-lim1ch}
\end{align}
Similarly, we derive the limit of the equations \eqref{intertao0-1} for the renormalized 
generators ${\textsf K}_{0}q^{-\mu}$ and ${\textsf K}_{1}q^{\mu}$:  
\begin{align}
q^{\pm \hs_{1}}\Kfc^{(1)}(x)=\Kfc^{(1)}(x) q^{\pm \hs_{1}}.
 \label{Cartlim1ch}
\end{align}
Then a solution
of \eqref{intertaolim1-1ch}, \eqref{intertaolim1-2ch} and \eqref{Cartlim1ch} is given
 by the limit of \eqref{sol6} or \eqref{sol6-2}:
\begin{align}
\begin{split}
 \Kfc^{(1)}(x) 
 &=
  x^{s_{0}\hs_{1}}    \left(-\frac{\epsilon_{-}}{\epsilon_{+}}x^{-s}q^{-\hs_{1}-1};q^{-2}
   \right)_{\infty}^{-1}
   \quad \text{for} \quad |q|>1, 
   \\[6pt]
   &=
  x^{s_{0}\hs_{1}}    \left(-\frac{\epsilon_{-}}{\epsilon_{+}}x^{-s}q^{-\hs_{1}+1};q^{2}
   \right)_{\infty}
   \quad \text{for} \quad |q|<1 . 
\end{split}
   \label{sol6limch}
\end{align} 

\paragraph{  The limit $q^{\mu} \to 0$ under the shift $x \to xq^{-\frac{\mu}{s}}$:}
Let us make a shift $x \to xq^{-\frac{\mu}{s}}$ on the spectral parameter in 
\eqref{intertao1-1}-\eqref{intertao1-2} and multiply 
the factors $x^{s_{1}\mu}q^{-\frac{\mu s_{1}H}{s}+\frac{2s_{0}\mu}{s}} $ 
(for \eqref{intertao1-1})  and 
$x^{s_{1}\mu}q^{-\frac{\mu s_{1}H}{s}+\frac{2s_{1}\mu}{s}} $ 
(for \eqref{intertao1-2}) 
from the left. 
We find that 
the limit $q^{\mu} \to 0$ for this in  $\pi^{+}_{\mu} $ produces 
\begin{align}
&\fs_{2}  (\epsilon_{-} x^{s_{1}} q^{-\hs_{2}+1}) \Kf^{(2)}(x)
=
 \Kf^{(2)}(x) \fs_{2} ( \epsilon_{+} x^{s_{0}} +
 \epsilon_{-} x^{-s_{1}} q^{-\hs_{2}+1}
), 
 \label{intertaolim2-1}
 \\[6pt]
& \es_{2}  (\epsilon_{+} x^{s_{0}}q^{\hs_{2}+1}+ \epsilon_{-} x^{-s_{1}}  
 )
 \Kf^{(2)}(x)
=
 \Kf^{(2)}(x)\es_{2} (\epsilon_{-} x^{s_{1}} 
 ),
 \label{intertaolim2-2}
\end{align}
where we define
\begin{align}
 \Kf^{(2)}(x)=
 \lim_{q^{\mu} \to 0} x^{s_{1}\mu} 
  \pi^{+}_{\mu} (q^{-\frac{s_{1}\mu H}{s}} \Kf(q^{-\frac{\mu}{s}}x) ) . 
 \label{Kmat-lim2}
 \end{align}
 Similarly, we derive the limit of the equations \eqref{intertao0-1} for the renormalized 
generators ${\textsf K}_{0}q^{-\mu}$ and ${\textsf K}_{1}q^{\mu}$:  
\begin{align}
q^{\pm \hs_{2}}\Kf^{(2)}(x)=\Kf^{(2)}(x) q^{\pm \hs_{2}}.
 \label{Cartlim2}
\end{align}
Then a 
solution of \eqref{intertaolim2-1}, \eqref{intertaolim2-2} and \eqref{Cartlim2} is given by 
the limit of \eqref{sol7} or \eqref{sol7-2} in  $\pi^{+}_{\mu} $:
\begin{align}
\begin{split}
 \Kf^{(2)}(x)
 &=
  x^{-s_{1}\hs_{2}}    \left(-\frac{\epsilon_{+}}{\epsilon_{-}}x^{s}q^{\hs_{2}-1};q^{-2}
   \right)_{\infty}
   \qquad \text{for} \quad |q|>1, 
   \\[6pt]
   &=
  x^{-s_{1}\hs_{2}}    \left(-\frac{\epsilon_{+}}{\epsilon_{-}}x^{s}q^{\hs_{2}+1};q^{2}
   \right)_{\infty}^{-1}
   \qquad \text{for} \quad |q|<1.
\end{split}
   \label{sol7lim}
\end{align} 

\paragraph{  The limit $q^{\mu} \to 0$ under the shift $x \to xq^{\frac{\mu}{s}}$:}
Let us make a shift $x \to xq^{\frac{\mu}{s}}$ on the spectral parameter in 
\eqref{intertao1-1}-\eqref{intertao1-2} and multiply 
the factor $x^{s_{1}\mu}q^{\frac{\mu s_{1}H}{s}+2\mu } $ (for \eqref{intertao1-1}) and 
$x^{s_{1}\mu}q^{\frac{\mu s_{1}H}{s} } $ (for \eqref{intertao1-2})  from the left. 
We find that 
the limit $q^{\mu} \to 0$ for this in  $\pi^{+}_{\mu} $ produces 
\begin{align}
&\fs_{2}  ( \epsilon_{+} x^{-s_{0}}+\epsilon_{-} x^{s_{1}}  q^{-\hs_{2}+1}) \Kfc^{(2)}(x)
=
 \Kfc^{(2)}(x) \fs_{2} (
 \epsilon_{-} x^{-s_{1}}  q^{-\hs_{2}+1}
), 
 \label{intertaolim2-1ch}
 \\[6pt]
& \es_{2}  ( \epsilon_{-} x^{-s_{1}} 
 )
 \Kfc^{(2)}(x)
=
 \Kfc^{(2)}(x)\es_{2} (\epsilon_{-} x^{s_{1}} +
 \epsilon_{+} x^{-s_{0}} q^{\hs_{2}+1}
 ),
 \label{intertaolim2-2ch}
\end{align}
where we define
\begin{align}
 \Kfc^{(2)}(x)=
 \lim_{q^{\mu} \to 0} x^{s_{1}\mu} 
  \pi^{+}_{\mu} (q^{\frac{s_{1}\mu H}{s}} \Kf(q^{\frac{\mu}{s}}x) ) . 
 \label{Kmat-lim2ch}
 \end{align}
 Similarly, we derive the limit of the equations \eqref{intertao0-1} for the renormalized 
generators ${\textsf K}_{0}q^{-\mu}$ and ${\textsf K}_{1}q^{\mu}$:  
\begin{align}
q^{\pm \hs_{2}}\Kfc^{(2)}(x)=\Kfc^{(2)}(x) q^{\pm \hs_{2}}.
 \label{Cartlim2ch}
\end{align}
Then a 
solution of \eqref{intertaolim2-1ch}, \eqref{intertaolim2-2ch} and \eqref{Cartlim2ch} is give by 
the limit of \eqref{sol7} or \eqref{sol7-2} in  $\pi^{+}_{\mu} $:
\begin{align}
\begin{split}
 \Kfc^{(2)}(x)
 &=
  x^{-s_{1}\hs_{2}}    \left(-\frac{\epsilon_{+}}{\epsilon_{-}}x^{-s}q^{\hs_{2}-1};q^{-2}
   \right)_{\infty}^{-1}
   \qquad \text{for} \quad |q|>1, 
   \\[6pt]
   &=
  x^{-s_{1}\hs_{2}}    \left(-\frac{\epsilon_{+}}{\epsilon_{-}}x^{-s}q^{\hs_{2}+1};q^{2}
   \right)_{\infty}
   \qquad \text{for} \quad |q|<1.
\end{split}
   \label{sol7limch}
\end{align} 

Renormalizing \eqref{refeq2} and taking the limits $q^{\mp \mu} \to 0$ 
 in $\pi^{+}_{\mu} \otimes 1$, we obtain the 
reflection equations
\footnote{
The limit does not change the form of the 
 reflection equation for the L-operators \eqref{refeq2}. 
 Making 
the shift $x \to xq^{\frac{\mu}{s}}$ on the spectral parameter in \eqref{refeq2}, and 
 multiplying 
$(\mathsf{ev}_{x^{-1}} \otimes \pi_{1}(y^{-1})) ( q^{-\mu h_{1}} \otimes q^{-\mu h_{1}} )
=q^{-\mu H} \otimes q^{-\mu (E_{11}-E_{22})}$ from the left and 
$\left(q^{\frac{s_{1}\mu H}{s}}\otimes 1 \right)$ from the right, 
we obtain 
\begin{multline}
\Lf \left(\frac{yq^{-\frac{\mu}{s}}}{x}\right) q^{-\frac{\mu \otimes \pi_{1}(1)(h_{1})}{2}}
 \left(q^{-\frac{s_{0}\mu H}{s}} \Kf(xq^{\frac{\mu}{s}}) \otimes 1 \right) 
 \left(q^{\frac{(s_{0}-s_{1})\mu H}{2s}}\otimes 1 \right) 
 \overline{\Lf} \left(q^{\frac{\mu}{s}}xy \right) 
  \left(q^{-\frac{(s_{0}-s_{1})\mu H}{2s}}\otimes 1 \right)
\\
\times 
  q^{-\frac{\mu \otimes \pi_{1}(1)(h_{1})}{2}} 
K_{2}(y) =
\\
=K_{2}(y)
 \Lf (\frac{q^{-\frac{\mu}{s}}}{xy}) q^{-\frac{\mu \otimes \pi_{1}(1)(h_{1})}{2}} 
  \left( q^{-\frac{s_{0}\mu H}{s}} \Kf(xq^{\frac{\mu}{s}}) \otimes 1 \right) 
 \left(q^{\frac{(s_{0}-s_{1})\mu H}{2s}}\otimes 1 \right)
 \overline{\Lf} \left(\frac{xq^{\frac{\mu}{s}}}{y}\right)
 \\
\times 
 \left(q^{-\frac{(s_{0}-s_{1})\mu H}{2s}}\otimes 1 \right)
   q^{-\frac{\mu \otimes \pi_{1}(1)(h_{1})}{2}}.
 \nonumber 
\end{multline}
Here we used the fact that $\pi_{1}(y^{-1})(h_{1})$ is independent of $y$, and 
the commutativity \eqref{com-Cartan}.
Multiplying $x^{-s_{0}\mu}$ and taking the limit $q^{-\mu} \to 0$
 in $\pi_{\mu}^{+} \otimes 1 $, we arrive at 
\eqref{refeqlim1st} for $a=1$.
Expanding \eqref{refeqlim1st} with respect to the spectral parameter $y$, 
one can reproduce \eqref{intertaolim1-1}, \eqref{intertaolim1-2} and \eqref{Cartlim1}. 
Similarly,   making 
the shift $x \to xq^{-\frac{\mu}{s}}$ on the spectral parameter in \eqref{refeq2}, 
multiplying 
$x^{-s_{0}\mu}q^{-2\mu} (\mathsf{ev}_{x^{-1}} \otimes \pi_{1}(y^{-1}))
(q^{-\frac{s_{1}\mu h_{1}}{s}}\otimes 1 )
 ( q^{\mu h_{1}} \otimes q^{\mu h_{1}} )$ from the left, 
 and taking the limit $q^{-\mu} \to 0$
 in $\pi_{\mu}^{+} \otimes 1 $, 
 we obtain  \eqref{refeqlim1stch} for $a=1$; 
making 
the shift $x \to xq^{-\frac{\mu}{s}}$ on the spectral parameter in \eqref{refeq2}, 
multiplying 
$x^{s_{1}\mu} (\mathsf{ev}_{x^{-1}} \otimes \pi_{1}(y^{-1}))
 ( q^{-\mu h_{1}} \otimes q^{-\mu h_{1}} )$ from the left 
 and $(q^{\frac{s_{0}\mu H}{s}}\otimes 1 ) $  from the right,  
and taking the limit $q^{\mu} \to 0$
 in $\pi_{\mu}^{+} \otimes 1 $, 
 we obtain  \eqref{refeqlim1st} for $a=2$; 
making 
the shift $x \to xq^{\frac{\mu}{s}}$ on the spectral parameter in \eqref{refeq2}, 
multiplying 
$x^{s_{1}\mu}q^{2\mu} (\mathsf{ev}_{x^{-1}} \otimes \pi_{1}(y^{-1}))
(q^{-\frac{s_{0}\mu h_{1}}{s}}\otimes 1 )
 ( q^{\mu h_{1}} \otimes q^{\mu h_{1}} )$ from the left, 
 and taking the limit $q^{\mu} \to 0$
 in $\pi_{\mu}^{+} \otimes 1 $, 
 we obtain \eqref{refeqlim1stch} for $a=2$.
} 
 for L-operators for Q-operators: 
\begin{align}
\Lf^{(a)} \left(\frac{y}{x}\right) \Kf^{(a)}_{1}(x) \overline{\Lf}^{(a)} \left( xy \right) 
 K_{2}(y) 
& =K_{2}(y) 
 \Lf^{(a)} \left(\frac{1}{xy} \right)  \Kf^{(a)}_{1}(x) \overline{\Lf}^{(a)} \left(\frac{x}{y}\right),
\label{refeqlim1st}
\\[6pt]
\Lfc^{(a)} \left(\frac{y}{x}\right) \Kfc^{(a)}_{1}(x) \Lfcb^{(a)} \left( xy \right) 
 K_{2}(y) 
& =K_{2}(y) 
 \Lfc^{(a)} \left(\frac{1}{xy} \right)  \Kfc^{(a)}_{1}(x) \Lfcb^{(a)} \left(\frac{x}{y}\right), 
 \quad a=1,2.
\label{refeqlim1stch}
\end{align}

\subsubsection{The second realization}

\paragraph{  The limit $q^{-\mu} \to 0$ under the shift $x \to xq^{\frac{\mu}{s}}$:}
Let us make a shift $x \to xq^{\frac{\mu}{s}}$ on the spectral parameter in 
\eqref{intertao1-1du}-\eqref{intertao1-2du} and multiply 
the factors $x^{-s_{0}\mu}q^{-\frac{\mu s_{0}H}{s} -\mu -\frac{2s_{1}\mu}{s}} $ 
(for \eqref{intertao1-1du}) and 
$x^{-s_{0}\mu}q^{-\frac{\mu s_{0}H}{s} -\mu - \frac{2s_{0}\mu}{s}} $ 
(for \eqref{intertao1-2du}) from the left. 
We find that 
the limit $q^{-\mu} \to 0$ for this in  $\pi^{+}_{\mu} $ produces 
\begin{align}
\es_{1} (\bar{\epsilon}_{+}q x^{s_{0}} q^{\hs_{1}+1} ) 
 \overline{\Kf}^{(1)}(x)^{t}
&=
\overline{\Kf}^{(1)}(x)^{t}
\es_{1}(\bar{\epsilon}_{+}q^{-1} x^{-s_{0}} q^{\hs_{1}+1}+ \bar{\epsilon}_{-} q x^{s_{1}}  ) , 
 \label{intertaolim1-1du}
 \\[6pt]
\fs_{1} ( \bar{\epsilon}_{+}q^{-1} x^{-s_{0}} +\bar{\epsilon}_{-}q x^{s_{1}}q^{-\hs_{1}+1}  ) 
 \overline{\Kf}^{(1)}(x)^{t}
&=
\overline{\Kf}^{(1)}(x)^{t}
\fs_{1} ( \bar{\epsilon}_{+}q x^{s_{0}}   ),
 \label{intertaolim1-2du}
\end{align}
where we set
\begin{align}
\overline{\Kf}^{(1)}(x)&=
 \lim_{q^{-\mu} \to 0} x^{-s_{0}\mu} 
 \pi^{+}_{\mu} ( \overline{\Kf}(q^{\frac{\mu}{s}}x) 
 q^{-\frac{s_{0}\mu H}{s}-\mu})
.
  \label{Kmat-lim1du}
\end{align}
Similarly, we derive the limit of the equations \eqref{intertao0-1du} for the renormalized 
generators $\overline{\textsf K}_{0}q^{\mu}$ and $\overline{\textsf K}_{1}q^{-\mu}$:  
\begin{align}
q^{\mp \hs_{1}}\Kfb^{(1)}(x)=\Kfb^{(1)}(x) q^{\mp \hs_{1}}.
 \label{Cartlim1du}
\end{align}

\paragraph{  The limit $q^{-\mu} \to 0$ under the shift $x \to xq^{-\frac{\mu}{s}}$:}
Let us make a shift $x \to xq^{-\frac{\mu}{s}}$ on the spectral parameter in 
\eqref{intertao1-1du}-\eqref{intertao1-2du} and multiply 
the factors $x^{-s_{0}\mu}q^{\frac{\mu s_{0}H}{s} -3\mu } $ 
(for \eqref{intertao1-1du}) and 
$x^{-s_{0}\mu}q^{\frac{\mu s_{0}H}{s} -\mu} $ 
(for \eqref{intertao1-2du}) from the left. 
We find that 
the limit $q^{-\mu} \to 0$ for this in  $\pi^{+}_{\mu} $ produces 
\begin{align}
\es_{1} (\bar{\epsilon}_{+}q x^{s_{0}} q^{\hs_{1}+1}
+\bar{\epsilon}_{-}q^{-1} x^{-s_{1}}  ) 
 \Kfcb^{(1)}(x)^{t}
&=
\Kfcb^{(1)}(x)^{t}
\es_{1}(\bar{\epsilon}_{+}q^{-1} x^{-s_{0}} q^{\hs_{1}+1}  ) , 
 \label{intertaolim1-1duch}
 \\[6pt]
\fs_{1} ( \bar{\epsilon}_{+}q^{-1} x^{-s_{0}}   ) 
 \Kfcb^{(1)}(x)^{t}
&=
\Kfcb^{(1)}(x)^{t}
\fs_{1} ( \bar{\epsilon}_{+}q x^{s_{0}}+\bar{\epsilon}_{-}q^{-1} x^{-s_{1}}q^{-\hs_{1}+1}   ),
 \label{intertaolim1-2duch}
\end{align}
where we set
\begin{align}
\Kfcb^{(1)}(x)&=
 \lim_{q^{-\mu} \to 0} x^{-s_{0}\mu}\pi^{+}_{\mu}
 ( \overline{\Kf}(q^{-\frac{\mu}{s}}x) q^{\frac{s_{0}\mu H}{s}-\mu}
) .
  \label{Kmat-lim1duch}
\end{align}
Similarly, we derive the limit of the equations \eqref{intertao0-1du} for the renormalized 
generators $\overline{\textsf K}_{0}q^{\mu}$ and $\overline{\textsf K}_{1}q^{-\mu}$:  
\begin{align}
q^{\mp \hs_{1}}\Kfcb^{(1)}(x)=\Kfcb^{(1)}(x) q^{\mp \hs_{1}}.
 \label{Cartlim1chdu}
\end{align}

\paragraph{  The limit $q^{\mu} \to 0$ under the shift $x \to xq^{-\frac{\mu}{s}}$:}
Let us make a shift $x \to xq^{-\frac{\mu}{s}}$ on the spectral parameter in 
\eqref{intertao1-1du}-\eqref{intertao1-2du} and multiply 
the factors $x^{s_{1}\mu}q^{-\frac{\mu s_{1}H}{s} +\frac{2s_{1}\mu}{s} +\mu} $ 
(for \eqref{intertao1-1du}) and 
$x^{s_{1}\mu}q^{-\frac{\mu s_{1}H}{s} +\frac{2s_{0}\mu}{s} +\mu } $
(for \eqref{intertao1-2du}) from the left. 
We find that 
the limit $q^{\mu} \to 0$ for this in  $\pi^{+}_{\mu} $ produces 
\begin{align}
\es_{2} (\bar{\epsilon}_{+}q x^{s_{0}} q^{\hs_{2}+1}
+\bar{\epsilon}_{-}q^{-1} x^{-s_{1}}  ) 
 \Kfb^{(2)}(x)^{t}
&=
\Kfb^{(2)}(x)^{t}
\es_{2}(\bar{\epsilon}_{-}q x^{s_{1}}  ) , 
 \label{intertaolim2-1du}
 \\[6pt]
\fs_{2} ( \bar{\epsilon}_{-}q x^{s_{1}}q^{-\hs_{2}+1}   ) 
 \Kfb^{(2)}(x)^{t}
&=
\Kfb^{(2)}(x)^{t}
\fs_{2} ( \bar{\epsilon}_{+}q x^{s_{0}}+\bar{\epsilon}_{-}q^{-1} x^{-s_{1}}q^{-\hs_{2}+1}   ),
 \label{intertaolim2-2du}
\end{align}
where we set
\begin{align}
\Kfb^{(2)}(x)&=
 \lim_{q^{\mu} \to 0} x^{s_{1}\mu}\pi^{+}_{\mu}
 (  \overline{\Kf}(q^{-\frac{\mu}{s}}x) q^{-\frac{s_{1}\mu H}{s}+\mu}
) .
  \label{Kmat-lim2du}
\end{align}
Similarly, we derive the limit of the equations \eqref{intertao0-1du} for the renormalized 
generators $\overline{\textsf K}_{0}q^{\mu}$ and $\overline{\textsf K}_{1}q^{-\mu}$:  
\begin{align}
q^{\mp \hs_{2}}\Kfb^{(2)}(x)=\Kfb^{(2)}(x) q^{\mp \hs_{2}}.
 \label{Cartlim2du}
\end{align}

\paragraph{  The limit $q^{\mu} \to 0$ under the shift $x \to xq^{\frac{\mu}{s}}$:}
Let us make a shift $x \to xq^{\frac{\mu}{s}}$ on the spectral parameter in 
\eqref{intertao1-1du}-\eqref{intertao1-2du} and multiply 
the factors $x^{s_{1}\mu}q^{\frac{\mu s_{1}H}{s} +\mu} $ 
(for \eqref{intertao1-1du}) and 
$x^{s_{1}\mu}q^{\frac{\mu s_{1}H}{s} +3\mu } $
(for \eqref{intertao1-2du}) from the left. 
We find that 
the limit $q^{\mu} \to 0$ for this in  $\pi^{+}_{\mu} $ produces 
\begin{align}
\es_{2} (\bar{\epsilon}_{-}q^{-1} x^{-s_{1}}  ) 
 \Kfcb^{(2)}(x)^{t}
&=
\Kfcb^{(2)}(x)^{t}
\es_{2}(\bar{\epsilon}_{+}q^{-1} x^{-s_{0}} q^{\hs_{2}+1}+\bar{\epsilon}_{-}q x^{s_{1}}  ) , 
 \label{intertaolim2-1duch}
 \\[6pt]
\fs_{2} ( \bar{\epsilon}_{+}q^{-1} x^{-s_{0}}+ \bar{\epsilon}_{-}q x^{s_{1}}q^{-\hs_{2}+1}   ) 
 \Kfcb^{(2)}(x)^{t}
&=
\Kfcb^{(2)}(x)^{t}
\fs_{2} (\bar{\epsilon}_{-}q^{-1} x^{-s_{1}}q^{-\hs_{2}+1}   ),
 \label{intertaolim2-2duch}
\end{align}
where we set
\begin{align}
\Kfcb^{(2)}(x)&=
 \lim_{q^{\mu} \to 0} x^{s_{1}\mu}\pi^{+}_{\mu}
 (  \overline{\Kf}(q^{\frac{\mu}{s}}x) q^{\frac{s_{1}\mu H}{s}+\mu}
) .
  \label{Kmat-lim2duch}
\end{align}
Similarly, we derive the limit of the equations \eqref{intertao0-1du} for the renormalized 
generators $\overline{\textsf K}_{0}q^{\mu}$ and $\overline{\textsf K}_{1}q^{-\mu}$:  
\begin{align}
q^{\mp \hs_{2}}\Kfcb^{(2)}(x)=\Kfcb^{(2)}(x) q^{\mp \hs_{2}}.
 \label{Cartlim2chdu}
\end{align}
Explicit expressions of the second intertwiner $\overline{\Kf}(x)$ 
are given by 
\eqref{id-du} with \eqref{sol6}-\eqref{sol7-2}. 
Thus explicit expressions of 
\eqref{Kmat-lim1du}, \eqref{Kmat-lim1duch}, \eqref{Kmat-lim2du} 
and \eqref{Kmat-lim2duch} are obtained via 
\begin{align}
\Kfb^{(a)}(x)
& 
  = \Kf^{(a)}(xq^{\frac{2}{s}})^{t} q^{-\frac{(s_{0}-s_{1}) \hs_{a}}{s}}
  |_{\epsilon_{\pm}=\bar{\epsilon}_{\pm}},
  \quad 
\Kfcb^{(a)}(x) 
  = \Kfc^{(a)}(xq^{\frac{2}{s}})^{t} q^{-\frac{(s_{0}-s_{1}) \hs_{a}}{s}}
    |_{\epsilon_{\pm}=\bar{\epsilon}_{\pm}}, 
  \quad 
  a=1,2, 
 \end{align}
 and \eqref{sol6lim}, \eqref{sol6limch}, \eqref{sol7lim} and \eqref{sol7limch}.
 
Renormalizing
\footnote{
Making 
the shift $x \to xq^{\frac{\mu}{s}}$ on the spectral parameter in \eqref{refeq2dual}, 
multiplying 
$x^{-s_{0}\mu}q^{-\mu} (\mathsf{ev}_{x} \otimes \pi_{1}(y))
((q^{\frac{s_{1}\mu h_{1}}{s}}\otimes 1 )
 ( q^{-\mu h_{1}} \otimes q^{-\mu h_{1}} ))$ from the right, 
 and taking the limit $q^{-\mu} \to 0$
 in $\pi_{\mu}^{+} \otimes 1 $, 
 we obtain  \eqref{refeqlim2nd} for $a=1$;
making 
the shift $x \to xq^{-\frac{\mu}{s}}$ on the spectral parameter in \eqref{refeq2dual}, 
multiplying 
$x^{-s_{0}\mu}q^{-2\mu} (\mathsf{ev}_{x^{-1}} \otimes \pi_{1}(y^{-1}))
( ( q^{\mu h_{1}} \otimes q^{\mu h_{1}} )(q^{-\frac{s_{1}\mu h_{1}}{s}}\otimes 1 ))$ 
from the left,  and taking the limit $q^{-\mu} \to 0$
 in $\pi_{\mu}^{+} \otimes 1 $, 
 we obtain  \eqref{refeqlim2ndch} for $a=1$;
 making 
the shift $x \to xq^{-\frac{\mu}{s}}$ on the spectral parameter in \eqref{refeq2dual}, 
multiplying 
$x^{s_{1}\mu}q^{\mu} (\mathsf{ev}_{x} \otimes \pi_{1}(y))
((q^{\frac{s_{0}\mu h_{1}}{s}}\otimes 1 )
 ( q^{-\mu h_{1}} \otimes q^{-\mu h_{1}} ))$ from the right, 
 and taking the limit $q^{\mu} \to 0$
 in $\pi_{\mu}^{+} \otimes 1 $, 
 we obtain  \eqref{refeqlim2nd} for $a=2$;
making 
the shift $x \to xq^{\frac{\mu}{s}}$ on the spectral parameter in \eqref{refeq2dual}, 
multiplying 
$x^{s_{1}\mu}q^{2\mu} (\mathsf{ev}_{x^{-1}} \otimes \pi_{1}(y^{-1}))
( ( q^{\mu h_{1}} \otimes q^{\mu h_{1}} )(q^{-\frac{s_{0}\mu h_{1}}{s}}\otimes 1 ))$ 
from the left,  and taking the limit $q^{\mu} \to 0$
 in $\pi_{\mu}^{+} \otimes 1 $, 
 we obtain  \eqref{refeqlim2ndch} for $a=2$.
}
\eqref{refeq2dual} and taking the limits $q^{\mp \mu} \to 0$ 
 in $\pi^{+}_{\mu} \otimes 1$, we obtain the dual 
reflection equations for L-operators for Q-operators: 
\begin{align}
& \Lf^{(a)} \left(\frac{y}{x}\right) \overline{\Kf}^{(a)}_{1}(x)^{t_{1}} 
g_{2}^{-1} \overline{\Lf}^{(a)}\left( xy q^{\frac{4}{s}}\right) g_{2}
\overline{K}_{2}(y)^{t_{2}} 
=
\nonumber \\
& \hspace{100pt} =
\overline{K}_{2}(y)^{t_{2}}
g_{2} \Lf^{(a)} \left(\frac{q^{-\frac{4}{s}}}{xy}\right)  g_{2}^{-1}
\overline{\Kf}^{(a)}_{1}(x)^{t_{1}} \overline{\Lf}^{(a)}\left(\frac{x}{y}\right),
\label{refeqlim2nd}
\\[6pt]
& \Lfc^{(a)} \left(\frac{y}{x}\right) \Kfcb^{(a)}_{1}(x)^{t_{1}} 
g_{2}^{-1} \Lfcb^{(a)}\left( xy q^{\frac{4}{s}}\right) g_{2}
\overline{K}_{2}(y)^{t_{2}} 
=
\nonumber \\
& \hspace{100pt} =\overline{K}_{2}(y)^{t_{2}}
g_{2} \Lfc^{(a)} \left(\frac{q^{-\frac{4}{s}}}{xy}\right)  g_{2}^{-1}
\Kfcb^{(a)}_{1}(x)^{t_{1}} \Lfcb^{(a)}\left(\frac{x}{y}\right), 
\quad a=1,2.
\label{refeqlim2ndch}
\end{align}
\subsubsection{Rational limit $q \to 1$}
One can take the rational limit $q \to 1$ of the formulas in this paper easily. 
The q-gamma function (see for example, \cite{AAR99}) is defined by 
\begin{align}
\Gamma_{q}(x)=\frac{(q;q)_{\infty}}{(q^{x};q)_{\infty}} (1-q)^{1-x} 
\qquad \text{for} 
\qquad |q| <1.
 \label{q-gamma}
\end{align}
This reduces to the normal gamma function in the rational limit. 
\begin{align}
\lim_{q \to 1} \Gamma_{q}(x)=\Gamma (x). 
 \label{gamma}
\end{align}
Let us define the rational limit of the generators of the 
q-oscillator algebra $\mathrm{Osc}_{1}$ by 
\begin{align}
\as=\lim_{q \to 1} \lambda \es_{1}, 
\qquad 
\ads=\lim_{q \to 1} \fs_{1}, 
\qquad 
\ns= \lim_{q \to 1}\frac{1-  \hs_{1}}{2} .
\end{align}
Then these generators satisfy $[\as, \ads]=1$, $\as \ads =\ns +\frac{1}{2}$ , 
$\ads \as  =\ns -\frac{1}{2}$.
Let $q^{2p}=- \epsilon_{-}/ \epsilon_{+}$. 
Then one can take the rational limit of
  renormalized versions of \eqref{sol6}-\eqref{sol7-2}, 
  \eqref{LQ1}, \eqref{LhQ1}, \eqref{K-mat1} and \eqref{sol6lim} as
\begin{align}
& \lim_{|q| \to 1+0}\Kf(q^{2u})(1-q^{-2})^{-2su}=
\frac{\Gamma(-p+su+\frac{H+1}{2})}{\Gamma(-p-su+\frac{H+1}{2})}
\quad \text{for} \quad \eqref{sol6}, 
\\[6pt]
& \lim_{|q| \to 1-0}\Kf(q^{2u})(1-q^{2})^{-2su}=\frac{\Gamma(p+su-\frac{H-1}{2})}{\Gamma(p-su-\frac{H-1}{2})} 
\quad \text{for} \quad \eqref{sol6-2}, 
\\[6pt]
& \lim_{|q| \to 1+0}\Kf(q^{2u})(1-q^{-2})^{-2su}=\frac{\Gamma(p+su-\frac{H-1}{2})}{\Gamma(p-su-\frac{H-1}{2})} 
\quad \text{for} \quad \eqref{sol7}, 
\\[6pt]
& \lim_{|q| \to 1-0}\Kf(q^{2u})(1-q^{2})^{-2su}=
\frac{\Gamma(-p+su+\frac{H+1}{2})}{\Gamma(-p-su+\frac{H+1}{2})}
\quad \text{for} \quad \eqref{sol7-2}, 
\\[6pt]
&  \lim_{q \to 1}\Lf^{(1)}(q^{2u}) (1 \otimes  (E_{11} - q^{-1}\lambda^{-1} E_{22}))
=
\begin{pmatrix}
1  & - \ads \\
\as & su -\ns
 \end{pmatrix} ,
 \label{LQ1-q1}
\\[6pt]
&  \lim_{q \to 1}\Lfb^{(1)}(q^{2u}) (1 \otimes  (E_{11} - q^{-1}\lambda^{-1} E_{22}))
=
\begin{pmatrix}
1  & - \ads \\
\as & -su -\ns
 \end{pmatrix} ,
 \label{LhQ1-q1}
\\[6pt]
&  \lim_{q \to 1}K(q^{2u})
(1-q^{-2})^{-2su} 
=
-\frac{\Gamma (su-p)}{\Gamma (1-su-p)}
\begin{pmatrix}
p-su  &0 \\
0 & p+su 
 \end{pmatrix} ,
 \label{K-mat1-q1}
\\[6pt]
&
\lim_{q \to 1} 
(q^{2};q^{2})_{\infty}(1-q^{2})^{-p-su+\frac{\hs_{1}+1}{2}}
\Kf^{(1)}(q^{2u}) =
\Gamma ( p+su +\ns ), 
\qquad 
u \in \mathbb{C}. 
 \label{K-op1-q1}
\end{align}
 It is important to note that 
the above limits keep the reflection equation \eqref{refeqlim1st} unchanged. 
This is because of the relation \eqref{com-Cartan} for 
$\xi=(1-q^{2})^{\frac{h_{1}}{2}}$
and 
\begin{align}
\rho^{(1)}_{x}((1-q^{2})^{\frac{h_{1}}{2}})=(1-q^{2})^{\frac{\hs_{1}}{2}},
\quad 
\pi_{1}(y)((1-q^{2})^{\frac{h_{1}}{2}})=(1-q^{2})^{\frac{1}{2}} (E_{11} - q^{-1}\lambda^{-1} E_{22}),
\end{align}
where $\rho^{(1)}_{x}$ is defined in \eqref{rho+}. 
The rational limit of 
all the other K- and L-operators can be taken in the same way. 
In this way, we have recovered 
 K-operators for Q-operators for rational (XXX-) models which are 
 similar to the ones in \cite{FS15}. 

\section{Concluding remarks}
 In the context of quantum groups and related coideal subalgebras, 
 finding a universal product formula for the K-matrix 
by analogy with the known product formula \eqref{UR-prod}
 for the universal R-matrix proposed in \cite{TK92} is an interesting problem. 
 In this direction, a  universal formulation of the reflection equation algebra 
 and related intertwining relations associated with a given coideal 
subalgebra are highly desirable (see recent progress in \cite{Ko17}). 
In the present paper, we have focused on homomorphic images of two different coideal subalgebras of $U_q(\widehat{sl_2})$ (onto $U_q({sl_2})$),  that are related with the augmented q-Onsager algebra first introduced
 by Ito and Terwilliger in \cite{IT} (see also \cite{BB1}).  Based on the intertwining relations, 
product formulae for the K-matrix solutions in terms of the generators of $U_q({sl_2})$ 
are derived. They solve certain reflection and dual reflection equations 
 associated with L-operators. 
In the second part of the paper, certain limits of these K-operators are
studied. For these limits, contracted versions of the augmented q-Onsager algebra 
 are considered and q-oscillator representations are constructed. Importantly, 
 these K-operators are the basic ingredient for the construction of Q-operators 
 that are relevant in the analysis of quantum integrable models with non-periodic diagonal boundary conditions. An interesting problem would be to extend the analysis presented here to the case of the q-Onsager algebra \cite{Ter03,BK}, which is isomorphic to the fixed point subalgebra of $U_q(\widehat{sl_2})$ under the action of the Chevalley involution \cite{Kolb12}.
A product formula  in this case is an open problem, that should find applications 
to the analysis of integrable models with non-diagonal boundary conditions. 

It is known that L-operators for Verma modules 
 of the quantum affine algebra (or the Yangian)
factorize with respect to L-operators for the Q-operators.
Examples for such factorization formulas appeared in a number of papers 
(see for example, \cite{Derkachov,BLMS10} and references therein).
In \cite{KT14}, such factorization formulas were 
 reconsidered in relation to properties of 
the universal R-matrix, and a universal factorization formula,
which is independent of the quantum space, was proposed.
One of our motivations was to generalize 
the universal factorization formula \cite{KT14} 
to the case of open boundary conditions in the light of the augmented q-Onsager algebra \cite{BB1}. 
 The main obstacle for this is that we do not have
the defining relations of the universal K-matrix
corresponding to the relations \eqref{R-def} for the
universal R-matrix of $U_q(\widehat{sl_2})$. This is a reason why we focused our discussions
only on one of the most essential objects, the K-matrices, without application to Q-operators
  and their properties. Still, in appendix G we suggest universal T- and Q- operators. For a class of representations, the commutativity is proven. It is desirable to reconsider the problem
  after formulation of the universal K-matrix for the  universal R-matrix of $U_q(\widehat{sl_2})$ in the future.
\section*{Acknowledgments} 
We thank S. Belliard for discussions. We thank the anonymous referee for comments. The research of Z.T. was supported by CNRS at Universit\'e de Tours 
 and is supported by 
 the European Research Council
(Programme ``Ideas'' ERC-2012-AdG 320769 AdS-CFT-solvable) 
at LPTENS. P.B. is supported by C.N.R.S. 

\appendix
\section{The universal R-matrix}
\label{ApA}
In this section, we briefly review the product expression of the universal R-matrix given by 
Khoroshkin and Tolstoy in \cite{TK92}. 
Their universal R-matrix was already reviewed by several authors \cite{ZG93,BGKNR10}. 
Here we basically follow these. 

Let $\{\alpha +k\delta \}_{k=0}^{\infty} \cup
 \{k\delta \}_{k=1}^{\infty}\cup
  \{\delta- \alpha +k\delta \}_{k=0}^{\infty} $ be a positive root system of $\widehat{sl_2}$ 
in the notation of \cite{TK92}. We choose the root ordering as 
 $\alpha +(k-1)\delta \prec \alpha +k\delta \prec 
 l\delta \prec (l+1)\delta \prec 
 \delta- \alpha +m\delta \prec \delta- \alpha +(m-1)\delta$ for any $k,l,m \in {\mathbb Z}_{\ge 1}$. 
 In this case, 
the universal R-matrix has the following expression:
\begin{align}
{\mathcal R}=\overline{\mathcal R}^{+} \, 
\overline{\mathcal R}^{0} \,
 \overline{\mathcal R}^{-}q^{\mathcal{K}},  
  \label{UR-prod}
\end{align}
\begin{align}
\overline{\mathcal R}^{+}&=\overrightarrow{\prod_{k=0}^{\infty}} \exp_{q^{-2}} 
\left( \lambda e_{\alpha + k \delta } \otimes f_{\alpha + k \delta } \right) ,
\\[6pt]
\overline{\mathcal R}^{0}&= \exp  
\left( \lambda   \sum_{k=1}^{\infty} \frac{k}{[2k]_{q}} e_{ k \delta } \otimes f_{ k \delta } \right) ,
\\[6pt]
\overline{\mathcal R}^{-}&=\overleftarrow{\prod_{k=0}^{\infty}} \exp_{q^{-2}} 
\left( \lambda e_{\delta-\alpha + k \delta } \otimes f_{\delta-\alpha + k \delta } \right) ,
\end{align}
where we use notations
\begin{align}
 \exp_{q}(x)& =1+\sum_{k=1}^{\infty} \frac{x^{k}}{(k)_{q} ! },
\qquad
(k)_{q}!  =(1)_{q}(2)_{q} \cdots (k)_{q} ,
\qquad
(k)_{q}=\frac{1-q^{k}}{1-q}.
 \nonumber
\end{align}
Let 
$e_{\alpha } =e_{1}$, $e_{\delta- \alpha } =e_{0}$, 
$f_{\alpha } =f_{1}$, $f_{\delta- \alpha } =f_{0}$. Then the other root vectors are defined by 
the following recursion relations:
\begin{align}
e_{\alpha +k \delta } &= [2]_{q}^{-1}[e_{\alpha +(k-1) \delta}, e_{\delta}^{\prime}], 
\\[6pt]
e_{k \delta }^{\prime } &= [e_{\alpha +(k-1) \delta}, e_{\delta -\alpha }]_{q^{-2}}, 
\\[6pt]
e_{\delta- \alpha +k \delta } &= [2]_{q}^{-1}[ e_{\delta}^{\prime}, e_{\delta- \alpha +(k-1) \delta}],
\\[6pt]
f_{\alpha +k \delta } &= [2]_{q}^{-1}[f_{\delta}^{\prime}, f_{\alpha +(k-1) \delta}], 
\\[6pt]
f_{k \delta }^{\prime } &= [f_{\delta -\alpha }, f_{\alpha +(k-1) \delta}]_{q^{2}}, 
\\[6pt]
f_{\delta- \alpha +k \delta } &= [2]_{q}^{-1}[ f_{\delta- \alpha +(k-1) \delta}, f_{\delta}^{\prime}],  
\qquad k \in {\mathbb Z}_{\ge 1} .
\end{align}
where 
the root vectors with prime are given by the following generating functions. 
\begin{align}
\lambda \sum_{k=1}^{\infty} e_{k \delta} z^{-k}&= 
\log \left(  1+ \lambda \sum_{k=1}^{\infty} e_{k \delta}^{\prime } z^{-k} \right),
\\[6pt]
-\lambda \sum_{k=1}^{\infty} f_{k \delta} z^{-k}&= 
\log \left(  1- \lambda \sum_{k=1}^{\infty} f_{k \delta}^{\prime } z^{-k} \right), 
\qquad z \in {\mathbb C}.
\end{align}
In general, root vectors contain many commutators. However, simplification occurs under the 
evaluation map.
\begin{align}
\mathsf{ev}_{x}(e_{\alpha +k \delta }) &=(-1)^{k}x^{ks+s_{1}}q^{-kH}E,
 \label{afev1}
\\[6pt]
\mathsf{ev}_{x}(e_{\delta- \alpha +k \delta }) &=(-1)^{k}x^{ks+s_{0}}Fq^{-kH}, 
 \label{afev2}
\\[6pt]
\mathsf{ev}_{x}(f_{\alpha +k \delta }) &=(-1)^{k}x^{-ks-s_{1}}Fq^{kH}, 
 \label{afev3}
\\[6pt]
\mathsf{ev}_{x}(f_{\delta- \alpha +k \delta }) &=(-1)^{k}x^{-ks-s_{0}}q^{kH}E \quad 
\text{for} 
\quad k \in {\mathbb Z}_{\ge 0}, \quad \text{and}
 \label{afev4}
\\[8pt]
\mathsf{ev}_{x}(e_{k \delta }^{\prime}) &=(-1)^{k-1}x^{ks}q^{-(k-1)H}[E,F]_{q^{-2k}}, 
 \label{afev5}
\\[6pt]
\mathsf{ev}_{x}(e_{k \delta }) &=
\frac{(-1)^{k-1}q^{-k}x^{ks}}{(q-q^{-1})k}\left(C_{k}-(q^{k}+q^{-k})q^{-kH}\right), 
 \label{afev6}
\\[6pt]
\mathsf{ev}_{x}(f_{k \delta }^{\prime}) &=(-1)^{k-1}x^{-ks}[E,F]_{q^{2k}}q^{(k-1)H},
 \label{afev7}
\\[6pt]
\mathsf{ev}_{x}(f_{k \delta }) &=-
\frac{(-1)^{k-1}q^{k}x^{-ks}}{(q-q^{-1})k}\left(C_{k}-(q^{k}+q^{-k})q^{kH}\right)
\quad \text{for} \quad  k \in {\mathbb Z}_{\ge 1},
 \label{afev8}
\end{align}
where the central elements $C_{k}$ are defined by 
\begin{align}
\sum_{k=1}^{\infty}\frac{(-1)^{k-1}C_{k}}{k}z^{-k}= \log(1+\lambda^{2} C z^{-1}+z^{-2}), 
\qquad z \in {\mathbb C}.
\label{highercas}
\end{align}
Inserting these\footnote{For the second component of the tensor product, 
one has to replace $x$ with $y$ beforehand.} into \eqref{UR-prod}, we obtain $\Rf(x,y)$. 
In order to obtain $\Rf_{21}(x,y)$, one has to swap the first and the second 
components of the tensor product in \eqref{UR-prod} beforehand. 
Based on these product formulas, one can check
\begin{align}
(\nu \otimes \nu )\Rf(x,y) =\Rf_{21}(x^{-1},y^{-1}). \label{Rf-auto}
\end{align}
\section{
Contraction of the quantum affine algebra
}
\label{ApB}
A systematic study of the asymptotic 
representation theory of the Borel subalgebras of quantum affine algebras 
was given in \cite{HJ11}. 
How to evaluate the universal R-matrix for the purpose of Q-operators 
was explained
 in detail in \cite{BGKNR10}. 
We nevertheless review the subject in the spirit of \cite{T12}, which 
is inspired by earlier discussions \cite{BLZ97,BHK02,Bp,talks,BT08}. 
We are interested in considering limits of representations of the whole 
quantum affine algebra rather than those of its Borel subalgebras. 
In particular, we will present the universal form of the intertwining relations 
for the L-operators for Q-operators \eqref{ituq1}-\eqref{ituq4}. 
\subsection{The contracted algebra $U_{q}(\widehat{sl}(2;I))$}
Let $I$ be a subset
\footnote{This came from a notation in \cite{T09}, where $2^{M+N}$
Q-functions for $U_{q}(\widehat{gl}(M|N))$ 
are classified in terms of all the subsets $I$ of the set $\{1,2,\dots, M+N\}$. 
The number `$0$' in $I$ 
corresponds to `$2$' (for $(M,N)=(2,0)$ case) 
in the  \cite{T09}. In our present paper, there are 
$2^2=4$ Q-operators. Two of them, which correspond to $I=\{0,1\},\emptyset $, 
are identity operators in 
the normalization of the universal R-matrix. 
} of the set $\{0,1\}$, and define $\theta(\text{True})=1, \theta(\text{False})=0$. 
The contracted algebra $\widetilde{U}_{q}(\widehat{sl}(2;I))$ is an algebra
generated by the generators
\footnote{In this paper, we do not use the derivation $d$.}
 $e_{i},f_{i},h_{i}$, where
$i \in \{0,1 \}$.
For $i,j \in \{0,1\}$, the
 defining relations 
  of the algebra $\widetilde{U}_{q}(\widehat{sl}(2;I))$ are
given by
\begin{align}
& [h_{i},h_{j}]=0, \quad [h_{i}, e_{j} ] =a_{ij} e_{j}, \quad
[h_{i}, f_{j} ] =-a_{ij} f_{j},
\label{cont-def1}
\\[6pt]
&[e_{i},f_{j}]=\delta_{ij} \frac{\theta(i+1 \in I)q^{h_{i}} -\theta(i \in I)q^{-h_{i}} }{q-q^{-1}},
\label{cont-def2}
\\[6pt]
&[e_{i},[e_{i},[e_{i},e_{j}]_{q^{2}}]]_{q^{-2}}=
[f_{i},[f_{i},[f_{i},f_{j}]_{q^{-2}}]]_{q^{2}}=0, 
\quad i \ne j, 
\label{cont-def2-2}
%
\end{align}
where $(a_{ij})_{0 \le i,j\le 1}$ is the
Cartan matrix of $\widehat{sl_2}$, and $2 \equiv 0$ in $I$.
Note that $\widetilde{U}_{q}(\widehat{sl}(2;\{0,1\}))$ coincides with $U_{q}(\widehat{sl_2})$. 
We use the following co-products
  $ \co , \co^{\prime}, \cob , \cob^{\prime} :
   \widetilde{U}_{q}(\widehat{sl}(2;I))\to  \widetilde{U}_{q}(\widehat{sl}(2;I))  \otimes U_{q}(\widehat{sl_2})$:
\begin{align}
\co (e_{i})&=e_{i} \otimes 1 + q^{-h_{i}} \otimes e_{i}
, \qquad 
\co^{\prime} (e_{i})=1\otimes e_{i}  + e_{i}  \otimes q^{-h_{i}} , \nonumber\\
\co (f_{i})&=f_{i} \otimes q^{h_{i}} + \theta(i \in I) (1 \otimes f_{i}), 
\qquad 
\co^{\prime} (f_{i})= \theta(i+1 \in I)  (q^{h_{i}} \otimes f_{i}) + f_{i} \otimes 1 , 
\label{copro-cont} \\
\co (h_{i})&=\co^{\prime} (h_{i})=h_{i} \otimes 1 + 1 \otimes h_{i}. \nonumber
\end{align}
and 
\begin{align}
\cob (e_{i})&=e_{i} \otimes 1 + \theta(i \in I)(q^{-h_{i}} \otimes e_{i})
, \qquad 
\cob^{\prime} (e_{i})=\theta(i+1 \in I)(1\otimes e_{i})  + e_{i}  \otimes q^{-h_{i}} , \nonumber\\
\cob (f_{i})&=f_{i} \otimes q^{h_{i}} +  1 \otimes f_{i}, 
\qquad 
\cob^{\prime} (f_{i})=   q^{h_{i}} \otimes f_{i} + f_{i} \otimes 1 , 
\label{copro-cont2} \\
\cob (h_{i})&=\cob^{\prime} (h_{i})=h_{i} \otimes 1 + 1 \otimes h_{i}. \nonumber
\end{align}
We define a smaller contracted algebra
 \footnote{
 The coproduct \eqref{copro-cont}
  (resp.\  \eqref{copro-cont2}) does not keep  \eqref{cont-def3} and \eqref{cont-def4} 
  (resp.\  \eqref{cont-def3-2} and \eqref{cont-def4-2}). Then 
 there is an option to consider algebras bigger than $U_{q}(\widehat{sl}(2;I))$, 
 where  \eqref{cont-def3-2} and \eqref{cont-def4-2} 
 with \eqref{copro-cont}, or 
 \eqref{cont-def3} and \eqref{cont-def4} 
 with \eqref{copro-cont2} 
 are imposed on  $\widetilde{U}_{q}(\widehat{sl}(2;I))$. 
 However, we focus on $U_{q}(\widehat{sl}(2;I))$ 
 since \eqref{cont-def3}-\eqref{cont-def4-2} always hold true for the q-oscillator representations 
 in this paper.}
 $U_{q}(\widehat{sl}(2;I))$  by imposing 
\begin{align}
& [e_{0}, [e_{0}, e_{1}]_{q^{2 } } ]=
[e_{1}, [e_{1}, e_{0}]_{q^{-2 } } ]=0
\qquad \text{for} \quad I=\{0\},
\label{cont-def3}
\\[6pt]
&[f_{0}, [f_{0}, f_{1}]_{q^{-2} } ]=
[f_{1}, [f_{1}, f_{0}]_{q^{2} } ]=0 
\qquad \text{for} \quad I=\{0\},
\label{cont-def3-2}
\\[6pt]
& [e_{0}, [e_{0}, e_{1}]_{q^{-2 } } ]=
[e_{1}, [e_{1}, e_{0}]_{q^{2 } } ]=0
\qquad \text{for} \quad I=\{1\},
\label{cont-def4}
\\[6pt]
&[f_{0}, [f_{0}, f_{1}]_{q^{2} } ]=
[f_{1}, [f_{1}, f_{0}]_{q^{-2} } ]=0 
\qquad \text{for} \quad I=\{1\},
\label{cont-def4-2}
\end{align}
on $U_{q}(\widehat{sl}(2;I))$. 
Note that \eqref{cont-def2-2} follows from \eqref{cont-def3}-\eqref{cont-def4-2}. 
$U_{q}(\widehat{sl}(2;\{0\}))$ and $U_{q}(\widehat{sl}(2;\{1\}))$ are sort of 
coupled q-oscillator algebras. 
The map \eqref{auto2}, which preserves the defining relations of $U_{q}(\widehat{sl}(2;\{0,1\}))$, 
 swaps the defining relations $U_{q}(\widehat{sl}(2;\{0\}))$  and $U_{q}(\widehat{sl}(2;\{1\}))$ 
one another. 
 The fact that Serre-type relations for q-oscillator representations for the Q-operators can be
  simpler than the original ones 
was pointed out first by Bazhanov et al. in \cite{BHK02} for  ${\mathcal B}_{+}$ of $U_{q}(\hat{sl}(3))$. 
 In \cite{T12}, this phenomenon was observed for the contracted algebras 
 $U_{q}(\widehat{gl}(M|N;I))$ of $U_{q}(\widehat{gl}(M|N))$.  
 It is desirable to study this systematically for the Drinfeld's second realization 
 of the quantum affine (super)algebras. 
 For the case of the Yangian $Y(sl(2))$, a degenerated algebra $\mathcal{A}$ for Q-operators 
 was studied in terms of the Drinfeld's second realization \cite{RT15}. Their co-products
  correspond to the case $\Delta :  \mathcal{A} \mapsto \mathcal{A} \otimes \mathcal{A}$. 
  On the other hand, our co-product \eqref{copro-cont}, which was used 
  for the intertwining relations for L-operators for Q-operators \cite{T12},  might be related to 
  the case $\Delta :  \mathcal{A} \mapsto \mathcal{A} \otimes Y(sl(2))$ if an appropriate
   rational limit 
  was taken. Thus, it will be interesting to consider the case 
  $\Delta : U_{q}(\widehat{sl}(2;I)) \mapsto U_{q}(\widehat{sl}(2;I))  \otimes U_{q}(\widehat{sl}(2;I)) $. 
 
%
The Borel subalgebras of 
$\widetilde{U}_{q}(\widehat{sl}(2;I))$  and $U_{q}(\widehat{sl}_{2})$ 
share the same defining relations. $U_{q}(\widehat{sl}(2;I))$ 
is a subalgebra of $\widetilde{U}_{q}(\widehat{sl}(2;I))$.
Taking note of this fact, we purposely use the same symbols for the generators 
of these algebras.  

The contracted algebra $U_{q}(\hat{sl}(2;I))$ has subalgebras $U_{q}(sl(2;I))$
generated by the generators $E,F,H$ obeying the 
 relations,
\begin{align}
& 
[H, E ] =2E , \quad
[H, F] =-2F,
\label{contfin-def1}
\\[6pt]
&[E,F]= \frac{\theta(0 \in I)q^{H} -\theta(1 \in I)q^{-H} }{q-q^{-1}}.
\label{contfin-def2}
\end{align}
This reduces to $U_{q}(sl(2))$ for $I=\{0,1\}$ and to 
q-oscillator algebras for $I=\{0\}$ or $I=\{1\}$. 
The q-oscillator algebra $\mathrm{Osc}_{1}$ (resp.\ $\mathrm{Osc}_{2}$) 
corresponds to $U_{q}(sl(2;\{0\}))$ (resp.\ $U_{q}(sl(2;\{1\}))$) with 
a fixed value of the central element $[E,F]_{q^{-2}}=1/(q-q^{-1})$ 
(resp.\  $[E,F]_{q^{2}}=-1/(q-q^{-1})$).
Contractions of a quantum algebra was previously 
discussed in \cite{Chaichian:1989rq}. 
Contracted quantum algebras 
in relation to L-operators for Q-operators 
 were previously discussed in \cite{Bp} and developed in
\cite{talks,T12}. 
\subsection{Universal $L$-operators for Q-operators and their intertwining relations}
The $L$-operators  \eqref{LQ1}-\eqref{LhQ2} can 
also be presented as homomorphic images of the universal $R$-matrix
under the homomorphism $\rho_x^{(i)}: \mathcal{B}_{+} \to
\mathrm{Osc}_{i}, i=1,2 $ defined by the relations
\begin{align}\label{rho+}
& \rho^{(i)}_{x}(e_{0})=x^{s_{0}}\fs_{i}, 
\quad \rho^{(i)}_{x}(e_{1})=x^{s_{1}}\es_{i}, 
\quad \rho^{(i)}_{x}(h_{0})=-\hs_{i}, \quad \rho^{(i)}_{x}(h_{1})=\hs_{i}, 
\end{align}
or  the homomorphism $\rho_x^{(i)}: \mathcal{B}_{-} \to
\mathrm{Osc}_{i}, i=1,2 $ defined by the relations
\begin{align}\label{rho-}
&\rho^{(i)}_{x}(f_{0})=x^{-s_{0}}\es_{i}, 
\quad \rho^{(i)}_{x}(f_{1})=x^{-s_{1}}\fs_{i}, 
\quad \rho^{(i)}_{x}(h_{0})=-\hs_{i}, \quad \rho^{(i)}_{x}(h_{1})=\hs_{i}. 
\end{align}

We remark that the maps $\rho^{(i)}_{x}$ cannot be straightforwardly
 extended to the whole algebra 
$U_{q}(\widehat{sl_2})$. 
$\rho^{(1)}_{x}$ (resp.\ $\rho^{(2)}_{x}$) should be regarded as a 
map from the contracted algebra
$U_{q}(\widehat{sl}(2;\{0 \}))$  or $\widetilde{U}_{q}(\widehat{sl}(2;\{0\}))$
 (resp.\ $U_{q}(\widehat{sl}(2;\{1 \}))$ or $\widetilde{U}_{q}(\widehat{sl}(2;\{ 1\}))$)
 to $\mathrm{Osc}_{1}$ (resp.\ $\mathrm{Osc}_{2}$). 
 Namely, they preserve the relations \eqref{cont-def1}-\eqref{cont-def2-2}, 
 \eqref{cont-def3}-\eqref{cont-def4-2},  
 \eqref{sl2h-def1} and \eqref{sl2h-def3}, 
 but do not keep the relation \eqref{sl2h-def2} unchanged. 
Let ${\mathcal N}_{+}$ (resp.\ ${\mathcal N}_{-}$ )  be
the nilpotent subalgebra of $U_{q}(\widehat{sl_2})$ 
generated by $e_{i}$ (resp. $f_{i}$), $i=0,1$.
The maps $\rho^{(1)}_{x}$ and $\rho^{(2)}_{x}$ can be realized as 
 limits of shifted 
  representations of $\mathcal{B}_{\pm}$ and 
 representations
 \footnote{Note that ${\mathcal N}_{\mp}$ is invariant under $\tau_{-\mu}$.}
  of  ${\mathcal N}_{\mp}$. 
 We are interested in the following realizations
 \footnote{To be precise, the letter $a$ denotes an element of the quantum 
 affine algebra on the right hand side, while that of a contracted 
 algebra on the left hand side.}:
\begin{align}
\rho^{(1)}_{x}(a)&=
\begin{cases}
\lim_{q^{-\mu} \to 0}\pi^{+}_{\mu} (xq^{-\frac{\mu}{s}})
\left(\tau_{-\mu}(a)\right) & \text{for} \quad a \in \mathcal{B}_{+}
\\[6pt]
\lim_{q^{-\mu} \to 0}\pi^{+}_{\mu} (xq^{-\frac{\mu}{s}})
 \left( q^{-\mathrm{deg}(a) \mu } a \right)
& \text{for} \quad a \in \mathcal{N}_{-} ,
\end{cases}
\label{rho1}
\\[6pt]
\rho^{(2)}_{x}(a)&=
\begin{cases}
\lim_{q^{\mu} \to 0}\pi^{+}_{\mu} (xq^{\frac{\mu}{s}})
\left(\tau_{-\mu}(a)\right) & \text{for} \quad a \in \mathcal{B}_{+}
\\[6pt]
\lim_{q^{\mu} \to 0}\pi^{+}_{\mu} (xq^{\frac{\mu}{s}})
\left(q^{\mathrm{deg}(a) \mu} a \right)
& \text{for} \quad a \in \mathcal{N}_{-} ,
\end{cases}
\label{rho2}
\end{align}
or
\footnote{Similarity transformations by the Cartan element are used to renormalize 
the generators: 
$\xi e_{0} \xi^{-1}=q^{-\frac{(s_{0}-s_{1})\mu }{s}}e_{0}$, 
$\xi e_{1} \xi^{-1}=q^{\frac{(s_{0}-s_{1})\mu }{s}}e_{1}$, 
$\xi f_{0} \xi^{-1}=q^{\frac{(s_{0}-s_{1})\mu }{s}}f_{0}$,
$\xi f_{1} \xi^{-1}=q^{-\frac{(s_{0}-s_{1})\mu }{s}}f_{1}$, 
where $\xi=q^{\frac{(s_{0}-s_{1})\mu h_{1}}{2s}}$.
}
%
\begin{align}
\rho^{(1)}_{x}(a)&=
\begin{cases}
\lim_{q^{-\mu} \to 0}\pi^{+}_{\mu} (xq^{\frac{\mu}{s}})
\left(
q^{-\mathrm{deg}(a) \mu+\frac{ (s_{0}-s_{1})\mu h_{1}}{2s}}
a \,
q^{-\frac{ (s_{0}-s_{1})\mu h_{1}}{2s}}
\right) & \text{for} \quad a \in \mathcal{N}_{+} 
\\[6pt]
\lim_{q^{-\mu} \to 0}\pi^{+}_{\mu} (xq^{\frac{\mu}{s}})
\left(
q^{\frac{ (s_{0}-s_{1})\mu h_{1}}{2s}}
\tau_{-\mu}(a) 
q^{-\frac{ (s_{0}-s_{1})\mu h_{1}}{2s}}
\right)
& \text{for} \quad a \in \mathcal{B}_{-} ,
\end{cases}
\label{rho1b}
\\[6pt]
\rho^{(2)}_{x}(a)&=
\begin{cases}
\lim_{q^{\mu} \to 0}\pi^{+}_{\mu} (xq^{-\frac{\mu}{s}})
\left(
q^{\mathrm{deg}(a)\mu+\frac{ (s_{1}-s_{0})\mu h_{1}}{2s}}
a \, 
q^{-\frac{ (s_{1}-s_{0})\mu h_{1}}{2s}}
\right)
 & \text{for} \quad a \in \mathcal{N}_{+} 
\\[6pt]
\lim_{q^{\mu} \to 0}\pi^{+}_{\mu} (xq^{-\frac{\mu}{s}})
\left(
q^{\frac{ (s_{1}-s_{0})\mu h_{1}}{2s}}
\tau_{-\mu}(a) 
q^{-\frac{ (s_{1}-s_{0})\mu h_{1}}{2s}}
\right)
& \text{for} \quad a \in \mathcal{B}_{-} ,
\end{cases}
\label{rhob2}
\end{align}
where $\mathrm{deg}$ is a linear operator which evaluates the degree of the monomials  
of the generators (for example, $\mathrm{deg}(f_{i})=1$, $\mathrm{deg}(e_{i}e_{j}e_{k})=3$).

We define
\footnote{
One may also define \eqref{uni-Lop2} as
$\check{\mathcal{L}}^{(i)}(x)=(\rho_x^{(i)}\circ S \otimes 1)
 \overline{\mathcal R}$, 
$\check{\overline{\mathcal{L}}}^{(i)}(x)=(\rho_x^{(i)}\circ S \otimes 1) {\mathcal R}$, 
where $S$ is the anti-pode satisfying $(S \otimes 1){\mathcal R}={\mathcal R}^{-1}=
(1 \otimes S^{-1}){\mathcal R}$.
}
 universal L-operators as homomorphic image of the universal R-matrix: 
\begin{align}
\mathcal{L}^{(i)}(x)&=(\rho_x^{(i)} \otimes 1){\mathcal R}, &
\overline{\mathcal{L}}^{(i)}(x)&=(\rho_x^{(i)} \otimes 1) \overline{\mathcal R}, 
 \label{uni-Lop1}
\\[6pt]
\check{\mathcal{L}}^{(i)}(x)&=(\rho_x^{(i)} \otimes 1)
 \overline{\mathcal R}^{-1}, &
\check{\overline{\mathcal{L}}}^{(i)}(x)&=(\rho_x^{(i)} \otimes 1) {\mathcal R}^{-1}.
 \label{uni-Lop2}
\end{align}
For example, one can calculate
\footnote{One can directly plug \eqref{rho+} into \eqref{UR-prod}, or apply 
 \eqref{rho1} to  \eqref{UR-prod} by way of \eqref{afev1},  \eqref{afev2},  \eqref{afev6}, 
 \eqref{highercas} with the help of \eqref{unRshift}, \eqref{limit1} and \eqref{caslimi1}. 
 As remarked in \cite{KT14}, $\mathcal{L}^{(1)}(x)$ contains only two q-exponentials, 
 while $\mathcal{L}^{(2)}(x)$ contains infinitely many. 
 One could use the universal version of \eqref{LQ2p}, 
 $ \mathcal{L}^{(2)\prime}(x)=
 (\rho_x^{(1)} \circ \sigma \otimes 1){\mathcal R}
 = (1 \otimes \sigma^{-1}) \mathcal{L}^{(1)}(x)$ 
 instead of \eqref{limUR2} to avoid the infinite product.} 
\begin{align}
\mathcal{L}^{(1)}(x)&= \exp_{q^{-2}} 
\left( \lambda x^{s_{1}} \es_{1} \otimes f_{\alpha } \right) 
\exp  
\left(  \sum_{k=1}^{\infty} \frac{(-1)^{k-1} x^{sk}}{[2k]_{q}} \otimes f_{ k \delta } \right) 
 \exp_{q^{-2}} 
\left( \lambda x^{s_{0}} \fs_{1} \otimes f_{\delta-\alpha  } \right) 
q^{\frac{1}{2}\hs_{1}\otimes h_{1}}, 
 \label{limUR1}
\\[6pt]
\mathcal{L}^{(2)}(x)&=
\overrightarrow{\prod_{k=0}^{\infty}} \exp_{q^{-2}} 
\left((-1)^{k} \lambda x^{ks+s_{1}}q^{-k\hs_{2}} \es_{2} \otimes f_{\alpha + k \delta } \right) 
\nonumber
\\[3pt]
& \qquad \times 
 \exp  
\left(  \sum_{k=1}^{\infty} \frac{(-1)^{k-1}q^{-k}x^{ks}}{[2k]_{q}}
 \left(q^{-k}-(q^{k}+q^{-k})q^{-k \hs_{2}}\right) \otimes f_{ k \delta } \right) 
\nonumber
\\[3pt]
& \qquad \times 
\overleftarrow{\prod_{k=0}^{\infty}} \exp_{q^{-2}} 
\left( (-1)^{k} \lambda x^{ks+s_{0}} \fs_{2}q^{-k\hs_{2}} \otimes f_{\delta-\alpha + k \delta } \right) ,
\label{limUR2}
\end{align}
where the relation
\begin{align}
\lim_{q^{\mp}\to 0} \pi^{+}_{\mu}(C_{k}q^{\mp k \mu})=
\left(\lambda^{2}\lim_{q^{\mp}\to 0} \pi^{+}_{\mu}(Cq^{\mp  \mu})\right)^{k}
=q^{\pm k} 
\quad \text{for} \quad k \in {\mathbb Z}_{\ge 1}, 
\end{align}
which follows from \eqref{highercas}, \eqref{caslimi1} and \eqref{caslimi2}, 
is used. 
Then the L-operators \eqref{LQ1}-\eqref{LhQ2} and 
\eqref{LQ1-an}-\eqref{LhQ2-an} are given by 
$\Lf^{(i)}(x)= \phi^{(i)}(x)( 1 \otimes \pi_{1}(1) ) {\mathcal L}^{(i)}(x) $,  
$\Lfb^{(i)}(x)= \phi^{(i)}(x^{-1})( 1 \otimes \pi_{1}(1) ) \overline{\mathcal L}^{(i)}(x) $, 
$\Lfc^{(i)}(x)= \Check{\phi}^{(i)}(x^{-1})( 1 \otimes \pi_{1}(1) ) \Check{\mathcal L}^{(i)}(x) $, 
$\Lfcb^{(i)}(x)= \Check{\phi}^{(i)}(x)( 1 \otimes \pi_{1}(1) )
\Check{\overline{\mathcal L}}^{(i)}(x) $ , 
where
\footnote{There is a useful identity
$  \phi^{(i)}(x) \Check{\phi}^{(i)}(x)=q^{-1}(q^{1-2\delta_{i,2}}-x^{-s})$, 
$i=1,2$.}
 $ \phi^{(i)}(x)=e^{-\Phi (x^{s}q^{-2\delta_{i,2}})} $, 
$ \Check{\phi}^{(i)}(x)=(-x^{-s}q^{-1})e^{-\Phi (x^{s}q^{2\delta_{i,1}})} $, 
$\Phi (x)=\sum_{k=1}^{\infty}\frac{1}{k(q^{k}+q^{-k})}x^{k}$.

The intertwining relations for these operators are given
\footnote{We may interpret the first space of these as 
a composition of natural evaluation maps 
$\widetilde{U}_{q}(\widehat{sl}(2;I)) \mapsto U_{q}(\widehat{sl}(2;I)) \mapsto 
\mathrm{Osc}_{i} $.
} 
by 
\begin{align}
((\rho_x^{(i)} \otimes 1) \co^{\prime }(a)) {\mathcal L}^{(i)}(x)
&=\mathcal{L}^{(i)}(x) ((\rho_x^{(i)} \otimes 1) \co (a)), 
\label{ituq1}
\\[6pt]
((\rho_x^{(i)} \otimes 1) \cob(a)) \overline{\mathcal{L}}^{(i)}(x)
&=\overline{\mathcal{L}}^{(i)}(x) ((\rho_x^{(i)} \otimes 1) \cob^{\prime } (a)),
\label{ituq2}
\\[6pt]
((\rho_x^{(i)} \otimes 1) \cob^{\prime }(a)) \Check{\mathcal{L}}^{(i)}(x)
&=\Check{\mathcal{L}}^{(i)}(x) ((\rho_x^{(i)} \otimes 1) \cob (a)), 
\label{ituq3}
\\[6pt]
((\rho_x^{(i)} \otimes 1) \co(a)) \Check{\overline{\mathcal L}}^{(i)}(x)
&= \Check{\overline{\mathcal L}}^{(i)}(x) ((\rho_x^{(i)} \otimes 1) \co^{\prime } (a)), 
\label{ituq4}
\\[6pt]
&
\text{for} \quad 
a \in \widetilde{U}_{q}(\widehat{sl}(2;I)), 
\quad 
I=\{i-1\},
\quad 
 i \in \{1,2\}.
 \nonumber 
\end{align}
The relation \eqref{ituq1} follows from \eqref{R-def}, \eqref{rho1} and  \eqref{rho2}; 
 \eqref{ituq2} follows from \eqref{Rb-def}, \eqref{rho1b} and \eqref{rhob2}; 
 \eqref{ituq3} follows from \eqref{Rb-def}, \eqref{rho1b} and  \eqref{rhob2}; 
  \eqref{ituq4} follows from \eqref{R-def}, \eqref{rho1} and  \eqref{rho2}. 
  For example, let us multiply $q^{-\mu-\frac{1}{2}\mu \otimes h_{1}} $ from the right of 
  the first equation in \eqref{R-def} for $f_{1}$: 
 \begin{align}
 \left( q^{h_{1}-\mu}  \otimes f_{1} +f_{1}q^{-\mu} \otimes  1 \right) {\mathcal R}q^{-\frac{1}{2}\mu \otimes h_{1}}
 ={\mathcal R}q^{-\frac{1}{2}\mu \otimes h_{1}}
 \left(f_{1}q^{-\mu} \otimes q^{h_{1}} +q^{-2\mu }\otimes f_{1} \right) .
 \end{align}
 Evaluating this for $\pi^{+}_{\mu}(xq^{-\frac{\mu}{s}}) \otimes 1$, we obtain 
  \begin{multline}
 \left( \pi^{+}_{\mu}(q^{H-\mu} ) \otimes f_{1} +
 x^{-s_{1}}\pi^{+}_{\mu}(Fq^{-\frac{s_{0}\mu}{s}}) \otimes  1 \right) 
 \left((\pi^{+}_{\mu}(xq^{-\frac{\mu}{s}})\tau_{-\mu} \otimes 1 )
 {\mathcal R}\right)
 \\
 =\left((\pi^{+}_{\mu}(xq^{-\frac{\mu}{s}})\tau_{-\mu} \otimes 1 )
 {\mathcal R}\right)
 \left( x^{-s_{1}}\pi^{+}_{\mu}(Fq^{-\frac{s_{0}\mu}{s}}) \otimes q^{h_{1}} +
 q^{-2\mu }\otimes f_{1} \right),
 \end{multline}
where \eqref{unRshift} for $c_{1}=-\mu$ is used. 
Then the limit $q^{-\mu} \to 0$ produces 
  \begin{align}
 \left( q^{\hs_{1}}  \otimes f_{1} +
 x^{-s_{1}}\fs_{1} \otimes  1 \right) 
 \left((\rho^{(1)}_{x} \otimes 1 )
 {\mathcal R}\right)
 = \left((\rho^{(1)}_{x} \otimes 1 )
 {\mathcal R}\right)
 \left( x^{-s_{1}}\fs_{1} \otimes q^{h_{1}} \right),
 \end{align}
 where \eqref{rho1} for ${\mathcal B}_{+} $ is applied to the first component of 
 the tensor product in 
 ${\mathcal R} \in {\mathcal B}_{+} \otimes  {\mathcal B}_{-}  $, 
 and \eqref{limit1} is used. Taking note on \eqref{copro-cont} and \eqref{rho-}, 
 we arrive at \eqref{ituq1} for $i=1$, $a=f_{1} \in \widetilde{U}_{q}(\widehat{sl}(2; \{0\}))$. 
 The other relations in \eqref{ituq1}-\eqref{ituq4} can be derived in a similar manner.
Evaluating these relations \eqref{ituq1}-\eqref{ituq4} for $1\otimes \pi_{1}(1)$, 
one can derive the intertwining relations for the L-operators \eqref{LQ1}-\eqref{LhQ2} and 
\eqref{LQ1-an}-\eqref{LhQ2-an}.
\section{Contraction of the augmented $q-$Onsager algebra}
\label{ApC}

Let $I$ be a subset of the set $\{0,1\}$, and define $\theta(\text{True})=1, \theta(\text{False})=0$. 
The contracted augmented $q-$Onsager algebra - denoted below $\widetilde{\cal O}(I)_q^{aug}$ -
 is generated by four generators ${\textsf K}_0, {\textsf K}_1,{\textsf Z}_1,\tilde{\textsf Z}_1$ subject to the defining relations:
\begin{align}
\begin{split}
[ {\textsf K}_0, {\textsf K}_1]&=0\ ,\\
 {\textsf K}_0{\textsf Z}_1&=
  q^{-2} {\textsf Z}_1{\textsf K}_0\ ,\qquad {\textsf K}_0\tilde{\textsf Z}_1=q^{2}\tilde{\textsf Z}_1{\textsf K}_0\ ,
\\
{\textsf K}_1{\textsf Z}_1&=
 q^{2}{\textsf Z}_1{\textsf K}_1\ ,\ \ 
 \qquad{\textsf K}_1\tilde{\textsf Z}_1=q^{-2}\tilde{\textsf Z}_1{\textsf K}_1\ ,
\\
\big[{\textsf Z}_1,\big[{\textsf Z}_1,\big[{\textsf Z}_1,\tilde{\textsf
Z}_1\big]_{q^{2}}\big]_{q^{-2}}\big]&=
\rho_{\mathrm{diag}}{\textsf Z}_1(\theta(1 \in I)\,{\textsf K}_1{\textsf K}_1-
\theta(0 \in I)
\,{\textsf K}_0{\textsf K}_0){\textsf Z}_1,
\\
\big[\tilde{\textsf Z}_1,\big[\tilde{\textsf Z}_1,\big[\tilde{\textsf Z}_1,{\textsf
Z}_1\big]_{q^{2}} \big]_{q^{-2}}\big]&=
  \rho_{\mathrm{diag}}\tilde{\textsf Z}_1(\theta(0 \in I) {\textsf K}_0{\textsf K}_0
  - \theta(1 \in I) {\textsf K}_1{\textsf K}_1)\tilde{\textsf Z}_1\ 
 \end{split}
 \label{contractedAO}
\end{align}
with
\beqa
\rho_{\mathrm{diag}}=\frac{(q^3-q^{-3})(q^2-q^{-2})^3}{q-q^{-1}}\  .\label{rhodiag2}
\eeqa
Note that $\widetilde{\cal O}(I)_q^{aug}$ coincides with ${\cal O}_q^{aug}$ for $I=\{0,1\}$.
This algebra can be embedded into coideal subalgebras of $U_{q}(\widehat{sl}_{2})$. 
We introduce the smaller contracted augmented q-Onsager algebra  ${\cal O}(I)_q^{aug}$ 
for $I=\{0\}, \{1\} $ by 
imposing the following additional relations on the the algebras $\widetilde{\cal O}(I)_q^{aug}$:  
\begin{align}
\begin{split}
& \big[{\textsf Z}_1,\big[{\textsf Z}_1,\tilde{\textsf
Z}_1\big]_{q^{2}}\big]=
\overline{\rho}_{\mathrm{diag}} \, q^{2}
{\textsf K}_0{\textsf K}_0 {\textsf Z}_1,
\quad  
\big[\tilde{\textsf Z}_1,\big[\tilde{\textsf Z}_1,{\textsf
Z}_1\big]_{q^{-2}}\big]= 
  \overline{\rho}_{\mathrm{diag}} \, \tilde{\textsf Z}_1 {\textsf K}_0{\textsf K}_0 
  \quad \text{for} \ I=\{0\},
\\[6pt]
& \big[{\textsf Z}_1,\big[{\textsf Z}_1,\tilde{\textsf
Z}_1\big]_{q^{-2}}\big]=
\overline{\rho}_{\mathrm{diag}} 
\, {\textsf Z}_1 {\textsf K}_1{\textsf K}_1 ,
\quad  
\big[\tilde{\textsf Z}_1,\big[\tilde{\textsf Z}_1,{\textsf
Z}_1\big]_{q^{2}}\big]= 
  \overline{\rho}_{\mathrm{diag}} \, q^{2}  {\textsf K}_1{\textsf K}_1 \tilde{\textsf Z}_1 
  \quad \text{for} \ I=\{1 \},
 \end{split}
 \label{contractedAOs}
\end{align}
with  
\begin{align}
\overline{\rho}_{\mathrm{diag}}=- \frac{q(q^2-q^{-2})^3}{q-q^{-1}}\  .\label{rhodiag3}
\end{align}
Note that the last two relations in \eqref{contractedAO} automatically hold
\footnote{One has to take care the relations of the form: 
$[A,[A,[A,B]_{q^2}]_{q^{-2}}]=[A,[A,[A,B]_{q^2}]]_{q^{-2}}=
A^{3}B-(q^{2}+1+q^{-2})A^{2}BA+(q^{2}+1+q^{-2})ABA^{2}-BA^{3}$, 
which are symmetric with respect to $q \leftrightarrow q^{-1}$.}
true under \eqref{contractedAOs}. 
This algebra can be embedded into
 $U_{q}(\widehat{sl}(2;I))$. 
Below, we will introduce four different realizations  of the algebra ${\cal O}(I)_q^{aug}$.  
\subsection{The first realization
}
In this subsection,  the contracted augmented $q$-Onsager algebra is embedded into 
$U_q(\widehat{sl}(2;I))$. 
We consider two types of realization of the contracted 
augmented $q-$Onsager algebra ${\cal O}_q^{aug}(I)$, 
as a 
subalgebra of $U_q(\widehat{sl}(2;I))$. 
The realizations of ${\cal O}_q^{aug}(\{0\})$  in terms of $U_q(\widehat{sl}(2;\{0\}))$ are given by 
\begin{align}
\begin{split}
{\textsf K}^{(1,-)}_0&= \epsilon_+q^{-h_0}  \ ,\qquad {\textsf K}^{(1,-)}_1= \epsilon_- q^{-h_1} \ , \\
{\textsf Z}^{(1,-)}_1&=      (q^2-q^{-2})\big(  \epsilon_+ e_0\big) \ ,
\\
\tilde{\textsf Z}^{(1,-)}_1&= (q^2-q^{-2})\big( \epsilon_- e_1  + \epsilon_+ qf_0 q^{-h_0}\big) \ ,
 \end{split}
 \label{realopaug1-}
\\[6pt]
\begin{split}
{\textsf K}^{(1,+)}_0&= \epsilon_+q^{-h_0}  \ ,\qquad {\textsf K}^{(1,+)}_1= \epsilon_- q^{-h_1} \ , \\
{\textsf Z}^{(1,+)}_1&=      (q^2-q^{-2})\big( \epsilon_-q f_1q^{-h_1}  + \epsilon_+ e_0\big) \ ,
\\
\tilde{\textsf Z}^{(1,+)}_1&= (q^2-q^{-2})\big( \epsilon_+ qf_0 q^{-h_0}\big) \ , 
 \end{split}
 \label{realopaug1+}
 \end{align}
 and the realizations of ${\cal O}_q^{aug}(\{1\})$  
 in terms of $U_q(\widehat{sl}(2;\{1\}))$ are given by 
\begin{align}
\begin{split}
{\textsf K}^{(2,-)}_0&= \epsilon_+q^{-h_0}  \ ,\qquad {\textsf K}^{(2,-)}_1= \epsilon_- q^{-h_1} \ , \\
{\textsf Z}^{(2,-)}_1&=      (q^2-q^{-2})\big( \epsilon_-q f_1q^{-h_1}  + \epsilon_+ e_0\big) \ ,
\\
\tilde{\textsf Z}^{(2,-)}_1&= (q^2-q^{-2})\big( \epsilon_- e_1 \big) \ ,
 \end{split}
 \label{realopaug2-}
\\[6pt]
\begin{split}
{\textsf K}^{(2,+)}_0&= \epsilon_+q^{-h_0}  \ ,\qquad {\textsf K}^{(2,+)}_1= \epsilon_- q^{-h_1} \ , \\
{\textsf Z}^{(2,+)}_1&=      (q^2-q^{-2})\big( \epsilon_-q f_1q^{-h_1} \big) \ ,
\\
\tilde{\textsf Z}^{(2,+)}_1&= (q^2-q^{-2})\big( \epsilon_- e_1  + \epsilon_+ qf_0 q^{-h_0}\big) \ .
 \end{split}
 \label{realopaug2+}
\end{align}
Here we attach symbols $(1,+),(1,-),(2,+),(2,-)$ on the generators
 to distinguish different realizations of the algebras. 
We remark that $\widetilde{\cal O}_q^{aug}(I)$ 
is  realized by \eqref{realopaug1-}-\eqref{realopaug2+} even if $\{e_{0},e_{1},f_{0},f_{1},h_{0},h_{1}\}$
are generators of $U_q(\widehat{sl}(2))$,  
while ${\cal O}_q^{aug}(I)$ 
is  realized only by the generators of $U_q(\widehat{sl}(2;I))$.

Limits of the renormalized generators of the augmented $q$-Onsager algebra in Verma modules 
are related to the images of the contracted augmented $q$-Onsager algebra under 
the maps \eqref{rho+}-\eqref{rho-} as follows:
\begin{align}
& \lim_{q^{-\mu}\to 0}\pi^{+}_{\mu}(xq^{\frac{\mu}{s}})({\textsf Z}_1 q^{-\frac{2s_{0}\mu}{s}})
=\rho^{(1)}_{x}({\textsf Z}^{(1,-)}_1), 
\quad
\lim_{q^{-\mu}\to 0}\pi^{+}_{\mu}(xq^{\frac{\mu}{s}})(\tilde{\textsf Z}_1 q^{-\frac{2s_{1}\mu}{s}})
=\rho^{(1)}_{x}(\tilde{\textsf Z}^{(1,-)}_1),  
 \label{Ons-rho1}
\\[6pt]
& \lim_{q^{-\mu}\to 0}\pi^{+}_{\mu}(xq^{-\frac{\mu}{s}})({\textsf Z}_1 )
=\rho^{(1)}_{x}({\textsf Z}^{(1,+)}_1), 
\quad
\lim_{q^{-\mu}\to 0}\pi^{+}_{\mu}(xq^{-\frac{\mu}{s}})(\tilde{\textsf Z}_1 q^{-2\mu})
=\rho^{(1)}_{x}(\tilde{\textsf Z}^{(1,+)}_1), 
\\[6pt]
& \lim_{q^{-\mu}\to 0}\pi^{+}_{\mu}(xq^{\pm \frac{\mu}{s}})(\tau_{-\mu }({\textsf K}_a ))
=\rho^{(1)}_{x}({\textsf K}^{(1,\mp)}_a), \quad a=0,1,
 \label{Ons-rho3}
\\[6pt]
& \lim_{q^{\mu}\to 0}\pi^{+}_{\mu}(xq^{-\frac{\mu}{s}})({\textsf Z}_1 q^{\frac{2s_{0}\mu}{s}})
=\rho^{(2)}_{x}({\textsf Z}^{(2,-)}_1), 
\quad
\lim_{q^{\mu}\to 0}\pi^{+}_{\mu}(xq^{-\frac{\mu}{s}})(\tilde{\textsf Z}_1 q^{\frac{2s_{1}\mu}{s}})
=\rho^{(2)}_{x}(\tilde{\textsf Z}^{(2,-)}_1), 
\\[6pt]
& \lim_{q^{\mu}\to 0}\pi^{+}_{\mu}(xq^{\frac{\mu}{s}})({\textsf Z}_1 q^{2\mu})
=\rho^{(2)}_{x}({\textsf Z}^{(2,+)}_1), 
\quad
\lim_{q^{\mu}\to 0}\pi^{+}_{\mu}(xq^{\frac{\mu}{s}})(\tilde{\textsf Z}_1 )
=\rho^{(2)}_{x}(\tilde{\textsf Z}^{(2,+)}_1), 
\\[6pt]
& \lim_{q^{\mu}\to 0}\pi^{+}_{\mu}(xq^{\pm \frac{\mu}{s}})(\tau_{-\mu }({\textsf K}_a ))
=\rho^{(2)}_{x}({\textsf K}^{(2,\pm)}_a), \quad a=0,1.
\end{align}
Based on these relations and the commutation relations \eqref{Tauggen}
, one can show
\footnote{For example, multiplying 
the equation in the 
4-th line in \eqref{Tauggen} by $q^{-2\mu-\frac{4s_{0}\mu}{s}}$, one obtains 
$\big[{\textsf Z}_1 q^{-\frac{2s_{0}\mu}{s}},
\big[{\textsf Z}_1 q^{-\frac{2s_{0}\mu}{s}},
\big[{\textsf Z}_1 q^{-\frac{2s_{0}\mu}{s}},
\tilde{\textsf Z}_1 q^{-\frac{2s_{1}\mu}{s}} \big]_{q^{2}}\big]_{q^{-2}}\big]=
\rho_{\mathrm{diag}}{\textsf Z}_1 q^{-\frac{2s_{0}\mu}{s}} 
(\,({\textsf K}_1q^{\mu})^{2}  q^{-4 \mu} -
\,({\textsf K}_0 q^{-\mu})^{2} )
{\textsf Z}_1 q^{-\frac{2s_{0}\mu}{s}}$. 
Then take the limit $q^{-\mu } \to 0$ 
in the representation $\pi^{+}_{\mu}(xq^{\frac{\mu}{s}})$. 
Thanks to \eqref{Ons-rho1} and \eqref{Ons-rho3}, 
one finds that the equation in the 4-th line in \eqref{contractedAO} 
for $I=\{0\}$ is satisfied under the map $\rho^{(1)}_{x}$.
 The other relations in \eqref{contractedAO} 
 can be checked in the same way.
}
 that the contracted commutation 
relations \eqref{contractedAO} hold 
 under the maps \eqref{rho+}-\eqref{rho-}. 
The limit of  the intertwining relations associated with the first realization of the 
augmented $q$-Onsager algebra given in the main text 
can now be compactly summarized, 
 in terms of  the contracted augmented $q$-Onsager algebra, as 
\begin{multline}
\rho^{(i)}_{x^{-1}}(a^{(i,+)}) \Kf^{(i)}(x) = 
 \Kf^{(i)}(x) \rho^{(i)}_{x}(a^{(i,-)})  ,
 \qquad 
\rho^{(i)}_{x^{-1}}(a^{(i,-)}) \Kfc^{(i)}(x) = 
 \Kfc^{(i)}(x) \rho^{(i)}_{x}(a^{(i,+)}) 
 \\
 \text{for any } 
 a \in 
 \{ \textsf{K}_{0},\textsf{K}_{1} ,\textsf{Z}_{1},\tilde{\textsf{Z}}_{1} \}. 
 \label{intertaocon0-1}
\end{multline}

\subsection{The second realization
}
In this subsection,  the contracted augmented $q$-Onsager algebra is also embedded into $U_q(\widehat{sl}(2;I))$. 
We consider two types of realization of the contracted 
augmented $q-$Onsager algebra ${\cal O}_q^{aug}(\overline{I})$ 
with $\overline{I}=\{0,1\} \setminus I$, 
as 
subalgebra of $U_q(\widehat{sl}(2;I))$. 
%
The realizations of ${\cal O}_q^{aug}(\{1\})$ 
in terms of $U_q(\widehat{sl}(2;\{0\}))$ are given by 
\begin{align}
\begin{split}
\overline{\textsf K}^{(1,-)}_0&= \tau({\textsf K}^{(2,-)}_0)= \overline{\epsilon}_-q^{h_0}  \ ,
\qquad \overline{\textsf K}^{(1,-)}_1= \tau({\textsf K}^{(2,-)}_1)  =\overline{\epsilon}_+ q^{h_1} ,
\\
\overline{\textsf Z}^{(1,-)}_1&= \tau({\textsf Z}^{(2,-)}_1) = (q^2-q^{-2})\big( \overline{\epsilon}_+ q e_1q^{h_1} + \overline{\epsilon}_- f_0 \big)  ,
\\
\tilde{\overline{\textsf Z}}^{(1,-)}_1&=  \tau(\tilde{\textsf Z}^{(2,-)}_1)  =  (q^2-q^{-2})\big( \overline{\epsilon}_+ f_1 \big)
 \  , 
 \end{split}
 \label{realopaug1-du}
\\[6pt]
\begin{split}
\overline{\textsf K}^{(1,+)}_0&= \tau({\textsf K}^{(2,+)}_0)= \overline{\epsilon}_-q^{h_0}  \ ,
\qquad \overline{\textsf K}^{(1,+)}_1= \tau({\textsf K}^{(2,+)}_1)  =\overline{\epsilon}_+ q^{h_1} ,
\\
\overline{\textsf Z}^{(1,+)}_1&= \tau({\textsf Z}^{(2,+)}_1) = (q^2-q^{-2})\big( \overline{\epsilon}_+ q e_1q^{h_1} \big)  ,
\\
\tilde{\overline{\textsf Z}}^{(1,+)}_1&=  \tau(\tilde{\textsf Z}^{(2,+)}_1)  =  (q^2-q^{-2})\big( \overline{\epsilon}_+ f_1 + \overline{\epsilon}_-q e_0q^{h_0} \big)
 \  , 
 \end{split}
 \label{realopaug1+du}
\end{align}
where $\tau(\epsilon_{\pm})=\overline{\epsilon}_{\mp}$ is assumed. 
 The realizations of ${\cal O}_q^{aug}(\{0\})$  
 in terms of $U_q(\widehat{sl}(2;\{1\}))$ are given by 
\begin{align}
\begin{split}
\overline{\textsf K}^{(2,-)}_0&= \tau({\textsf K}^{(1,-)}_0)= \overline{\epsilon}_-q^{h_0}  \ ,
\qquad \overline{\textsf K}^{(2,-)}_1= \tau({\textsf K}^{(1,-)}_1)  =\overline{\epsilon}_+ q^{h_1} ,
\\
\overline{\textsf Z}^{(2,-)}_1&= \tau({\textsf Z}^{(1,-)}_1) = (q^2-q^{-2})\big(  \overline{\epsilon}_- f_0 \big)  ,
\\
\tilde{\overline{\textsf Z}}^{(2,-)}_1&=  \tau(\tilde{\textsf Z}^{(1,-)}_1)  =  (q^2-q^{-2})\big( \overline{\epsilon}_+ f_1 + \overline{\epsilon}_-q e_0q^{h_0} \big)
 \  , 
 \end{split}
 \label{realopaug2-du}
\\[6pt]
\begin{split}
\overline{\textsf K}^{(2,+)}_0&= \tau({\textsf K}^{(1,+)}_0)= \overline{\epsilon}_-q^{h_0}  \ ,
\qquad \overline{\textsf K}^{(2,+)}_1= \tau({\textsf K}^{(1,+)}_1)  =\overline{\epsilon}_+ q^{h_1} ,
\\
\overline{\textsf Z}^{(2,+)}_1&= \tau({\textsf Z}^{(1,+)}_1) = (q^2-q^{-2})\big( \overline{\epsilon}_+ q e_1q^{h_1} + \overline{\epsilon}_- f_0 \big)  ,
\\
\tilde{\overline{\textsf Z}}^{(2,+)}_1&=  \tau(\tilde{\textsf Z}^{(1,+)}_1)  =  (q^2-q^{-2})\big(  \overline{\epsilon}_-q e_0q^{h_0} \big)
 \  .
 \end{split}
 \label{realopaug2+du}
\end{align}
We remark that $\widetilde{\cal O}_q^{aug}(\overline{I})$ 
is  realized by \eqref{realopaug1-du}-\eqref{realopaug2+du} even if $\{e_{0},e_{1},f_{0},f_{1},h_{0},h_{1}\}$
are generators of $U_q(\widehat{sl}(2))$,  
while ${\cal O}_q^{aug}(\overline{I})$ 
is  realized only by the generators of $U_q(\widehat{sl}(2;I))$.

Limits of the renormalized generators of the augmented $q$-Onsager algebra in the Verma modules 
are related to the images of the contracted augmented $q$-Onsager algebra under 
the maps \eqref{rho+}-\eqref{rho-} as follows:
\begin{align}
& \lim_{q^{-\mu}\to 0}\overline{\pi}^{+}_{\mu}(xq^{\frac{\mu}{s}})(\overline{\textsf Z}_1 q^{-\frac{2s_{1}\mu}{s}})
=\overline{\rho}^{(1)}_{x}(\overline{\textsf Z}^{(1,-)}_1), 
\quad
\lim_{q^{-\mu}\to 0}\overline{\pi}^{+}_{\mu}(xq^{\frac{\mu}{s}})(\tilde{\overline{\textsf Z}}_1 q^{-\frac{2s_{0}\mu}{s}})
= \overline{\rho}^{(1)}_{x}(\tilde{\overline{\textsf Z}}^{(1,-)}_1), 
\\[6pt]
& \lim_{q^{-\mu}\to 0}\overline{\pi}^{+}_{\mu}(xq^{-\frac{\mu}{s}})(\overline{\textsf Z}_1  q^{-2\mu})
=\overline{\rho}^{(1)}_{x}(\overline{\textsf Z}^{(1,+)}_1), 
\quad
\lim_{q^{-\mu}\to 0}\overline{\pi}^{+}_{\mu}(xq^{-\frac{\mu}{s}})(\tilde{\overline{\textsf Z}}_1)
=\overline{\rho}^{(1)}_{x}(\tilde{\overline{\textsf Z}}^{(1,+)}_1), 
\\[6pt]
& \lim_{q^{-\mu}\to 0}\overline{\pi}^{+}_{\mu}(xq^{\pm \frac{\mu}{s}})
(\tau_{-\mu }(\overline{\textsf K}_a ))
=\overline{\rho}^{(1)}_{x}(\overline{\textsf K}^{(1,\mp)}_a), \quad a=0,1,
\\[6pt]
& \lim_{q^{\mu}\to 0}\overline{\pi}^{+}_{\mu}(xq^{-\frac{\mu}{s}})
(\overline{\textsf Z}_1 q^{\frac{(s_{1}-s_{0})\mu}{s}})
=\overline{\rho}^{(2)}_{x}(\overline{\textsf Z}^{(2,-)}_1), 
\quad
\lim_{q^{\mu}\to 0}\overline{\pi}^{+}_{\mu}(xq^{-\frac{\mu}{s}})(\tilde{\overline{\textsf Z}}_1 q^{\frac{(s_{0}-s_{1})\mu}{s}})
=\overline{\rho}^{(2)}_{x}(\tilde{\overline{\textsf Z}}^{(2,-)}_1), 
\\[6pt]
& \lim_{q^{\mu}\to 0}\overline{\pi}^{+}_{\mu}(xq^{\frac{\mu}{s}})(\overline{\textsf Z}_1 q^{-\mu})
=\overline{\rho}^{(2)}_{x}(\overline{\textsf Z}^{(2,+)}_1), 
\quad
\lim_{q^{\mu}\to 0}\overline{\pi}^{+}_{\mu}(xq^{\frac{\mu}{s}})(\tilde{\overline{\textsf Z}}_1 q^{\mu})
=\overline{\rho}^{(2)}_{x}(\tilde{\overline{\textsf Z}}^{(2,+)}_1), 
\\[6pt]
& \lim_{q^{\mu}\to 0}\overline{\pi}^{+}_{\mu}(xq^{\pm \frac{\mu}{s}})(\tau_{-\mu }(\overline{\textsf K}_a ))
=\overline{\rho}^{(2)}_{x}(\overline{\textsf K}^{(2,\pm)}_a), \quad a=0,1,
\end{align}
 where 
$\overline{\pi}^{+}_{\mu}(x)=\pi^{+}_{\mu}(x)|_{(s_{0},s_{1})\mapsto (-s_{1},-s_{0})}$ 
and 
$\overline{\rho}^{(i)}_{x}=\rho^{(i)}_{x}|_{(s_{0},s_{1})\mapsto (-s_{1},-s_{0})}$.\vspace{1mm}

The limit of the  intertwining relations associated with the first realization of the 
augmented $q$-Onsager algebra in the main text 
can now be compactly summarized, 
 in terms of  the contracted augmented $q$-Onsager algebra, as 
\begin{multline}
\overline{\rho}^{(i)}_{x^{-1}q^{-\frac{2}{s}}}(a^{(i,+)}) \mathsf{g}^{(i)}\Kfb^{(i)}(x)^{t} = 
 \mathsf{g}^{(i)}\Kfb^{(i)}(x)^{t} \overline{\rho}^{(i)}_{xq^{\frac{2}{s}}}(a^{(i,-)})  ,
 \\
\overline{\rho}^{(i)}_{x^{-1}q^{-\frac{2}{s}}}(a^{(i,-)}) \mathsf{g}^{(i)}\Kfcb^{(i)}(x)^{t} = 
 \mathsf{g}^{(i)}\Kfcb^{(i)}(x)^{t} \overline{\rho}^{(i)}_{xq^{\frac{2}{s}}}(a^{(i,+)}) 
  \\
 \text{for any } 
 a \in 
 \{ \overline{\textsf{K}}_{0},\overline{\textsf{K}}_{1} ,
 \overline{\textsf{Z}}_{1},\tilde{\overline{\textsf{Z}}}_{1} \},
 \label{intertaocon0-1du}
\end{multline}
where $\mathsf{g}^{(i)}=q^{\hs_{i}(s_{0}-s_{1})/s} \in \mathrm{Osc}_{i}$. 
\section{Inversion relations}
\label{ApC}
\begin{multline}
g_{1}^{-1}R(xq^{\frac{4}{s}})^{t_{1}}g_{1} (R(x)^{-1})^{t_{1}}=
(R(x)^{-1})^{t_{1}}g_{1}^{-1} R(xq^{\frac{4}{s}})^{t_{1}}g_{1}=
\\[6pt]
=
g_{2}R(xq^{\frac{4}{s}})^{t_{2}}g_{2}^{-1} (R(x)^{-1})^{t_{2}}=
(R(x)^{-1})^{t_{2}}g_{2} R(xq^{\frac{4}{s}})^{t_{2}}g_{2}^{-1}=
\frac{(x^{s}-1)(q^{4}x^{s}-1)}{(x^{s}-q^{2})(x^{s}-q^{-2})},
\\[6pt]
g_{1}=\begin{pmatrix}
 q^{\frac{s_{0}-s_{1}}{s}} & 0 \\
 0 &  q^{-\frac{s_{0}-s_{1}}{s}}
\end{pmatrix}
\otimes
\begin{pmatrix}
 1 & 0 \\
 0 &  1
\end{pmatrix}
,
\quad
g_{2}=
\begin{pmatrix}
 1 & 0 \\
 0 &  1
\end{pmatrix}
\otimes
\begin{pmatrix}
 q^{\frac{s_{0}-s_{1}}{s}} & 0 \\
 0 &  q^{-\frac{s_{0}-s_{1}}{s}}
\end{pmatrix}
.
\end{multline}
\section{Various expressions of the solutions}\label{ApC}
\label{ApD}
Up to an overall factor
\footnote{The overall factor has the form $f(H)$, where $f(x)$ is a function 
of $x\in \mathbb{C} $ with $f(x+2)=f(x)$.}, 
a formal solution of \eqref{intertao1-1} and  \eqref{intertao1-2} is given by 
\begin{align}
 \Kf(x)= \prod_{j=0}^{\infty} 
   \left(
   \epsilon_{-} x^{-s_{1}} q^{-\frac{H-1}{2}+j} + \epsilon_{+} x^{s_{0}} q^{\frac{H-1}{2}-j}
   \right)
   \left (
   \epsilon_{-} x^{s_{1}} q^{-\frac{H+1}{2}-j} + \epsilon_{+} x^{-s_{0}} q^{\frac{H+1}{2}+j}
   \right) .  
    \label{sol1}
\end{align} 
In order to make this converge, we have to rewrite this in various different form 
with different prefactors. In addition to the ones in the main text, we find 
the following expressions
%

.\begin{align}
 \Kf(x)&=  
  \left(\frac{\epsilon_{-}}{\epsilon_{+}}x^{s_{0}-s_{1}}q^{-\frac{H}{2}} \right)^{\frac{H}{2}}
   \left(-\frac{\epsilon_{+}}{\epsilon_{-}}x^{s}q^{H-1};q^{-2}
   \right)_{\infty} 
      \left(-\frac{\epsilon_{-}}{\epsilon_{+}}x^{s}q^{-H-1};q^{-2}
   \right)_{\infty}
   \qquad \text{for} \quad |q|>1, 
   \label{sol2}
   \\[6pt]
   &=  
 \left(\frac{\epsilon_{-}}{\epsilon_{+}}x^{s_{0}-s_{1}}q^{-\frac{H}{2}} \right)^{\frac{H}{2}}
   \left(-\frac{\epsilon_{+}}{\epsilon_{-}}x^{s}q^{H+1};q^{2}
   \right)_{\infty}^{-1} 
      \left(-\frac{\epsilon_{-}}{\epsilon_{+}}x^{s}q^{-H+1};q^{2}
   \right)_{\infty}^{-1} 
   \quad \text{for} \quad |q|<1, 
   \label{sol4}
\end{align} 
and 
\begin{align}
 \Kf(x)&=  
       \left(\frac{\epsilon_{+}}{\epsilon_{-}}x^{s_{0}-s_{1}}q^{\frac{H}{2}} \right)^{\frac{H}{2}}
   \left(-\frac{\epsilon_{-}}{\epsilon_{+}}x^{-s}q^{-H-1};q^{-2}
   \right)_{\infty}^{-1}  
      \left(-\frac{\epsilon_{+}}{\epsilon_{-}}x^{-s}q^{H-1};q^{-2}
   \right)_{\infty}^{-1} 
   \quad \text{for} \quad |q|>1,
   \label{sol3}
   \\[6pt]
   &=  
     \left(\frac{\epsilon_{+}}{\epsilon_{-}}x^{s_{0}-s_{1}}q^{\frac{H}{2}} \right)^{\frac{H}{2}}
   \left(-\frac{\epsilon_{-}}{\epsilon_{+}}x^{-s}q^{-H+1};q^{2}
   \right)_{\infty} 
      \left(-\frac{\epsilon_{+}}{\epsilon_{-}}x^{-s}q^{H+1};q^{2}
   \right)_{\infty}
   \qquad \text{for} \quad |q|<1.
   \label{sol5}
\end{align}
%
\section{
General scalar K-matrices
}
\label{ApE}
Let $k_\pm,\overline{k}_\pm,\epsilon_\pm,\overline{\epsilon}_\pm$ be scalars. The most general solutions of the reflection equation \eqref{refeq0} and the dual one \eqref{refeqdual0-2} are given, respectively,  by (see \cite{DeV,GZ,MN})
\begin{align}
K (x)& =
\begin{pmatrix}
x^{s_{0}}\epsilon_{+} + x^{-s_{1}} \epsilon_{-} & 
\frac{k_{+}(x^{s}-x^{-s})}{q-q^{-1}} \\
\frac{k_{-}(x^{s}-x^{-s})}{q-q^{-1}}  & 
x^{-s_{0}}\epsilon_{+} + x^{s_{1}} \epsilon_{-}
 \end{pmatrix} ,
 \label{K-mat1-genfun}
\\[6pt]
\overline{K} (x)& =
\begin{pmatrix}
qx^{s_{0}}\overline{\epsilon}_{+} + q^{-1}x^{-s_{1}} \overline{\epsilon}_{-} & 
\frac{\overline{k}_{+}(q^{2}x^{s}-q^{-2}x^{-s})}{q-q^{-1}} \\
\frac{\overline{k}_{-}(q^{2}x^{s}-q^{-2}x^{-s})}{q-q^{-1}}  & 
q^{-1}x^{-s_{0}}\overline{\epsilon}_{+} +q x^{s_{1}} \overline{\epsilon}_{-}
 \end{pmatrix} .
 \label{K-matdual1-genfun}
 \end{align}
Note that \eqref{K-matdual1-genfun} is related  to \eqref{K-mat1-genfun} via 
\begin{align}
\overline{K}(x) = K^t\left(xq^{2/s}\right)g^{-1}|_{\epsilon_{\pm} =\overline{\epsilon}_{\pm}, 
\ 
k_{\pm}q^{\mp \frac{s_{0}-s_{1}}{s}} =\overline{k}_{\mp}}.\nonumber
\end{align}
Moreover, 
 \eqref{K-mat1-genfun} and 
 \eqref{K-matdual1-genfun}  reduce to 
  \eqref{K-mat1} and \eqref{Ksoldual} 
  at $k_{\pm}=\overline{k}_{\pm}=0$, respectively. 
\section{Universal T- and Q-operators}
\label{ApG}
In this appendix, we propose several version of operators 
in terms of L-and K-operators in the main text, which are 
candidate of universal T- and Q-operators for integrable systems 
with open boundary conditions 
associated with $U_{q}(\widehat{sl}_{2})$, and 
mention merit and demerit of them. 
We only sketch our idea on the definition of them and do not discuss 
convergence of the trace, explicit rational limit, 
functional relations among T- and Q-operators, 
Bethe equations, etc, which we prefer to consider in a separate 
publication (if there is an opportunity). 

We define universal L-operators as 
\begin{align}
{\mathcal L} (x)&= 
(\mathsf{ev}_{x} \otimes 1 ) \mathcal{R}, 
\qquad 
\overline{\mathcal L} (x)= 
(\mathsf{ev}_{x} \otimes 1 ) \overline{\mathcal R}, 
\quad 
x \in {\mathbb C}, 
\end{align}
and  define universal dressed K-operators as
\footnote{As remarked in subsection 2.1, 
$\overline{\mathcal R}^{-1}={\mathcal R}_{21}^{-1}$ also satisfies the  relation \eqref{R-def}. Thus one may also define a 
universal dressed K-operator as 
${\mathcal K}(x)=\overline{\mathcal L}^{-1}(x^{-1}) (\Kf(x)\otimes 1) \overline{\mathcal L}(x)$, 
where $\overline{\mathcal L} (x)= 
(\mathsf{ev}_{x} \otimes 1 ) \overline{\mathcal R}$. 
In this case, ${\mathcal K}(x)$ 
 is an element of $U_{q}(sl_{2}) \otimes {\mathcal B}_{+}$.}
\begin{align}
{\mathcal K}(x)&={\mathcal L}(x^{-1}) (\Kf(x)\otimes 1) {\mathcal L}^{-1}(x).
 \label{dressed-UK}
 \\[6pt]
\tilde{\mathcal K}(x)&={\mathcal L}(x^{-1}) (\Kf(x)\otimes 1) \overline{\mathcal L}(x) 
 \label{dressed-UKti}
%
%
%
\end{align}
Note that \eqref{dressed-UK} is an element of $U_{q}(sl_{2}) \otimes {\mathcal B}_{-}$ and 
\eqref{dressed-UKti} is an element of $U_{q}(sl_{2}) \otimes  U_{q}(\widehat{sl}_{2})$ . 
One can show
\footnote{One has to use 
$\Rf_{12}(x,y){\mathcal L}_{13}(x){\mathcal L}_{23}(y) =
{\mathcal L}_{23}(y){\mathcal L}_{13}(x)\Rf_{12}(x,y)$ and 
$\Rf_{21}(x,y){\mathcal L}_{23}(y){\mathcal L}_{13}(x) =
{\mathcal L}_{13}(x){\mathcal L}_{23}(y)\Rf_{21}(x,y)$, which follow from \eqref{YBE}, 
and \eqref{un-refeq0}.}
 that \eqref{dressed-UK} satisfies the following dressed reflection equation 
\begin{align}
\Rf_{12}(x^{-1},y^{-1}) {\mathcal K}_{13}(x) \Rf_{21}(x,y^{-1}){\mathcal K}_{23}(y)
={\mathcal K}_{23}(y) \Rf_{12}(x^{-1},y)  {\mathcal K}_{13}(x) \Rf_{21}(x,y) ,
\label{refeq-unv}
\end{align}
under \eqref{un-refeq0}.
In contrast, we have no proof (or disproof) that \eqref{dressed-UKti} satisfies 
\begin{align}
\Rf_{12}(x^{-1},y^{-1}) \tilde{\mathcal K}_{13}(x) \Rf_{21}(x,y^{-1})\tilde{\mathcal K}_{23}(y)
=\tilde{\mathcal K}_{23}(y) \Rf_{12}(x^{-1},y)  \tilde{\mathcal K}_{13}(x) \Rf_{21}(x,y) ,
\label{refeq-unvti}
\end{align}
 even if we assume \eqref{un-refeq0} 
 since we do not have an analogue of \eqref{LL=C} 
 for the universal L-operators.
%
 %
%
%
%
Evaluating
\footnote{Here we abuse notation and use the same expression for both 
the  image of  ${\mathcal K}_{23}(y)$ for 
$1 \otimes \pi_{1}\otimes 1 $ and the original one.}
 \eqref{refeq-unv} and \eqref{refeq-unvti} for $1 \otimes \pi_{1}\otimes 1 $, we obtain the following 
dressed reflection equations for the L-operators
\begin{align}
\Lf_{12} \left(x^{-1} y \right){\mathcal K}_{13}(x) \overline{\Lf}_{12} \left( xy \right) 
{\mathcal K}_{23}(y) 
&={\mathcal K}_{23}(y)
 \Lf_{12} \left(x^{-1}y^{-1} \right) {\mathcal K}_{13}(x) \overline{\Lf}_{12} \left(x y^{-1} \right).
\label{RE-dressL}
\\[6pt]
\Lf_{12} \left(x^{-1} y \right) \tilde{\mathcal K}_{13}(x) \overline{\Lf}_{12} \left( xy \right) 
\tilde{\mathcal K}_{23}(y) 
&=\tilde{\mathcal K}_{23}(y)
 \Lf_{12} \left(x^{-1}y^{-1} \right) \tilde{\mathcal K}_{13}(x) \overline{\Lf}_{12} \left(x y^{-1} \right).
\label{RE-dressLti}
\end{align}
One can also prove 
 \eqref{RE-dressL} 
independent of \eqref{refeq-unv} since 
\eqref{RE-dressL} is a  
 dressed version of \eqref{refeq2}. 
 As for \eqref{RE-dressLti}, we can only prove the image of it for 
  (tensor product of) the fundamental representation in 
the 3rd space 
 based on the relation \eqref{LL=C}. 
We define universal T-operators  as 
\begin{align}
\mathcal{T}_{\pi}(x) & = 
(\mathrm{tr}_{\pi} \otimes 1)
\left(
(\overline{\mathbf K}(q^{-\frac{4}{s}}x^{-1}) \g^2 \otimes 1)
{\mathcal K}(x)
\right) , 
\label{T-op-fin?}
\\[6pt]
\tilde{\mathcal T}_{\pi}(x) & = 
(\mathrm{tr}_{\pi} \otimes 1)
\left(
(\overline{\mathbf K}(q^{-\frac{4}{s}}x^{-1}) \g^2 \otimes 1)
\tilde{\mathcal K}(x)
\right) , 
\label{T-op-fin?ti}
%
%
\end{align}
where $\pi $ is any representation of $U_{q}(sl_{2})$ for which 
the trace converges. 
Thanks to the relation \eqref{quotient}, the 
T-operators for a finite dimensional representation are expressed in terms of those for Verma modules: 
\begin{align}
\mathcal{T}_{\pi_{\mu}}(x)&= \mathcal{T}_{\pi^{+}_{\mu}}(x) - \mathcal{T}_{\pi^{+}_{-\mu-2}}(x) 
\quad \text{for} \ \mu \in \mathbb{Z}_{\ge 0}, 
\\[6pt]
\tilde{\mathcal T}_{\pi_{\mu}}(x)&= \tilde{\mathcal T}_{\pi^{+}_{\mu}}(x) - \tilde{\mathcal T}_{\pi^{+}_{-\mu-2}}(x) 
\quad \text{for} \ \mu \in \mathbb{Z}_{\ge 0}.
\end{align}
In a similar way as for T-operators, we define universal L-operators for Q-operators as 
\begin{align}
{\mathcal L}^{(a)} (x)&= 
(\mathsf{\rho}^{(a)}_{x} \otimes 1 ) \mathcal{R}, 
\qquad
\overline{\mathcal L}^{(a)} (x)= 
(\mathsf{\rho}^{(a)}_{x} \otimes 1 ) \overline{\mathcal R}, 
\quad
x \in {\mathbb C}, \quad a=1,2,
\end{align}
and  universal dressed K-operators as
\begin{align}
{\mathcal K}^{(a)}(x)&={\mathcal L}^{(a)}(x^{-1}) (\Kf^{(a)}(x)\otimes 1) 
{{\mathcal L}^{(a)}(x)}^{-1}.
 \label{dressed-UK-Q}
 \\[6pt]
\tilde{\mathcal K}^{(a)}(x)&={\mathcal L}^{(a)}(x^{-1}) (\Kf^{(a)}(x)\otimes 1) 
\overline{\mathcal L}^{(a)}(x), \quad a=1,2.
 \label{dressed-UK-Qti}
\end{align}
One can prove that \eqref{dressed-UK-Q} satisfy dressed reflection equations 
\begin{align}
\Lf^{(a)}_{12} \left(x^{-1} y \right){\mathcal K}^{(a)}_{13}(x) 
\overline{\Lf}^{(a)}_{12} \left( xy \right) 
{\mathcal K}_{23}(y) 
={\mathcal K}_{23}(y)
 \Lf^{(a)}_{12} \left(x^{-1}y^{-1} \right) {\mathcal K}^{(a)}_{13}(x) 
 \overline{\Lf}^{(a)}_{12} \left(x y^{-1} \right),
 \quad a=1,2.
\label{RE-dressL-Q}
\end{align}
In contrast, we have no proof (or disproof) 
that \eqref{dressed-UK-Qti} satisfy
\begin{align}
\Lf^{(a)}_{12} \left(x^{-1} y \right)\tilde{\mathcal K}^{(a)}_{13}(x) 
\overline{\Lf}^{(a)}_{12} \left( xy \right) 
\tilde{\mathcal K}_{23}(y) 
=\tilde{\mathcal K}_{23}(y)
 \Lf^{(a)}_{12} \left(x^{-1}y^{-1} \right) \tilde{\mathcal K}^{(a)}_{13}(x) 
 \overline{\Lf}^{(a)}_{12} \left(x y^{-1} \right),
 \quad a=1,2.
\label{RE-dressL-Qti}
\end{align}
As 
\eqref{RE-dressL-Qti} are limit of \eqref{RE-dressLti}, 
 we can prove \eqref{RE-dressL-Qti} 
only for (tensor product of) the fundamental representation in 
the 3rd space at the moment. 
We also define 
universal Q-operators
\footnote{Another possible definition will be
$\check{\mathcal Q}^{(a)} (x) = 
(\mathrm{tr}_{W_{a}} \otimes 1)
\left(
(\overline{\mathbf K}^{(a)}(q^{-\frac{4}{s}}x^{-1}) (\g^{(a)})^2 \otimes 1)
\check{\mathcal K}^{(a)}(x) 
\right)$, 
$\check{\mathcal K}^{(a)}(x)={\check{\mathcal L}}^{(a)}(x^{-1}) (\Kfc^{(a)}(x)\otimes 1) 
{{\check{\mathcal L}}^{(a)}(x)}^{-1}$,  $a =1,2$.}
  as 
\begin{align}
{\mathcal Q}^{(a)} (x) & = 
(\mathrm{tr}_{W_{a}} \otimes 1)
\left(
(\check{\overline{\mathbf K}}^{(a)}(q^{-\frac{4}{s}}x^{-1}) (\g^{(a)})^2 \otimes 1)
{\mathcal K}^{(a)}(x) 
\right) , 
\label{Q-op}
\\[6pt]
\tilde{\mathcal Q}^{(a)} (x) & = 
(\mathrm{tr}_{W_{a}} \otimes 1)
\left(
(\check{\overline{\mathbf K}}^{(a)}(q^{-\frac{4}{s}}x^{-1}) (\g^{(a)})^2 \otimes 1)
\tilde{\mathcal K}^{(a)}(x) 
\right) 
\quad \text{for} \ a =1,2,
\label{Q-opti}
\end{align}
where $\g^{(a)}=q^{\frac{(s_{0}-s_{1})\hs_{a}}{s}}$ and 
$W_{a}$ are Fock spaces generated by $\mathrm{Osc}_{a}$. 
Note that \eqref{Q-op} are elements of ${\mathcal B}_{-}$ and 
\eqref{Q-opti} are elements of $U_{q}(\widehat{sl}_{2})$. 
We remark that \eqref{Q-opti} are limit of the universal T-operator \eqref{T-op-fin?ti}:
\begin{align}
\tilde{\mathcal Q}^{(1)} (x) & = 
\lim_{q^{-\mu} \to 0} \tilde{\mathcal T}_{\pi^{+}_{\mu}}(q^{\frac{\mu}{s}}x)q^{\mu-\mu h_{1}},
\label{Q-optil1}
\\[6pt]
\tilde{\mathcal Q}^{(2)} (x) & = 
\lim_{q^{\mu} \to 0} \tilde{\mathcal T}_{\pi^{+}_{\mu}}(q^{-\frac{\mu}{s}}x)q^{-\mu-\mu h_{1}},
\label{Q-optil2}
\end{align}
where \eqref{rho1}-\eqref{rhob2} are used. 
In contrast, we can not interpret \eqref{Q-op} as a straightforward
%
%
limit of \eqref{T-op-fin?}. 
Evaluating these for various representations of ${\mathcal B}_{-}$ or $U_{q}(\widehat{sl}_{2})$, 
we obtain a wide class of  T-and Q-operators. 
For example, T-and Q-operators acting on 
$({\mathbb C}^{2})^{\otimes L}$ are given by 
\begin{multline}
{\mathbf T}_{\pi_{\mu} }(x) = \left( \pi_{1}(\xi_{1}) \otimes \dots \otimes \pi_{1}(\xi_{L}) \right )
 \Delta^{\otimes (L-1)}  \mathcal{T}_{\pi_{\mu} }(x) ,
\\[6pt]
=
\Psi(x,\{\xi_{i}\})
(\mathrm{tr}_{\pi_{\mu}} \otimes 1^{\otimes L})
\Bigl(
\left(\overline{\mathbf K}
\left(q^{-\frac{4}{s}}x^{-1}\right) \g^{2} \otimes 1^{\otimes L}
\right) 
{\mathbf L}_{0L}\left(x^{-1}\xi_{L}^{-1}\right)
\cdots 
{\mathbf L}_{01}\left(x^{-1}\xi_{1}^{-1}\right)
\\
\times 
\left({\mathbf K}(x) \otimes 1^{\otimes L} \right)
\overline{\mathbf L}_{01}\left(x \xi_{1}^{-1}\right)
\cdots 
\overline{\mathbf L}_{0L}\left(x \xi_{L}^{-1} \right)
\Bigr), 
\label{T-op-lat}
\end{multline}
\begin{multline}
\tilde{\mathbf T}_{\pi_{\mu} }(x) = \left( \pi_{1}(\xi_{1}) \otimes \dots \otimes \pi_{1}(\xi_{L}) \right )
 \Delta^{\otimes (L-1)}  \tilde{\mathcal T}_{\pi_{\mu} }(x) ,
\\[6pt]
=
\tilde{\Psi}(x,\{\xi_{i}\})
(\mathrm{tr}_{\pi_{\mu}} \otimes 1^{\otimes L})
\Bigl(
\left(\overline{\mathbf K}
\left(q^{-\frac{4}{s}}x^{-1}\right) \g^{2} \otimes 1^{\otimes L}
\right) 
{\mathbf L}_{0L}\left(x^{-1}\xi_{L}^{-1}\right)
\cdots 
{\mathbf L}_{01}\left(x^{-1}\xi_{1}^{-1}\right)
\\
\times 
\left({\mathbf K}(x) \otimes 1^{\otimes L} \right)
\overline{\mathbf L}_{01}\left(x \xi_{1}^{-1}\right)
\cdots 
\overline{\mathbf L}_{0L}\left(x \xi_{L}^{-1} \right)
\Bigr), 
\label{T-op-latti}
\end{multline}
\begin{multline}
{\mathbf Q}^{(a)}(x) = \left( \pi_{1}(\xi_{1}) \otimes \dots \otimes \pi_{1}(\xi_{L}) \right )
 \Delta^{\otimes (L-1)}  \mathcal{Q}^{(a)}(x) ,
\\[6pt]
=
\Psi^{(a)}(x,\{\xi_{i}\})
(\mathrm{tr}_{W_{a}} \otimes 1^{\otimes L})
\Bigl(
\left(\check{\overline{\mathbf K}}^{(a)}
\left(q^{-\frac{4}{s}} x^{-1} \right) (\g^{(a)})^{2} \otimes 1^{\otimes L}
\right) 
{\mathbf L}^{(a)}_{0L}\left(x^{-1} \xi_{L}^{-1} \right)
\cdots 
{\mathbf L}^{(a)}_{01}\left(x^{-1} \xi_{1}^{-1} \right)
\\
\times 
\left({\mathbf K}^{(a)}(x) \otimes 1^{\otimes L} \right)
\check{\overline{\mathbf L}}^{(a)}_{01}\left(x \xi_{1}^{-1} \right)
\cdots 
\check{\overline{\mathbf L}}^{(a)}_{0L}\left(x \xi_{L}^{-1} \right)
\Bigr), 
\quad a=1,2, 
 \label{Q-op-fun}
\end{multline}
\begin{multline}
\tilde{\mathbf Q}^{(a)}(x) = 
\left( \pi_{1}(\xi_{1}) \otimes \dots \otimes \pi_{1}(\xi_{L}) \right )
 \Delta^{\otimes (L-1)} \tilde{\mathcal{Q}}^{(a)}(x) ,
\\[6pt]
=
\tilde{\Psi}^{(a)}(x,\{\xi_{i}\})
(\mathrm{tr}_{W_{a}} \otimes 1^{\otimes L})
\Bigl(
\left(\check{\overline{\mathbf K}}^{(a)}
\left(q^{-\frac{4}{s}} x^{-1} \right) (\g^{(a)})^{2} \otimes 1^{\otimes L}
\right) 
{\mathbf L}^{(a)}_{0L}\left(x^{-1} \xi_{L}^{-1} \right)
\cdots 
{\mathbf L}^{(a)}_{01}\left(x^{-1} \xi_{1}^{-1} \right)
\\
\times 
\left({\mathbf K}^{(a)}(x) \otimes 1^{\otimes L} \right)
\overline{\mathbf L}^{(a)}_{01}\left(x \xi_{1}^{-1} \right)
\cdots 
\overline{\mathbf L}^{(a)}_{0L}\left(x \xi_{L}^{-1} \right)
\Bigr), 
\quad a=1,2, 
 \label{Q-op'-fun}
\end{multline}
where $\xi_{1},\dots ,\xi_{L} \in {\mathbb C} \setminus \{0 \}$ are inhomogeneities on the spectral parameter 
in the quantum space; 
the trace is taken over the auxiliary space (denoted as $0$); 
the relations \eqref{LL=C}, \eqref{LLcb1=c} and \eqref{LLcb2=c} are applied. 
The overall factors
\footnote{These factors cannot be taken outside of the trace if the representations are not irreducible.}
 are given by 
\begin{align}
\Psi(x,\{\xi_{i}\})&=\prod_{k=1}^{L} 
\pi_{\mu}
\left(
\frac{\phi(x \xi^{-1}_{k})}{\varphi(x \xi^{-1}_{k}) \phi(x^{-1} \xi^{-1}_{k})}
\right), 
\\
\tilde{\Psi}(x,\{\xi_{i}\})&=\prod_{k=1}^{L} 
\pi_{\mu}
\left(
\frac{1}{\phi(x^{-1} \xi^{-1}_{k}) \phi(x^{-1} \xi_{k})}
\right), 
\qquad 
\varphi(x):=q^{-1}(\lambda^2C-x^{s}-x^{-s}), 
\\[6pt]
\Psi^{(a)}(x,\{\xi_{i}\})&=\prod_{k=1}^{L} 
\left(
\frac{\phi^{(a)}(x \xi^{-1}_{k})}{\varphi^{(a)}(x \xi^{-1}_{k}) \phi^{(a)}(x^{-1} \xi^{-1}_{k})}
\right), 
\qquad 
\varphi^{(a)}(x):=q^{-1}(q^{3-2a}-x^{-s}), 
\\[6pt]
\tilde{\Psi}^{(a)}(x,\{\xi_{i}\})&=\prod_{k=1}^{L} 
\left(
\frac{1}{\phi^{(a)}(x^{-1} \xi^{-1}_{k}) \phi^{(a)}(x^{-1} \xi_{k})}
\right), 
\qquad a=1,2.
\end{align}
The T-operators \eqref{T-op-lat} and \eqref{T-op-latti} are essentially the same object 
if the representation for the auxiliary space is irreducible, while 
the corresponding Q-operators \eqref{Q-op-fun} and \eqref{Q-op'-fun}
 are substantially different each other. 
We expect that \eqref{Q-op-fun} or \eqref{Q-op'-fun} give Q-operators for the XXZ-model. 
In fact our Q-operators
\eqref{Q-op'-fun} 
reduce to Q-operators for the XXX-model similar to 
the ones in \cite{FS15} in the rational limit $q \to 1$. 
In contrast, we can not take the rational limit of \eqref{Q-op-fun} straightforwardly
\footnote{One may have to generalize \eqref{Q-op-fun} to interpolate \eqref{Q-op-fun} 
and a rational analogue of it 
(to use renormalized operators, which appear for example in 
the left hand side of \eqref{K-op1-q1}). 
This remains to be clarified.}. 
In this sense,  \eqref{Q-op'-fun} might be more promising than \eqref{Q-op-fun}. 
However, \eqref{Q-op-fun} still deserve further study as they have good properties on
 commutativity of operators. 

We expect that these operators constitute mutually commuting family of operators.
In fact, we have proven commutativity of the universal T-operators \eqref{T-op-fin?}
 for the fundamental representation 
in the auxiliary space and the universal Q-operators \eqref{Q-op}:
\begin{align}
\mathcal{T}_{\pi_{1}}(x)\mathcal{T}_{\pi_{1}}(y)& =\mathcal{T}_{\pi_{1}}(y)\mathcal{T}_{\pi_{1}}(x), 
\quad 
\mathcal{Q}^{(a)}(x)\mathcal{T}_{\pi_{1}}(y)=
\mathcal{T}_{\pi_{1}}(y)\mathcal{Q}^{(a)}(x), 
\quad 
a=1,2, 
\ 
x,y \in {\mathbb C}. 
 \label{commutativity}
\end{align}
At the moment, we do not have a proof of commutativity among T-operators 
for the generic representations in the auxiliary space and commutativity among 
Q-operators. 
As for \eqref{T-op-fin?ti} and \eqref{Q-opti}, we have proof
\footnote{The second relation also follows from limit of 
$ \tilde{\mathbf T}_{\pi^{+}_{\mu}}(x)\tilde{\mathbf T}_{\pi_{1}}(y)=
\tilde{\mathbf T}_{\pi_{1}}(y) \tilde{\mathbf T}_{\pi^{+}_{\mu}}(x)$.}
 of commutativity for 
only a particular representation in the quantum space. 
\begin{align}
\tilde{\mathbf T}_{\pi_{1}}(x)\tilde{\mathbf T}_{\pi_{1}}(y)&
 =\tilde{\mathbf T}_{\pi_{1}}(y)\tilde{\mathbf T}_{\pi_{1}}(x), 
\quad 
\tilde{\mathbf Q}^{(a)}(x)\tilde{\mathbf T}_{\pi_{1}}(y)=
\tilde{\mathbf T}_{\pi_{1}}(y)\tilde{\mathbf Q}^{(a)}(x), 
\quad 
a=1,2, 
\ 
x,y \in {\mathbb C}. 
 \label{commutativityti}
\end{align}
A proof of the second relation in \eqref{commutativity} is given as follows
\footnote{The first relation in \eqref{commutativity} can be proven 
similarly (easier than the second relation).}. 
First, we rewrite the reflection equation \eqref{refeqlim2ndch} in terms of 
$\overline{\mathbf K}^{\prime}(x)=\overline{\mathbf K}(q^{-\frac{4}{s}}x) g^2$ and 
$\check{\overline{\mathbf K}}^{(a)\prime}(x)=\check{\overline{\mathbf K}}^{(a)}(q^{-\frac{4}{s}}x) (\g^{(a)})^2$ 
as 
\begin{align}
& \Lfc^{(a)} \left(\frac{y}{x}\right) \Kfcb^{(a)\prime}_{1}(x)^{t_{1}} 
g_{2} \Lfcb^{(a)}\left( xy q^{-\frac{4}{s}}\right) g_{2}^{-1}
\overline{K}^{\prime}_{2}(y)^{t_{2}} 
=
\nonumber \\
& \hspace{100pt} =\overline{K}^{\prime}_{2}(y)^{t_{2}}
g_{2}^{-1} \Lfc^{(a)} \left(\frac{q^{\frac{4}{s}}}{xy}\right)  g_{2}
\Kfcb^{(a)\prime}_{1}(x)^{t_{1}} \Lfcb^{(a)}\left(\frac{x}{y}\right), 
\quad a=1,2.
\label{refeqlim2ndch-pr}
\end{align}
Any object we consider under the trace is a linear combination of elements of 
the form ${\mathcal J}_{m,n}=E^{m}F^{n}q^{\xi H} \in U_{q}(sl_{2})$  
(or ${\mathcal J}_{m,n}=\es_{a}^{m}\fs_{a}^{n}q^{\xi \hs_{a}} \in \mathrm{Osc}_{a}$, $a=1,2$)
for $m,n \in {\mathbb Z}_{\ge 0}$, $\xi \in {\mathbb C}$. 
In particular, only the terms for $m=n$ (${\mathcal J}:={\mathcal J}_{m,m}$) contribute to the trace. 
Then $\mathrm{Tr}{\mathcal J}^{t}=\mathrm{Tr}{\mathcal J}$ holds if 
the trace converges and the cyclicity of the trace holds. 
In the following, we assume
\footnote{This is not a trivial issue in particular for infinite dimensional representations.}
 this. 
One can check the following relations for the L-operators
\footnote{The anti-automorphism $^{t}$ defined by \eqref{t-sl2} or 
\eqref{t-osc} might not be enough for more general R-operators. 
A relation corresponding to \eqref{tt}
(similar to \eqref{Rf-auto}) may not hold true in general situation. 
However, it is enough for the L-operators discussed here.}: 
\begin{align}
\Lf(x)^{t_{1}t_{2}}&=\Lfb(x^{-1}), &
\Lfb(x)^{t_{1}t_{2}}&=\Lf(x^{-1}), 
\nonumber
\\[6pt]
\Lf^{(a)}(x)^{t_{1}t_{2}}&=\Lfb^{(a)}(x^{-1}), &
\Lfb^{(a)}(x)^{t_{1}t_{2}}&=\Lf^{(a)}(x^{-1}), 
\nonumber \\[6pt]
\Lfc^{(a)}(x)^{t_{1}t_{2}}&=\Lfcb^{(a)}(x^{-1}), &
\Lfcb^{(a)}(x)^{t_{1}t_{2}}&=\Lfc^{(a)}(x^{-1}),
\quad a=1,2. 
\label{tt}
\end{align}
Let $\varphi^{(a)}(x)=q^{-1}(q^{3-2a}-x^{-s})$,
 $\tilde{\varphi}^{(a)}(x)=q(q^{3-2a}-q^{-2}x^{-s})$ for $a=1,2$. 
 Then we use a refinement of the Sklyanin's method \cite{Skly}:  
\begin{align}
& \mathcal{Q}^{(a)}(x)\mathcal{T}_{\pi_{1}}(y)=
\mathrm{tr}_{1}
\left(
\check{\overline{\mathbf K}}^{(a)}_{1}(q^{-\frac{4}{s}}x^{-1}) (\g^{(a)}_{1})^2
{\mathcal K}^{(a)}_{13}(x) 
\right)
\mathrm{tr}_{2}
\left(
\overline{\mathbf K}_{2}(q^{-\frac{4}{s}}y^{-1}) \g_{2}^2 
{\mathcal K}_{23}(y)
\right) 
\nonumber  \\[6pt]
&=
\mathrm{tr}_{1}
\left(
\check{\overline{\mathbf K}}^{(a)\prime}_{1}(x^{-1})
{\mathcal K}^{(a)}_{13}(x) 
\right)
\mathrm{tr}_{2}
\left(
\overline{\mathbf K}^{\prime}_{2}(y^{-1}) 
{\mathcal K}_{23}(y)
\right) 
\nonumber  \\[6pt]
&=
\mathrm{tr}_{1}
\left(
\check{\overline{\mathbf K}}^{(a)\prime}_{1}(x^{-1})^{t_{1}}
{\mathcal K}^{(a)}_{13}(x)^{t_{1}}
\right)
\mathrm{tr}_{2}
\left(
\overline{\mathbf K}^{\prime}_{2}(y^{-1}) 
{\mathcal K}_{23}(y)
\right) 
\nonumber  \\[6pt]
&=
\mathrm{tr}_{12}
\left(
\check{\overline{\mathbf K}}^{(a)\prime}_{1}(x^{-1})^{t_{1}}
\overline{\mathbf K}^{\prime}_{2}(y^{-1}) 
{\mathcal K}^{(a)}_{13}(x)^{t_{1}}
{\mathcal K}_{23}(y)
\right) 
\nonumber  \\[6pt]
&=\tilde{\varphi}^{(a)}\left( x^{-1}y^{-1} q^{-\frac{4}{s}}\right)^{-1}
\mathrm{tr}_{12}
\left(
\check{\overline{\mathbf K}}^{(a)\prime}_{1}(x^{-1})^{t_{1}}
g_{2} \Lfcb_{12}^{(a)}\left( x^{-1}y^{-1} q^{-\frac{4}{s}}\right) g^{-1}_{2}
\overline{\mathbf K}^{\prime}_{2}(y^{-1}) ^{t_{2}}
\right)^{t_{2}}
\nonumber \\
& \qquad \times 
\left(
{\mathcal K}^{(a)}_{13}(x)
\Lf_{12}^{(a)}\left( x^{-1}y^{-1}\right)^{t_{1}t_{2}}
{\mathcal K}_{23}(y)
\right)^{t_{1}} 
\qquad \text{[by \eqref{LcbL1s=c}, \eqref{LcbL2s=c}]}
\nonumber  \\[6pt]
&=\tilde{\varphi}^{(a)}\left( x^{-1}y^{-1} q^{-\frac{4}{s}}\right)^{-1}
\mathrm{tr}_{12}
\left(
\check{\overline{\mathbf K}}^{(a)\prime}_{1}(x^{-1})^{t_{1}}
g_{2} \Lfcb_{12}^{(a)}\left( x^{-1}y^{-1} q^{-\frac{4}{s}}\right) g^{-1}_{2}
\overline{\mathbf K}^{\prime}_{2}(y^{-1}) ^{t_{2}}
\right)^{t_{1}t_{2}}
\nonumber \\
& \qquad \times 
\left(
{\mathcal K}^{(a)}_{13}(x)
\Lfb_{12}^{(a)}\left( xy\right)
{\mathcal K}_{23}(y)
\right) 
\qquad \text{[by \eqref{tt}]}
\nonumber  \\[6pt]
&=\varphi^{(a)}\left( x^{-1}y\right)^{-1}
\tilde{\varphi}^{(a)}\left( x^{-1}y^{-1} q^{-\frac{4}{s}}\right)^{-1}
\nonumber \\
& \qquad  \times 
\mathrm{tr}_{12}
\left(
 \Lfcb_{12}^{(a)}\left( x^{-1}y\right)^{t_{1}t_{2}}
\check{\overline{\mathbf K}}^{(a)\prime}_{1}(x^{-1})^{t_{1}}
g_{2} \Lfcb_{12}^{(a)}\left( x^{-1}y^{-1} q^{-\frac{4}{s}}\right) g^{-1}_{2}
\overline{\mathbf K}^{\prime}_{2}(y^{-1}) ^{t_{2}}
\right)^{t_{1}t_{2}}
\nonumber \\
& \qquad \times 
\left(
\Lf_{12}^{(a)}\left( x^{-1}y\right)
{\mathcal K}^{(a)}_{13}(x)
\Lfb_{12}^{(a)}\left( xy\right)
{\mathcal K}_{23}(y)
\right) 
\qquad \text{[by \eqref{LLcb1=c}, \eqref{LLcb2=c}]}
\nonumber  \\[6pt]
&=\varphi^{(a)}\left( x^{-1}y\right)^{-1}
\tilde{\varphi}^{(a)}\left( x^{-1}y^{-1} q^{-\frac{4}{s}}\right)^{-1}
\nonumber \\
& \qquad  \times 
\mathrm{tr}_{12}
\left(
 \Lfc_{12}^{(a)}\left( xy^{-1}\right)
\check{\overline{\mathbf K}}^{(a)\prime}_{1}(x^{-1})^{t_{1}}
g_{2} \Lfcb_{12}^{(a)}\left( x^{-1}y^{-1} q^{-\frac{4}{s}}\right) g^{-1}_{2}
\overline{\mathbf K}^{\prime}_{2}(y^{-1}) ^{t_{2}}
\right)^{t_{1}t_{2}}
\nonumber \\
& \qquad \times 
\left(
\Lf_{12}^{(a)}\left( x^{-1}y\right)
{\mathcal K}^{(a)}_{13}(x)
\Lfb_{12}^{(a)}\left( xy\right)
{\mathcal K}_{23}(y)
\right) 
\qquad \text{[by \eqref{tt}]}
\nonumber  \\[6pt]
&=\varphi^{(a)}\left( x^{-1}y\right)^{-1}
\tilde{\varphi}^{(a)}\left( x^{-1}y^{-1} q^{-\frac{4}{s}}\right)^{-1}
\nonumber \\
& \qquad  \times 
\mathrm{tr}_{12}
\left(
\overline{\mathbf K}^{\prime}_{2}(y^{-1}) ^{t_{2}}
g^{-1}_{2}
 \Lfc_{12}^{(a)}\left( xy q^{\frac{4}{s}}\right) 
 g_{2}
 \check{\overline{\mathbf K}}^{(a)\prime}_{1}(x^{-1})^{t_{1}}
 \Lfcb_{12}^{(a)}\left( x^{-1}y\right)
\right)^{t_{1}t_{2}}
\nonumber \\
& \qquad \times 
\left(
{\mathcal K}_{23}(y)
\Lf_{12}^{(a)}\left( x^{-1}y^{-1}\right)
{\mathcal K}^{(a)}_{13}(x)
\Lfb_{12}^{(a)}\left( xy^{-1}\right)
\right) 
\qquad \text{[by \eqref{RE-dressL-Q}, \eqref{refeqlim2ndch-pr}]}
\nonumber  \\[6pt]
&=\varphi^{(a)}\left( x^{-1}y\right)^{-1}
\tilde{\varphi}^{(a)}\left( x^{-1}y^{-1} q^{-\frac{4}{s}}\right)^{-1}
\nonumber \\
& \qquad  \times 
\mathrm{tr}_{12}
 \Lfc_{12}^{(a)}\left( xy^{-1}\right)
\left(
\overline{\mathbf K}^{\prime}_{2}(y^{-1}) ^{t_{2}}
g^{-1}_{2}
 \Lfc_{12}^{(a)}\left( xy q^{\frac{4}{s}}\right) 
 g_{2}
 \check{\overline{\mathbf K}}^{(a)\prime}_{1}(x^{-1})^{t_{1}}
\right)^{t_{1}t_{2}}
\nonumber \\
& \qquad \times 
\left(
{\mathcal K}_{23}(y)
\Lf_{12}^{(a)}\left( x^{-1}y^{-1}\right)
{\mathcal K}^{(a)}_{13}(x)
\Lfb_{12}^{(a)}\left( xy^{-1}\right)
\right) 
\qquad \text{[by \eqref{tt}]}
\nonumber \\[6pt]
&=
\tilde{\varphi}^{(a)}\left( x^{-1}y^{-1} q^{-\frac{4}{s}}\right)^{-1}
\mathrm{tr}_{12}
\left(
\overline{\mathbf K}^{\prime}_{2}(y^{-1}) ^{t_{2}}
g^{-1}_{2}
 \Lfc_{12}^{(a)}\left( xy q^{\frac{4}{s}}\right) 
 g_{2}
 \check{\overline{\mathbf K}}^{(a)\prime}_{1}(x^{-1})^{t_{1}}
\right)^{t_{2}}
\nonumber \\
& \qquad \times 
\left(
{\mathcal K}_{23}(y)
\Lf_{12}^{(a)}\left( x^{-1}y^{-1}\right)
{\mathcal K}^{(a)}_{13}(x)
\right)^{t_{1}}
\qquad \text{[by \eqref{LcLb1=c}, \eqref{LcLb2=c}]}
\nonumber \\[6pt]
&=
\tilde{\varphi}^{(a)}\left( x^{-1}y^{-1} q^{-\frac{4}{s}}\right)^{-1}
\mathrm{tr}_{12}
\Bigl(
g_{2}
 \Lfc_{12}^{(a)}\left( xy q^{\frac{4}{s}}\right)^{t_{2}}
 g^{-1}_{2}
\overline{\mathbf K}^{\prime}_{2}(y^{-1})
 \check{\overline{\mathbf K}}^{(a)\prime}_{1}(x^{-1})^{t_{1}}
\nonumber \\
& \qquad \times 
{\mathcal K}_{23}(y)
{\mathcal K}^{(a)}_{13}(x)^{t_{1}}
\Lfb_{12}^{(a)}\left( xy \right)^{t_{2}}
\Bigr)
\qquad \text{[by \eqref{tt}]}
\nonumber  \\[6pt]
&=
\mathrm{tr}_{12}
\left(
\overline{\mathbf K}^{\prime}_{2}(y^{-1})
 \check{\overline{\mathbf K}}^{(a)\prime}_{1}(x^{-1})^{t_{1}}
{\mathcal K}_{23}(y)
{\mathcal K}^{(a)}_{13}(x)^{t_{1}}
\right)
\qquad \text{[by \eqref{LbLc1s=c},  \eqref{LbLc2s=c}]}
\nonumber  \\[6pt]
&=
\mathrm{tr}_{2}
\left(
\overline{\mathbf K}^{\prime}_{2}(y^{-1})
{\mathcal K}_{23}(y)
\right)
\mathrm{tr}_{1}
\left(
 \check{\overline{\mathbf K}}^{(a)\prime}_{1}(x^{-1})^{t_{1}}
{\mathcal K}^{(a)}_{13}(x)^{t_{1}}
\right)
\nonumber \\[6pt]
&=
\mathcal{T}_{\pi_{1}}(y)\mathcal{Q}^{(a)}(x).
\end{align}

\end{document}